\documentclass[a4paper,12pt]{article}
\usepackage{HECAPdoc}

\usepackage[toc, acronym]{glossaries}
\makeglossaries

\newacronym{ghg}{GHG}{Greenhouse gas}
\newacronym{sdg}{SDG}{UN Sustainable Development Goal}
\newacronym{tco2}{tCO$_2$}{Metric tonne \CdO}
\newacronym{hl-lhc}{HL-LHC}{High-luminosity upgrade to the Large Hadron Collider at CERN, expected to be operational from 2029}
\newacronym{hpc}{HPC}{High-Performance Computing}
\newacronym{htc}{HTC}{High-Throughput Computing}
\newacronym{pue}{PUE}{Power Usage Effectiveness, a measure of the overhead energy costs of an IT facility.  It is defined as the ratio of the total power used by the facility to the energy used by the IT equipment}
\newacronym{co2e}{CO$_2$e}
{\CdO\ equivalent.  For any \acrshort{ghg}, it is the mass of CO$_2$ that would result in an equivalent amount of warming.  It is computed by multiplying the total mass of a GHG by its global warming potential, \acrshort{gwp}}
\newacronym{tco2e}{tCO$_2$e}{Metric tonne \acrshort{co2e}}
\newacronym{gwp}{GWP}{Global Warming Potential.  Factor used to quantify the warming potential due to a certain mass of \acrshort{ghg} relative to that of CO$_2$.  Defined as the ratio of the infrared radiation absorbed by a certain mass of GHG to that absorbed by the same mass of CO$_2$ over a chosen time frame, usually 20, 100 or 500 years}

\begin{document}
\bstctlcite{bstctl:nodash} 

\RaggedRight
\sloppy
\begin{titlepage}
    \thispagestyle{titlestyle}

    \doublespacing    
    \begin{flushleft}
    \textbf{\huge Environmental sustainability in basic research}\\
    \textbf{\Large A perspective from HECAP+}\\

    \bigskip

    {\textbf{Sustainable HECAP+ Initiative}}
    \end{flushleft}
    \singlespacing

    \vspace{24em}

    \noindent {\bf Abstract}\\

    \noindent The climate crisis and the degradation of the world's ecosystems require humanity to take immediate action. The international scientific community has a responsibility to limit the negative environmental impacts of basic research. The \textbf{HECAP+ communities} (\textbf{High Energy Physics, Cosmology, Astroparticle Physics, and Hadron and Nuclear Physics}) make use of common and similar experimental infrastructure, such as accelerators and observatories, and rely similarly on the processing of big data. Our communities therefore face similar challenges to improving the sustainability of our research. This document aims to reflect on the environmental impacts of our work practices and research infrastructure, to highlight best practice, to make recommendations for positive changes, and to identify the opportunities and challenges that such changes present for wider aspects of social responsibility.\\
           
    \begin{flushright}
        \textbf{Version 3.0, 21 September 2026}\\
        This is an author-prepared post-print of \href{https://doi.org/10.1088/1748-0221/20/03/P03012}{JINST {\bfseries 20} (2025) P03012}, published by IOP Publishing Ltd under the terms of the \href{http://creativecommons.org/licenses/by/4.0/}{CC BY 4.0 license}.\\
        \textbf{\textcolor{Pythongreen}{Please consider reading this document via the least environmentally impactful method available to you. Thank you.}}
    \end{flushright}

\end{titlepage}

\newpage

\setcounter{page}{2}

\thispagestyle{empty}

~

\vspace{7em}
\RaggedRight

\noindent {\bfseries Conflicts of interest}

\noindent The authors have not declared any financial, commercial or other interests in any private companies that are referenced explicitly or are otherwise referenced by implication in this document.

\noindent {\bfseries Licence}

\noindent This work is licensed under the Creative Commons Attribution 4.0 International License (CC-BY 4.0). To view a copy of this license, visit \href{http://creativecommons.org/licenses/by/4.0/}{http://creativecommons.org/licenses/by/4.0/} or send a letter to Creative Commons, PO Box 1866, Mountain View, CA 94042, USA.

\noindent {\bfseries Citing this document}

\noindent Please cite this document as:
\begin{quotation}
    SustainableHECAP+ Initiative, ``Environmental sustainability in basic research:\ A perspective from HECAP+'',
    \href{https://doi.org/10.1088/1748-0221/20/03/P03012}{JINST {\bfseries 20} (2025) P03012} [\href{https://doi.org/10.48550/arXiv.2306.02837}{arXiv:2306.02837}, also available at: \href{https://www.hecap.eco/}{https://www.hecap.eco/}].
\end{quotation}

\noindent {\bfseries Endorsement}

\noindent To register to and view the list of endorsers, please visit \href{https://indico.cern.ch/event/1293733}{https://indico.cern.ch/event/1293733/}. To make an institutional endorsement, please make direct contact. 

\newpage

\thispagestyle{plain}

~

\vspace{2em}
\RaggedRight

\noindent The acronym {\bf HECAP} was adopted in the early stages of this initiative, otherwise in common use to refer to High Energy Physics, Cosmology and Astroparticle Physics.\footnote{Examples include the HECAP Research Section of the Abdus Salam International Centre for Theoretical Physics (see \href{https://www.ictp.it/hecap}{https://www.ictp.it/hecap}); the Latin American Association for High Energy, Cosmology and Astroparticle Physics (see \href{https://www.ictp-saifr.org/laa-hecap/}{https://www.ictp-saifr.org/laa-hecap/}); by the Latin American Giant Observatory (see \href{http://lagoproject.net/lasf4ri20.html}{http://lagoproject.net/lasf4ri20.html} [link no longer active]; and in the Latin American Strategy for Research Infrastructures for High Energy, Cosmology, Astroparticle Physics (available at \href{https://arxiv.org/pdf/2104.06852.pdf}{https://arxiv.org/pdf/2104.06852.pdf}).} With the subsequent inclusion of key contributions from the \textbf{Hadron and Nuclear Physics} community, with whom HECAP share common research infrastructure and common challenges in the pursuit of improved environmental sustainability, this acronym was modified to {\bf HECAP+}. This modification is also intended to emphasise that many of the issues highlighted in this document apply broadly to all members of the basic research community.

\vspace{2em}

\noindent This document has been typeset in LaTeX using Atkinson Hyperlegible to maximise readability (see \href{https://tug.org/FontCatalogue/atkinsonhyperlegible/}{https://tug.org/FontCatalogue/atkinsonhyperlegible/}).

\noindent An HTML version of this document is available at:~\href{https://www.hecap.eco/}{https://www.hecap.eco/}.  The forewords to Version 1 of this document can be found in Appendix~\ref{sec:Forewords}.

\noindent The content of this publication has not been approved by the United Nations and does not reflect the views of the United Nations or its officials or Member States.  Details of the UN Sustainable Development Goals, referenced in this work, can be found at Refs.~\cite{UNwebsite,UNSustainableGoalsFigure}.

\newpage

{\bfseries Changes in version 3.0:}\ Prajval Shastri has been added to the authorship, and a statement on conflicts of interest has been added above. The brief statements relating to carbon offsetting have been removed from the Executive Summary and Best Practice 5.2, and a more extensive discussion has been added in Section 1.3; it has been emphasised that the move towards plant-based catering relies on the equitable availability of alternative sources of nutrition, both in the Executive Summary and the recommendations of Section 4, where the need for local audits is also identified; a paragraph highlighting climate injustice has been included in the Introduction; the energy goals for the FCC (ee and hh) reported in Section 1.2.2 were updated at the time of publication (March 2025) --- these are now outdated, and plans for the FCC continue to evolve following its prioritisation in the update to the European Strategy for Particle Physics; the additional challenges of addressing sustainability within the broader research ecosystem are now acknowledged in Section 1.3; Figure 1.4 and its caption have been updated; SESAME and Green ILC best practice insets have been added to Section 3; the discussion of UHVDC cable projects has been extended in Section 3, alongside additional emphasis on the need to avoid neo-colonialism; Figure 3.2 on mitigation potential has been replaced by the full figure as presented in the IPCC Sixth Assessment Report; a paragraph on remote shift operations has been added to Section 5; the partial LCA for the production phase of a monocrystalline silicone wafer and the associated discussion in Section 6 have been clarified; a Cool Copper Collider best practice inset has been added to Section 6; the need to eliminate premature or planned obsolence has been added in Section 7; the acknowledgements have been extended; and a number of more minor additions, clarifications, and revisions of wording have been made throughout. Finally, the Outlook has been extended beyond the revisions already made in the published version (March 2025) to emphasise the many positive steps that have and are being taken by the community to embed environmental sustainability in its research practices.

\newpage

{\bfseries Changes in version 2.0:}\ Clarifications have been added in relation to the authorship of this document on the title page and in the Statement of Intent; the inset on personal emissions and the figure reproduced in full from Ref.~[10] have been removed in favour of a text-based discussion of the elements of the study~[10] that give relevant context to research-related emissions; the text beneath the right-hand chart of Figure 1.3 has been corrected to indicate per-researcher rather than per-capita numbers, and clarifications have been added to the caption to Figure 1.4 and the accompanying text in relation to ``per-researcher'' emissions; corrections and clarifications have been made in the Computing section in relation to the High-Luminosity phase of the Large Hadron Collider; a correction was made to the left-most bar of Figure 4.1; and the paragraphs on low-carbon nuclear energy and the discussions on energy import and export via HVDC cables in Section~3.1.2 and Case Study 3.3 of the Energy section have been revised. Other minor typographical and grammatical changes have been made.

\newpage

\section*{Statement of Intent}
    
        This reflective document was developed by a small group of concerned members of the HECAP+ communities (see~\nameref{AuthorsList}), beginning as a grassroots initiative {\it Striving towards Environmental Sustainability in High Energy Physics, Cosmology and Astroparticle Physics}.
        
        Its focus is not to stipulate the research that our communities should undertake, nor to debate its intrinsic value.  Rather, it is intended to be a synthesis of current data, best practices, and research in climate science and sustainability, as applied to our fields to the best of our ability as physicists,  and a reflection on the roles that our communities can play in limiting negative environmental impacts due to our research work and scientific culture. 
        
        The scope of the document is inspired by the holistic approach of annual environmental reports of major institutes~\cite{Environment:2737239,FermilabEnvReport2019}, which include emissions directly related to research and collateral emissions, such as from personal commutes and institutional catering. Any imbalance in its content, in part, reflects imbalances in the availability of reliable data and resources relating to the environmental impact of aspects of our communities' activities.
        Redressing this imbalance will require input from across our communities, in particular to identify the technical challenges of limiting the environmental impacts of our current and future research infrastructure.

        While this document is primarily framed from the perspectives of high energy physics, cosmology, astroparticle physics, and hadron and nuclear physics (HECAP+), much of its discussion applies to basic research more generally. Its broad scope is intended to provide a first step toward greater coordination across the community in efforts to address environmental sustainability, and it is hoped that it may serve as a useful reference for our and other fields.

        \noindent \textbf{Thank you for taking the time to read this document.}

\newpage

\tableofcontents

\RaggedRight
\sloppy
\newpage


\section*{Executive Summary}
\label{sec:Executive_Summary}
\addcontentsline{toc}{section}{Executive Summary}

Humanity's impact upon the world's climate and ecosystems is now as unequivocal as it is extreme~\cite{IPCC2021reportSPM, IPBESIPCC}. Averting this catastrophe and its growing and inequitable human cost must be a critical concern for all global citizens at this pivotal time in world history.

\textbf{High Energy Physics}, \textbf{Cosmology}, \textbf{Astroparticle Physics}, and \textbf{Hadron and Nuclear Physics} (\textbf{\ACR}) research has direct impacts on the environment.  Our research infrastructure, including accelerators, detectors, telescopes and computing resources, requires enormous power generation and, in many cases, contributes directly to greenhouse gas (\acrshort{ghg}) emissions. Our work practices give rise to additional emissions, e.g., from procurement, business travel and commuting, and our industry generates various forms of waste that are harmful to the environment.

As scientists working in \ACR\ and related disciplines, our responsibilities to limit and mitigate our impact on the world's climate and ecosystems are manifold. Our opportunities and training have given us the science capital to appreciate the evidence that has been collated over many years by climate and environmental science. We must use our  unique and privileged platform as professional scientists to impel positive changes in, as well as educate and advocate on, environmental sustainability and the connected issues of social justice. Moreover, as a community focused on basic scientific research, we should be no less accountable for our impacts on the world's climate and ecosystems than any other industry, and we should anticipate that our activities will come under increasing scrutiny from the public, governments and funders. We have moral and pragmatic obligations to act.

This document assesses the environmental impacts of \ACR\ research across six areas:\ computing, energy, food, mobility, research infrastructure and technology, and resources and waste, also within the larger context of global emissions. Specific recommendations are made for each of these areas, but the overarching message is simple:
\begin{quotation}
{\bfseries Assessing, reporting on, defining targets for, and undertaking coordinated efforts to limit our negative impacts on the world's climate and ecosystems must become an integral part of how we plan and undertake all aspects of our research.}
\end{quotation}
This requires urgent action at an individual level, at a group level (including research groups, collaborations and organising committees), and at an institutional level (including universities, research institutes, funding agencies, and professional societies). Moreover, it requires systematic positive changes in everything from our day-to-day activities and the ways we interact as a global community through to the design and running of the `big science' infrastructure on which \ACR\ research depends.

We urge all members of the \ACR\ and related communities to take individual actions and push for group- and institution-level changes that:
\begin{itemize}
    \item Establish community-wide formal and coordinated efforts to assess and improve the environmental sustainability of basic research, which calls for standardised reporting and data sharing.
      \item Consider the environmental cost of computational infrastructure and algorithms in decision making, and prioritise the development of common and reusable software solutions across \ACR.
     \item Prioritise the use of sustainable and renewable energy to power our workspaces and research infrastructure; increase their energy efficiency and recovery, and energy storage capacity.
    \item Move towards plant-based catering at conferences and in cafeterias, immediately reducing the provision of carbon-intensive foods, such as ruminant meats and dairy products, where alternative sources of nutrition are equitably available.
    \item Prioritise environmentally sustainable modes of transport for commuting where possible.
    \item Prioritise responsible business travel that balances in-person and online meetings, acknowledges the benefits of virtual and hybrid meetings for inclusivity, and considers the disproportionate impact of changes to travel culture on different groups, \eg early career researchers and those who are geographically isolated.
    \item Mandate comprehensive life cycle analysis for all proposed research infrastructure projects that critically assesses the environmental impact of all project stages, including design and approval, construction, commissioning, operation, maintenance, decommissioning, and removal.
     \item Prioritise environmentally  and socially sustainable sourcing of raw materials for experiments and infrastructure.
     \item Propagate and expand the culture of ``reduce, reuse, repair, recycle'', including the implementation of life-cycle awareness and end-of-life planning for hardware.
    \item Educate and advocate on issues of environmental sustainability and social justice,  and engage more broadly with policy makers to push for wider change, \eg the improvement and decarbonisation of local transport infrastructure.
\end{itemize}

\RaggedRight
\sloppy


\section*{Outline}
\addcontentsline{toc}{section}{Outline}

The aims of this document are:
\begin{itemize}
    \item To improve awareness of the impact that high energy physics, cosmology and astroparticle physics, and hadron and nuclear physics (\ACR) has on the environment.
    \item To provide suggestions and encourage immediate action on ways that we, as a community, can play our part in limiting further degradation of the world's climate and ecosystems.
    \item To provide impetus for ongoing and collective discussions of how we can make positive changes to our community's work practices, in terms of environmental sustainability and for the issues of social justice from which climate change and environmental degradation cannot be disentangled.
\end{itemize}

The discussions are divided into seven sections. Sections 2 through 7 cover the topics of Computing, Energy, Food, Mobility, Research Infrastructure and Technology, and Resources and Waste. Each of these sections contains a set of recommendations, for individuals, groups and institutions, and these are followed by longer discussions that include case studies and best practice examples, which can be read independently of the surrounding material. 
As a consequence, a certain level of repetition is unavoidable. Collated lists of acronyms and abbreviations, best practices, case studies, and figures and tables are included at the end of this document.

\sref{sec:Introduction}, Preliminaries, begins by acknowledging the climate crisis and the environmental impacts of HECAP+ research. It provides a summary of the United Nations (UN) Sustainable Development Goals (\acrshort{sdg}s) and how these relate to \ACR\ research, and briefly reviews similar and complementary documents.


\newpage

\section{Preliminaries}
\label{sec:Introduction}

\subsection{Introduction}

The 2021 report of the Intergovernmental Panel on Climate Change (IPCC)~\cite{IPCC2021report} is emphatic in its statements about the current status of the climate and the damaging impact that humanity continues to have upon it~\cite{IPCC2021reportSPM}:
\begin{quotation}
"It is unequivocal that human influence has warmed the atmosphere, ocean and land. Widespread and rapid changes in the atmosphere, ocean, cryosphere and biosphere have occurred.\ [\dots]\ Human-induced climate change is already affecting many weather and climate extremes in every region 
across the globe. Evidence of observed changes in extremes such as heatwaves, heavy precipitation, droughts, 
and tropical cyclones, and, in particular, their attribution to human influence, has strengthened since [the Fifth Assessment Report in 2014]."
\end{quotation}
It is also clear on the consequences of further inaction~\cite{IPCC2021reportSPM}:
\begin{quotation}
"Global surface temperature will continue to increase until at least mid-century under all emissions scenarios
considered. Global warming of 1.5\degree C and 2\degree C will be exceeded during the 21$^\text{st}$ century unless deep reductions 
in $\mathrm{CO_2}$ and other greenhouse gas emissions occur in the coming decades.\ [\dots]\ Many changes in the climate system become larger in direct relation to increasing global warming. They 
include increases in the frequency and intensity of hot extremes, marine heatwaves, heavy precipitation, 
and, in some regions, agricultural and ecological droughts; an increase in the proportion of intense tropical 
cyclones; and reductions in Arctic sea ice, snow cover and permafrost."
\end{quotation}

\begin{figure}
    \centering
    \includegraphics[width=\textwidth]{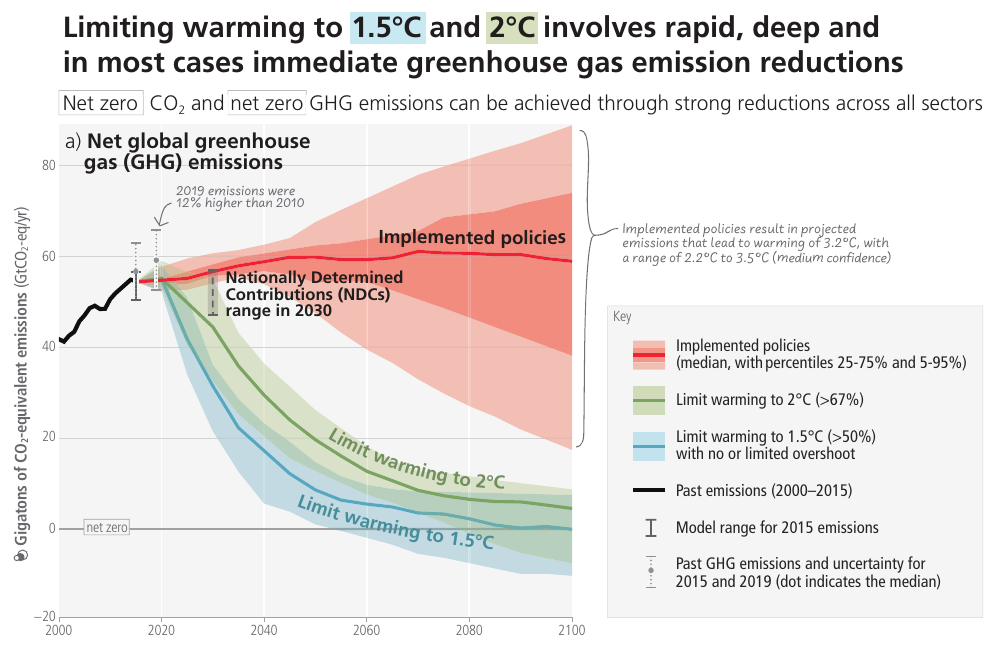}
    \caption[Necessary reduction in global emissions]%
        {Figure excerpted from the IPCC 2023 Synthesis Report, Ref.~\cite{IPCC2023SynthesisSPM}.  Original caption: Global emissions pathways consistent with implemented policies and mitigation strategies. Panel (a) [...] shows the development of global GHG [...] emissions in modelled pathways [...]. Coloured ranges denote the 5th to 95th percentile across the global modelled pathways falling within a given category as described in Box SPM~1 of Ref.~\cite{IPCC2023SynthesisSPM}. The red ranges depict emissions pathways assuming policies that were implemented by the end of 2020. Ranges of modelled pathways that limit warming to 1.5°C (>50\%) with no or limited overshoot are shown in light blue (category C1) and pathways that limit warming to 2°C (>67\%) are shown in green (category C3). Global emission pathways that would limit warming to 1.5°C (>50\%) with no or limited overshoot and also reach net zero GHG in the second half of the century do so between 2070-2075.
        \label{fig:ene-netco2}}
    \end{figure}

Net global \acrshort{co2e} emissions must at least be halved before 2030 to fulfill the Paris Climate Agreement. Without this, we are unlikely to meet the  target of limiting global warming to 1.5\degree C in order to avoid fatal tipping points in the global biosphere (see \fref{fig:ene-netco2})~\cite{IPCC19policy}. Pledged policy changes by nations party to the Paris Agreement, known as Nationally Determined Contributions, are insufficiently far-reaching, and ``make it {\it likely} that warming will exceed 1.5\degree\ during the 21st century''~\cite{IPCC2023SynthesisSPM} (original emphasis).  This is also true despite more recent legislation, such as the Inflation Reduction Act in the United States~\cite{REPEAT}.  Demand-side mitigation, including changes in infrastructure use and social and behavioural practices, can reduce global \acrshort{ghg} emissions in end-use sectors by 40--70\% by 
2050~\cite{IPCC2022reportSPM}.

The climate crisis and the degradation of the world's ecosystems are already causing immense human suffering. Moreover, those who are and will endure the worst consequences of these impacts are those who have contributed the least (see below). This climate injustice is compounded by the continued exploitation of resource rich countries (by other nations), upon whom the technological changes cited for addressing the climate crisis will only increase pressure. Together with the increasing politicisation of the climate and environment, the technological and societal changes that the world chooses to make cannot be disentangled from their social and political implications, and no less so in a world founded on the exploitation of one people by another.


\begin{figure}
    \centering
    \includegraphics[width=\textwidth]{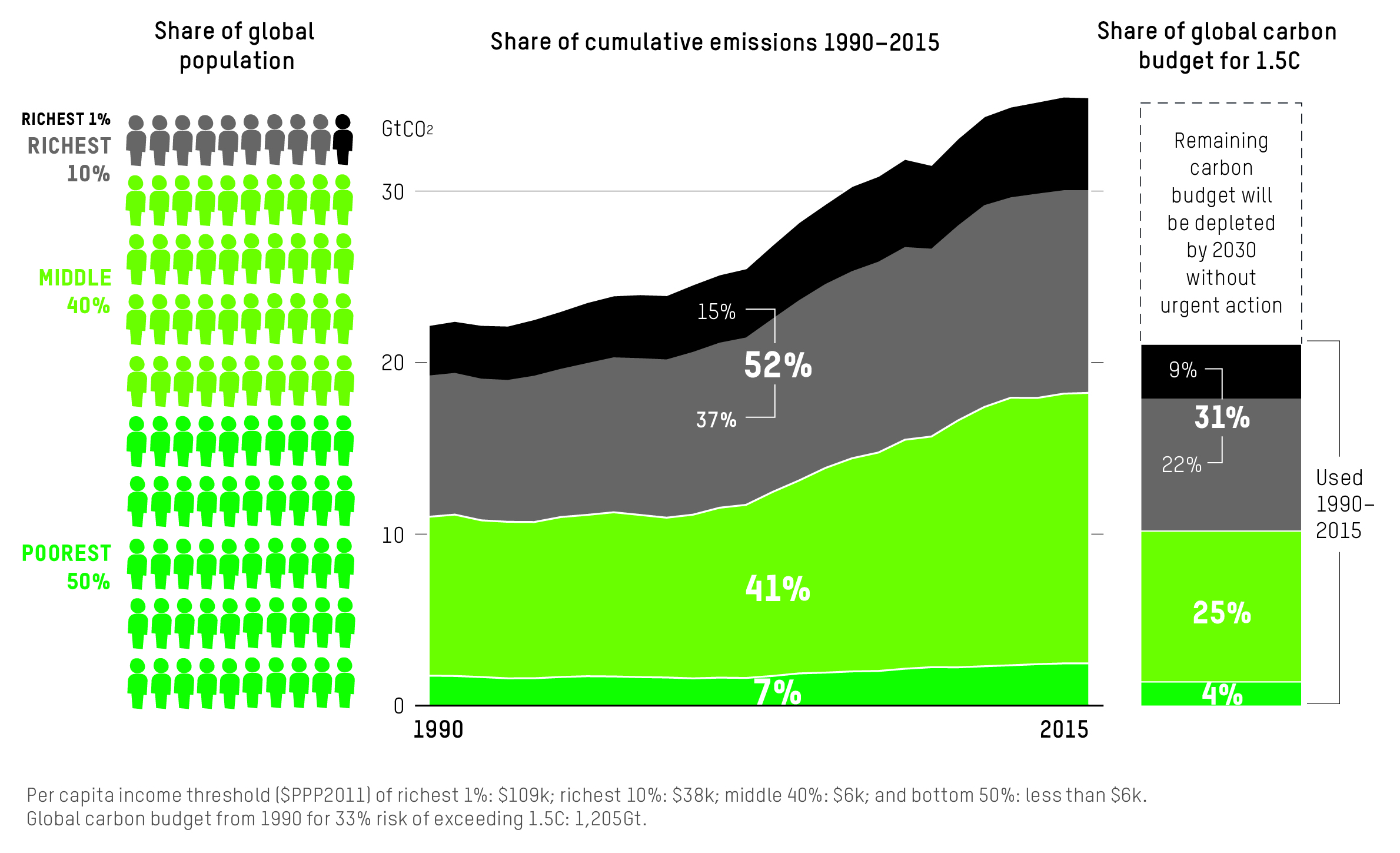}
    \caption[Global emissions (1990--2015) by income groups]{Share of cumulative emissions from 1990 to 2015 and use of the global carbon budget for 1.5\degree C linked to consumption by different global income groups. Figure reproduced from Ref.~\cite{Oxfam2020} with the permission of Oxfam.\footnote{Oxfam House, John Smith Drive, Cowley, Oxford OX4 2JY, UK, \url{https://www.oxfam.org.uk/}.  Oxfam does not necessarily endorse any text or activities that accompany the materials.}}
    \label{fig:OxfamFig}
\end{figure}

Oxfam's recent study "Confronting carbon inequality"~\cite{Oxfam2020} notes a strong correlation between GHG emissions and income level. While the use of a single global income scale obscures the impact of high emitters in low- and middle-income countries, which can also be significant~\cite{BHAR2024108104}, the study concludes that those earning an average annual income of over \euro{34,000}, placing them in the top 10\% of earners globally, account for over half of cumulative global emissions (see \fref{fig:OxfamFig}). This income bracket was identified by the IPCC as having ``the greatest potential for emissions reductions, \eg as citizens, investors, consumers, role models, and professionals''~\cite{IPCC2022reportSPM}, and includes many \ACR\ physicists (notwithstanding significant disparities in income within our communities depending on location and career stage).  A recent meta-study from Lund University~\cite{Wynes_2017} concluded that the most impactful individual climate actions include: living car free (country-averaged range of 1.4 -- 3 \acrshort{tco2e} reduction per year); avoiding one transatlantic flight (1.6 \tCdOe); purchasing green energy (country-averaged range of < 0.1 -- 2.5 \tCdOe\ reduction per year); and eating a plant based diet (country-averaged range of 0.4 -- 1 \tCdOe\ reduction per year). Note that the numbers quoted were based on “average conditions in developed countries”, so miss the substantial differences between emissions levels in developing and developed nations. They also neglect the effects of future climate policy. Even so, they give indicative scales that place HECAP+ researchers' work-related emissions in context, cf.~Figure~\ref{fig:Intro-ComparativeEmissions}. The study also highlights the disconnect between the low- and moderate-impact measures that consumers are commonly encouraged to take, and the high-impact measures that require more significant lifestyle changes and pertain to more complex and nuanced issues.

\begin{figure}[!tb]
    \centering
    \includegraphics[width=0.98\linewidth]{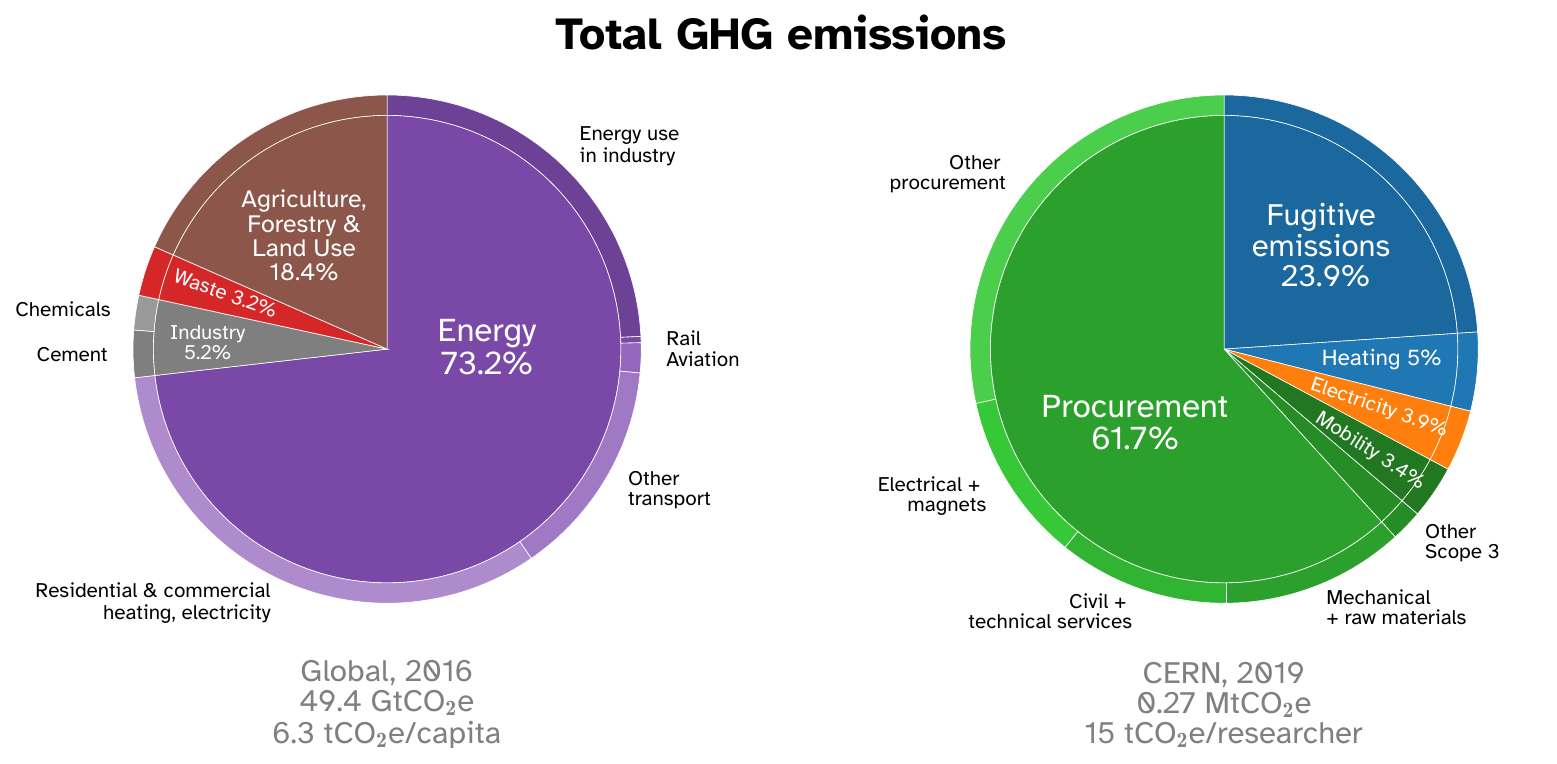}
    \caption[2016 global GHG emissions vs 2019 CERN emissions]%
        {Distribution of 2016 global GHG emissions by sector, compared with CERN emissions for 2019, during LHC shutdown. Data are taken from Ref.~\cite{owidco2andothergreenhousegasemissions} and CERN Environmental Reports~\cite{Environment:2737239,CERN:2723123,CERNTownHall}. \label{fig:intro-GlobalEmissions}}
    \end{figure}

\begin{figure}[!tb]
    \centering
    \includegraphics[width=\textwidth]{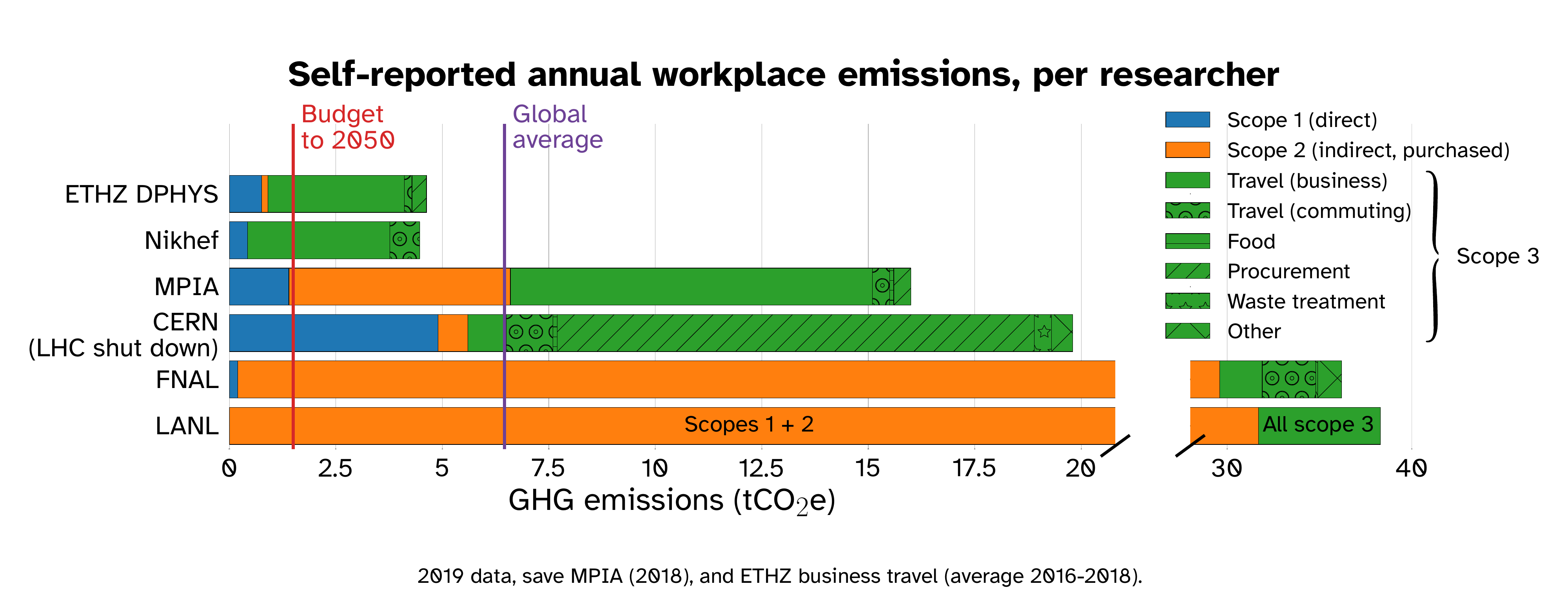}
    \caption[Reported workplace GHG emissions for researchers at \ACR\ institutions]{Reported workplace GHG emissions, distributed among researchers at five different \ACR\ institutions, with the global per-capita average and remaining carbon "budget" to stay within the Paris Climate Accord limit of 1.5\degree C of warming shown for comparison.
    CERN data for 2019 is taken from Refs.~\cite{Environment:2737239,CERN-HR-STAFF-STAT-2019,CERN:2723123,CERN-HR-STAFF-STAT-2019}, MPIA data for 2019 from Ref.~\cite{Jahnke2020}, ETHZ DPhys data from 2018 taken from Ref.~\cite{Beisert2020}, Nikhef data from 2019 from Refs.~\cite{Nikhef,NikhefStats}, Fermilab (FNAL) data from Refs.~\cite{FermilabEnvReport2019,FNALPrivate,FNALPrivate2}, and Los Alamos National Laboratory (LANL) data from Refs.~\cite{LANLEmissions,LANLEmployeeDemo} (more granular data publicly unavailable).  Although FNAL's 2019 electricity consumption was half that of CERN, its scope 2 emissions were an order of magnitude larger, reflecting local differences in carbon intensity of fuel sources for electricity generation~\cite{Environment:2737239,FermilabEnvReport}. Two-thirds of FNAL's purchased electricity is used to run its high-energy accelerators~\cite{FermilabEnvReport}.   Scope 3 estimates are incomplete for all but CERN.  To estimate emissions per researcher, each individual emissions category was divided by the nominal number of users for that resource, see Appendix~\ref{sec:DataforFig1.4} for details.  
    \label{fig:Intro-ComparativeEmissions}}
\end{figure}

\fref{fig:intro-GlobalEmissions} presents a breakdown of 2016 global GHG emissions by sector, with the emissions at the European Organization for Nuclear Research (CERN) for 2019, during the Large Hadron Collider (LHC) shutdown, shown as a proxy for research emissions.\footnote{Note that direct and indirect emissions more than double when the LHC is operational \cite{Environment:2737239}.}  CERN, like other \ACR\ institutions, categorises its emissions by scope rather than sector, making a direct comparison difficult.  Instead, we consider total
per-researcher emissions. A similar per-employee metric is used by the French National Centre for Scientific Research (CNRS) to quantify its climate impact~\cite{CNRSEmissions}; this is also a default output of the Labos1point5 research emissions assessment tool~\cite{labos1p5web} (see also~\bpref{BP:l1p5}).  Dividing CERN's emissions equally among its 17,000 Users (researchers involved in CERN-based experiments), we obtain roughly 15 \tCdOe\ per researcher per annum, or twice the global average of 6.3 \tCdOe.  Note that this does not include personal household emissions for researchers, which may exceed their workplace emissions.  It also incorrectly apportions emissions due to CERN's smaller cohort of direct personnel, \eg due to CERN-funded travel or personal computing equipment, to its entire User base.  
%
%
For a fairer accounting, see \fref{fig:Intro-ComparativeEmissions}.  The latter chart shows reported work-related emissions in each sector divided by the `true' consumers of each resource, for researchers at CERN, as well as five other \ACR\ institutions:\ the Max Planck Institute for Astronomy (MPIA) in Heidelberg, Germany, the Department of Physics (DPhys) at the Swiss Federal Institute of Technology in Zurich (ETH Zurich), the National Institute for Subatomic Physics (NIKHEF) in Amsterdam, the Netherlands, Fermi National Accelerator Laboratory in Batavia, USA (FNAL), and Los Alamos National Laboratory in New Mexico, USA (LANL).  The emissions are categorised by scope, with scope 1 emissions corresponding to direct emissions that take place on-site, \eg local heating; scope 2 emissions are indirect emissions due to purchased energy, \eg electricity; scope 3 emissions are indirect emissions occurring either upstream of the institution, due to purchased goods or services, or downstream, due to use of its outputs, or actions of its employees.  Granular data is unavailable for LANL. For raw data and details of underlying methodology, see Appendix~\ref{sec:DataforFig1.4}.  Note that these institutions are somewhat self-selected, counting among the minority that have published quantitative estimates of their environmental footprint.  Since CERN is the only institution on this list that attempts a full accounting of its scope 3 emissions, the numbers in this chart should be taken as indicative only, and caution should be employed in making comparisons across institutions on this basis.\footnote{CERN procurement data were estimated with a spend-based method~\cite{SpendBased} using the ecoinvent database~\cite{ecoinvent}. The results therefore have a large margin of error and should be interpreted with care.}   It is nevertheless evident that work-related emissions for many  \ACR\ researchers far exceed our remaining budget to stay within 
1.5\degree C of warming~\cite{IPCC2021reportSPM}.\footnote{The per-capita budget is computed using an emissions budget of 420 G\tCdOe\, and an average world population of 8.8 billion.} 

Figure \ref{fig:intro-GlobalEmissions} also indicates that the primary sources of emissions from HECAP+ research infrastructure differ from the primary sources of emissions globally. It is for this reason that the reduction of GHG emissions or other environmental impacts within each of the areas identified in this report should be prioritised by the community, whatever the comparative scale of these areas within the emissions profile of HECAP+ research.

Our work-related emissions are due to choices we make, as individuals, collaborations, or institutions.  They could be a direct consequence of \ACR\ research, such as the choice of detector design; computing setup or software pipeline for simulation and analysis; or how we collaborate or communicate the results of our work.  Alternatively they could be peripherally related to the science we do:~e.g., how we commute between our home and workplace, or the food we consume while at work; or how our offices are powered, heated and ventilated. 
Historically, many of these choices have been made prioritising cost or convenience over environmental and social impact.  However, the rapid and systemic societal change needed to keep to our climate change goals requires system-wide engagement at all levels of academia.  We can impel positive change throughout the academic research system by re-assessing these choices, by determining how central they are to our primary function as scientists, and by striving to limit our negative environmental impacts.

This process has already begun. Universities and other institutions are including sustainability in buildings planning (see \bpref{BP:Nikhef}) and engaging with voluntary assessments of the environmental sustainability of their research facilities (see, \eg Refs.~\cite{Environment:2737239, Jahnke2020, Beisert2020, Nikhef, FermilabEnvReport2019}). Examples focussed on the environmental sustainability of laboratories include France's Labos1point5 (see \bpref{BP:l1p5}) and University College London’s Laboratory Efficiency Assessment Framework (LEAF) initiative~\cite{LEAF_framework, LEAF_take_part}, a standard being adopted across an increasing number of universities~\cite{LEAF_impact}. The LEAF initiative, which awards three levels of certification to participating laboratories, is structured around online tools that promote best practice, and aid calculations of impact and reporting, as well as additional resources and training opportunities for staff and students. 


\begin{bestpractice}[Nikhef renovation and sustainability plan\label{BP:Nikhef}]{Nikhef renovation and sustainability plan}%
    Nikhef is the Dutch National Institute for Subatomic Physics in the Netherlands. It is both a consortium of universities and an institution with a building in Amsterdam. The total \tCdOe\ footprint of Nikhef was 1,082 \tCdOe\ in 2019, three quarters of which is due to flying to conferences and laboratories, and 15\% is due to heating the building with natural gas~\cite{Nikhef}. The building underwent a major renovation in 2021--2023, which removed the need for gas for heating. Instead, the heat from the nearby data centre will be used, in addition to better thermal insulation of the building.

    The Nikhef sustainability roadmap~\cite{Nikhef} covers all sources of direct and indirect carbon emissions. For instance, by 2030, air travel should be reduced by 50\% and daily commuting should be climate neutral.  Intermediate targets for 2025 are also set and yearly emissions will be monitored and reported.
\end{bestpractice}


\newpage

\subsection{Previous and Parallel Initiatives}
\label{sec:other_initiatives}

It is important to acknowledge the attention that the topics of environmental sustainability and associated social justice issues are rightly being given across our communities. This includes not only parallel tracks devoted to sustainability, and equity, equality, diversity, inclusivity and accessibility, but also plenary talks and panel discussions at large physics conferences. This section provides a brief review of other documents with similar and complementary focuses on environmental and wider social responsibilities.


\subsubsection{ALLEA, Towards Climate Sustainability of the Academic System in Europe and Beyond}

The All European Academies (ALLEA) Working Group on Climate Sustainability in the Academic System published a report in May 2022~\cite{ALLEA}, the aim of which is "to  assess current practices and to critically examine current and proposed measures." 
The document urges stakeholders --- either individual (researchers and students) or structural (universities, funding bodies, conference organisers, ranking agencies, and
policy makers) --- to know their roles and responsibilities toward a climate-sustainable academic system.
After summarizing available data on GHG emissions from various stakeholders and reviewing
the current practices aimed at reducing those
emissions, the report outlines recommendations for individual and group stakeholders. 
Dimensions of social justice and equity are among the principles underlying all recommendations, as well as the opportunity for the academic system to be a role model in the matter. 
While all group stakeholders are advised to embed sustainability in their strategies, individual ones differ:\ students and academic members are encouraged to hold university management accountable, to demand divestment and to generate awareness. 
The importance of the development of an evidence base is emphasised, and stakeholders are pushed to allocate funding to the decarbonisation of the academic system.


\subsubsection{Snowmass Contribution, Climate Impacts of Particle Physics}

The report "Climate impacts of particle physics"~\cite{Bloom:2022gux}, submitted to the proceedings of the US Community Study on the Future of Particle Physics (Snowmass 2021) focuses on facility construction, detector gases, computing, and GHG emissions from particle physics laboratories. 
The report highlights two key motivations for addressing the ecological and climate impacts of particle physics: (i) that the particle physics community has a moral obligation to do so and (ii) that its professional activities will be under increasing scrutiny from a number of stakeholders. 
The latter means that the community will be under increasing pressure to justify its carbon emissions against its relative size, compared to other industries, and its societal benefits.

As a concrete example, the report focuses on the Future Circular Collider (FCC) --- the proposed 90 to 360 GeV electron-positron (and later 100 TeV hadron) collider. 
The authors estimate that the construction of the roughly 100 km circumference tunnel alone would lead to \CdO\ emissions at the level of a few hundred kilotons, several times more than other large US building projects.
This corresponds to "per physicist" emissions 80 times larger than their estimated 1.1 \acrshort{tco2}\ per capita per year limit needed to keep global warming to less that 1.5\degree C. The authors emphasise the significant impact of the GHGs used in particle physics detectors and for cooling, which can have warming potential that exceeds that of \CdO\ by as much as four orders of magnitude (in the case of ${\rm SF}_6$).  
The report then highlights a number of avenues for reducing the GHG emissions due to the electricity consumption of computing, which is pivotal to running and exploiting such facilities. It also discusses the additional emissions due to collaborative research activities, with a particular emphasis on taking careful steps to reduce air travel that capitalise on potential benefits for social justice while minimising unintended negative consequences for members of the community.

The report's recommendations stress the need for reporting on planned emissions and energy usage for new facilities; standardised reporting of emissions across the sector, and community-wide engagement to tackle the negative climate impacts of particle physics research through dedicated research time.

\subsubsection{Recommendations by the yHEP association in Germany}

The young High Energy Physicists (yHEP) association in Germany published the "yHEP recommendations on improvement of environmental sustainability in science"~\cite{yHEP1} and its Addendum~\cite{yHEP2} in December 2020 and 2021, respectively. The documents, which were the result of proposals from the yHEP community, including HECAP, hadron and nuclear, and accelerator physicists, contain ideas for improving the environmental sustainability of basic research. They take a qualitative approach on a broad range of topics, including, but not limited to, travel, conferences, computing and infrastructure, resource management and financing, and green energy. 


\subsection{Impelling Positive Change}
\label{sec:impelling}

The aim of this document is to provide as comprehensive a discussion as practicable of the various impacts of \ACR\ research, from our day-to-day activities through to the large infrastructure projects on which our science depends. The discussions presented here have much in common with those of the documents described in \sref{sec:other_initiatives}. This document is, however, intended to have broad scope and, through case studies and best practice, to illustrate potential actions that can be implemented at individual, group and institutional levels to limit the impacts of \ACR\ research on the world's climate and ecosystems.

The emphasis of this document is on addressing the sources of emissions from HECAP+ research directly. The integrity of carbon and biodiversity offsets (e.g., through afforrestation) remains under scrutiny. Concerns relate to additionality, reversibility, and the negative effects that such offsets can have on self-determination, equity and, indeed, biodiversity~\cite{IPBESIPCC}. Offsetting should therefore be seen as a last resort, used only once all other options for reducing the \CdO\ equivalent emissions have been exhausted and to offset any residual CO$_2$e output. Even then, close consideration should be given to the sources of emissions that should be offset, the method of offsetting (with long-term preference towards carbon removal with durable storage~\cite{SmithSchoolOffsetting}) and its integrity, so as to avoid unintended consequences and to avoid contributing to the lack of urgency in dealing directly with GHG emissions~\cite{IPBESIPCC}.

Thus, if the \ACR\ community is to succeed in improving the sustainability of its working practices, then the environment and related issues of social justice must be recognised as integral parts of the planning and management of our research activities.

Addressing sustainability within scientific collaborations presents a unique set of challenges, especially when these collaborations are deeply integrated within a broad ecosystem of partners, such as international agencies, universities, and research laboratories. The primary challenge arises not from a reluctance to adopt environmental stewardship but from the delicate task of harmonizing sustainability initiatives with the policies and expectations of a diverse array of partners. Ensuring that these efforts align without overstepping or conflicting with any partner's established policies requires careful navigation and consensus-building. This intricacy underscores the importance of establishing dedicated sustainability task forces capable of adeptly managing these considerations, fostering a unified approach to sustainability that respects the individual constraints and guidelines of each entity involved.

With this in mind, we collect below a list of recommendations for structural changes to the organisation of our community, our training and our professional development. These recommendations complement those listed in the discussions of specific sources of environmental impacts of \ACR\ research on which the bulk of this document focuses. Together, these provide concrete suggestions of ways in which the \ACR\ community can act to reduce its negative climate and ecological impacts, and address issues of social justice in line with the United Nations Sustainable Development Goals, discussed in the next subsection.


\clearpage
\begin{minipage}{\textwidth}
\begin{reco2}{\currentname}
{
\begin{itemize}[leftmargin=6 mm]
\setlength\itemsep{\recskip}
\item Consider the environmental impact of work practices.

\item Be proactive in seeking best practice.

\item Make and model positive change in research activities.

\item Drive positive group and institutional actions.
\end{itemize}
}
{
\begin{itemize}[leftmargin=6 mm]
\setlength\itemsep{\recskip}
\item  Include critical assessment of the environmental impact of all activities during planning stages.

\item  Monitor, assess, report on and set targets in relation to the environmental impacts of research activities.

\item Drive institutional actions, and encourage, support and incentivise individual actions, e.g., through training.

\end{itemize}
}
{
\begin{itemize}[leftmargin=6 mm]
\setlength\itemsep{\recskip}
\item Require funding applications to outline plans for monitoring, reporting and minimising adverse environmental impacts, and for ensuring that research is undertaken in line with principles of social justice.

\item Allow flexibility in policies and procedures \eg budget allocation, that enable environmentally sustainable choices to be made.

\item Ensure that degree programmes include a focus on global citizenship, encompassing environmental sustainability and associated social justice implications.

\item Acknowledge focus on environmental sustainability and social justice in the accreditation of degrees by governments and professional bodies.

\item Encourage, support and incentivise individual and group actions, \eg by considering them in professional development and appraisal processes.

\end{itemize}
}

\end{reco2}
\end{minipage}


\newpage
\subsection{United Nations Sustainable Development Goals}


\begin{figure}[!ht]
     \centering
     \includegraphics[width=0.75\linewidth]{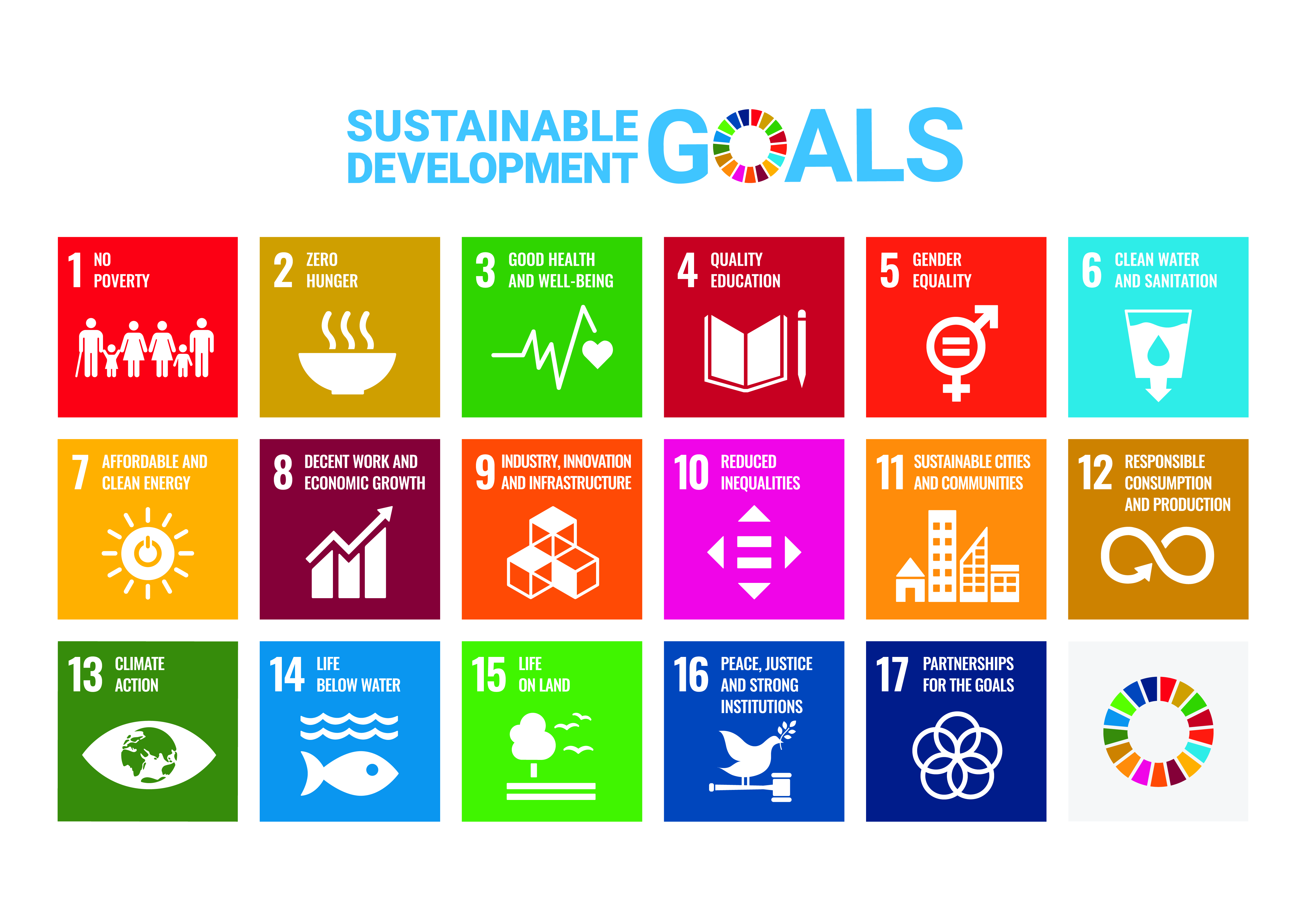}
     \caption[United Nations Sustainable Development Goals]{The seventeen United Nations Sustainable Development Goals~\cite{UNSustainableGoalsFigure,UNwebsite}. The content of this publication has not been approved by the United Nations and does not reflect the views of the United Nations or its officials or Member States.\label{fig:17_UN_SustGoals}}
\end{figure}


As a global research community, \ACR\ has an impact on society all over the world. We contribute to basic scientific knowledge, drive innovation and promote international collaboration. Our institutions are large employers, large consumers of goods and services, and a key resource in the training and development of national skills bases. This places our institutions in a position to influence policy decisions, drive investment in local infrastructure, and leverage wider improvements to social and environmental standards. For these reasons, the \ACR\ community is in a strong position to support the UN Sustainable Development Goals (\acrshort{sdg}s), summarised in~\fref{fig:17_UN_SustGoals}.

The topics discussed in this document are meant to support a multiplicity of these goals, and we aim to signpost the influence of our work in all aspects. The goals are listed below, with examples of how each is impacted by the HECAP+ research community and its work. The SDGs are defined in UN resolution A/RES/70/1 in detail~\cite{UNResolutionARES701}. It is impossible to cover all aspects in this document, but the manifold impact of the \ACR\ community on sustainable development is clear from this non-exhaustive list. 

\bigskip


\newcommand{\SDGscale}{0.05}
\newcommand{\iconskip}{-2em}
\newcommand{\SDGleft}{0.07}
\newcommand{\SDGright}{0.89}
\renewcommand*{\arraystretch}{3}
\noindent\begin{longtable*}{l l}
\parbox[t]{\SDGleft\textwidth}{\raisebox{\iconskip}{\includegraphics[scale=\SDGscale]{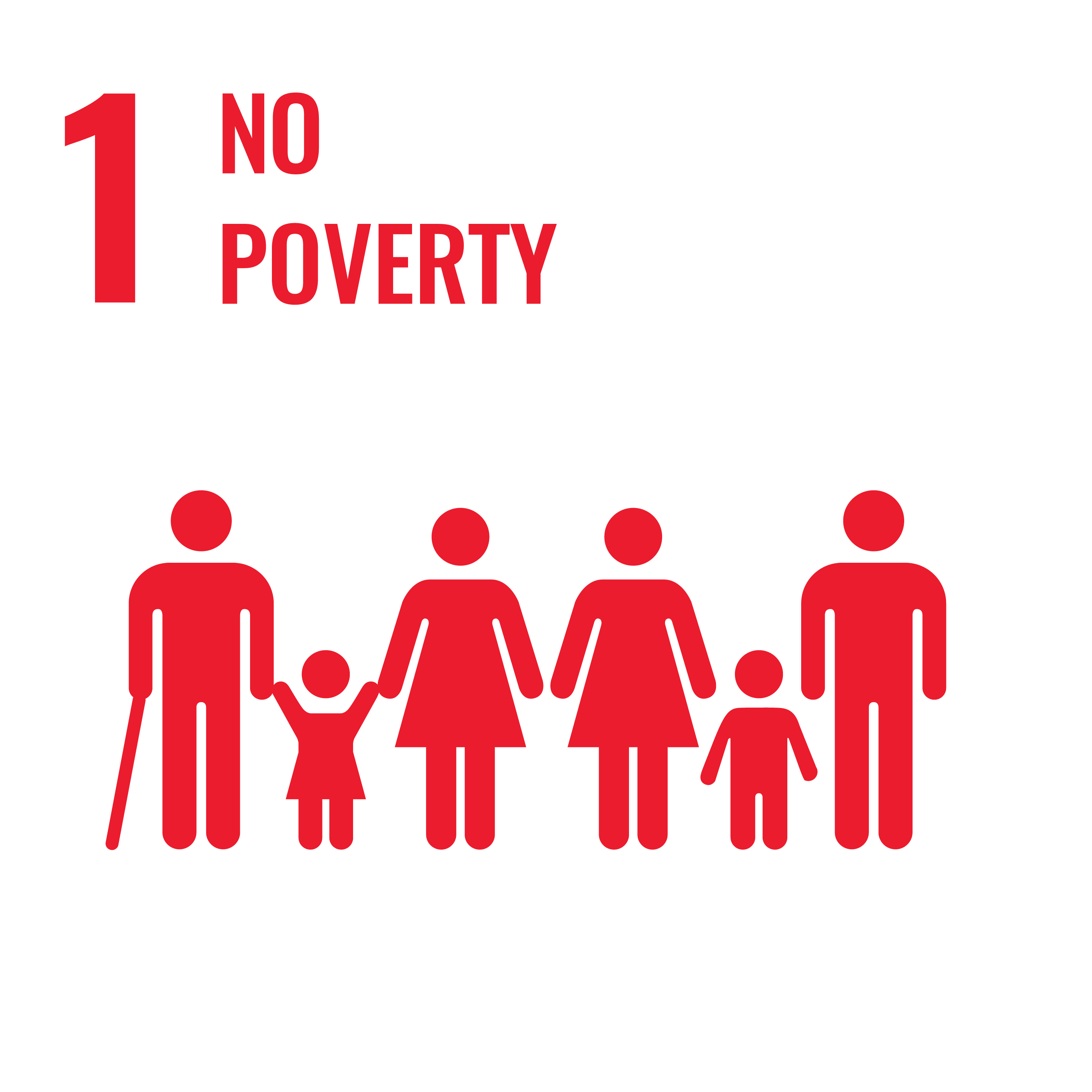}}} & \parbox[t]{\SDGright\textwidth}{\textbf{Goal 1:\ No poverty --- End poverty in all its forms everywhere}
\vspace{\recskip}
\begin{itemize}[leftmargin=20pt]
\setlength{\itemsep}{\recskip}
\item The contractual and payment standards in employment contracts of institutes and collaborations influence their employees' lives.
\item The terms of contract with external companies influence the working and living conditions of their employees.
\end{itemize}}\\
\parbox[t]{\SDGleft\textwidth}{\raisebox{\iconskip}{\includegraphics[scale=\SDGscale]{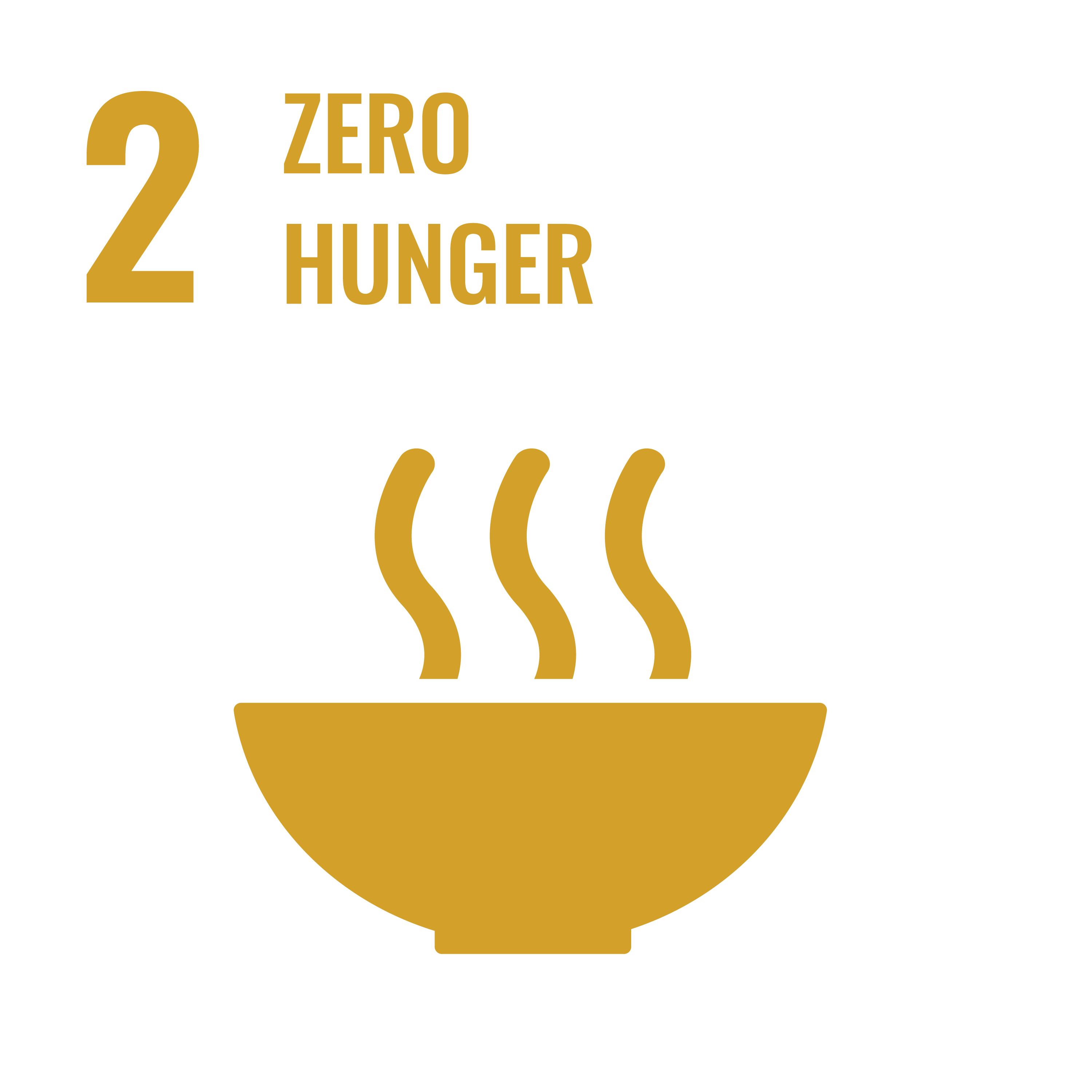}}} & \parbox[t]{\SDGright\textwidth}{\textbf{2 Zero hunger:\ End hunger, achieve food security and improved nutrition and promote sustainable agriculture}
\vspace{\recskip}
\begin{itemize}[leftmargin=20pt]
\setlength{\itemsep}{\recskip}
\item The food consumed at institutes and events has an effect on the behaviour of the food market/industry from which it is purchased.
\end{itemize}}\\

\parbox[t]{\SDGleft\textwidth}{\raisebox{\iconskip}{\includegraphics[scale=\SDGscale]{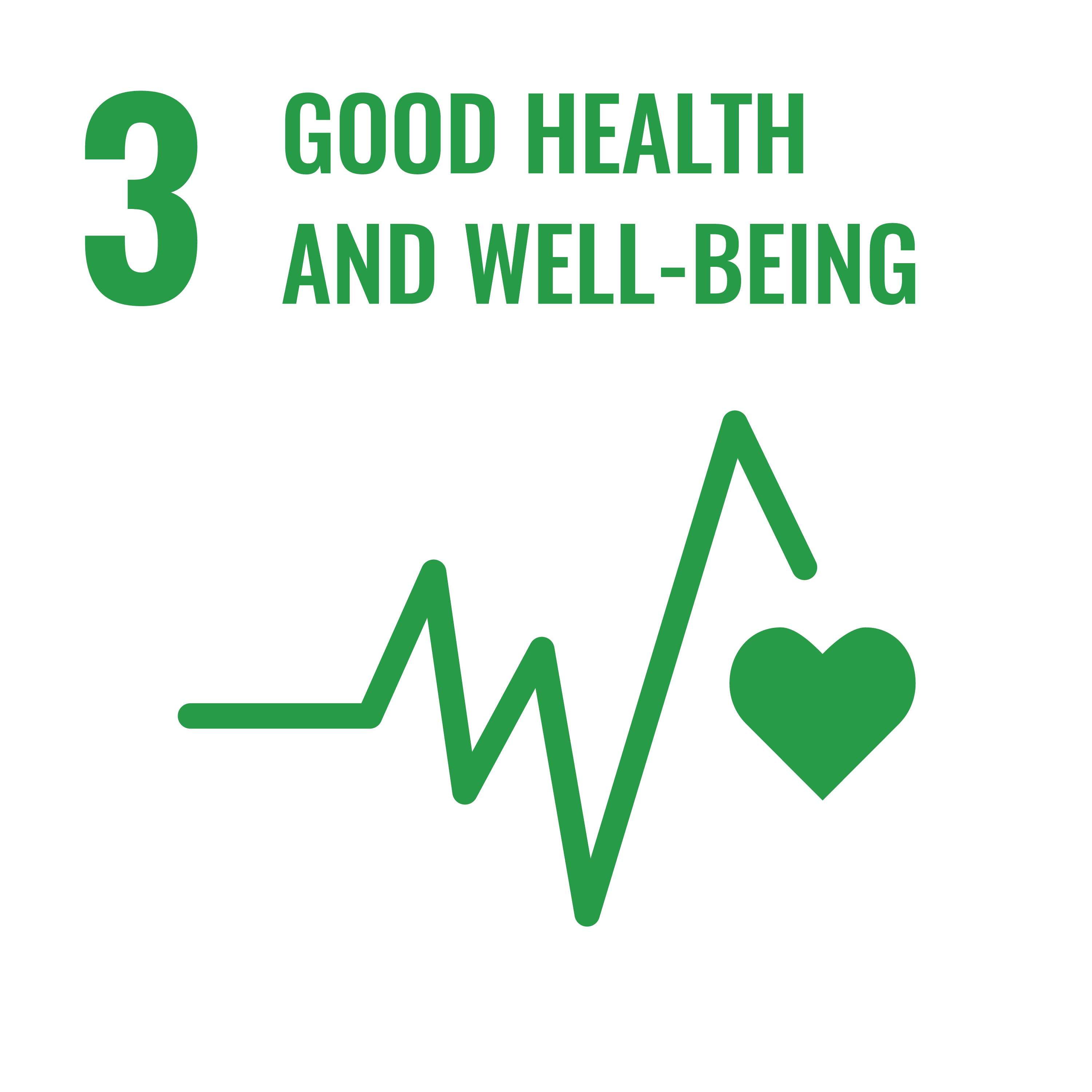}}} & \parbox[t]{\SDGright\textwidth}{\textbf{3 Good health and well-being:\ Ensure healthy lives and promote well-being for all at all ages}
\vspace{\recskip}
\begin{itemize}[leftmargin=20pt]
\setlength{\itemsep}{\recskip}
\item HECAP+ research helps to develop medical diagnostics and treatments, \eg for cancer.
\item The working culture practised every day has an impact on the mental health of ourselves and co-workers.
\item The design of experimental setups has an effect on (work) safety issues.
\item Food served and consumed at institutes and events has an impact on the health and well-being of the consumers.
\end{itemize}}\\

\parbox[t]{\SDGleft\textwidth}{\raisebox{\iconskip}{\includegraphics[scale=\SDGscale]{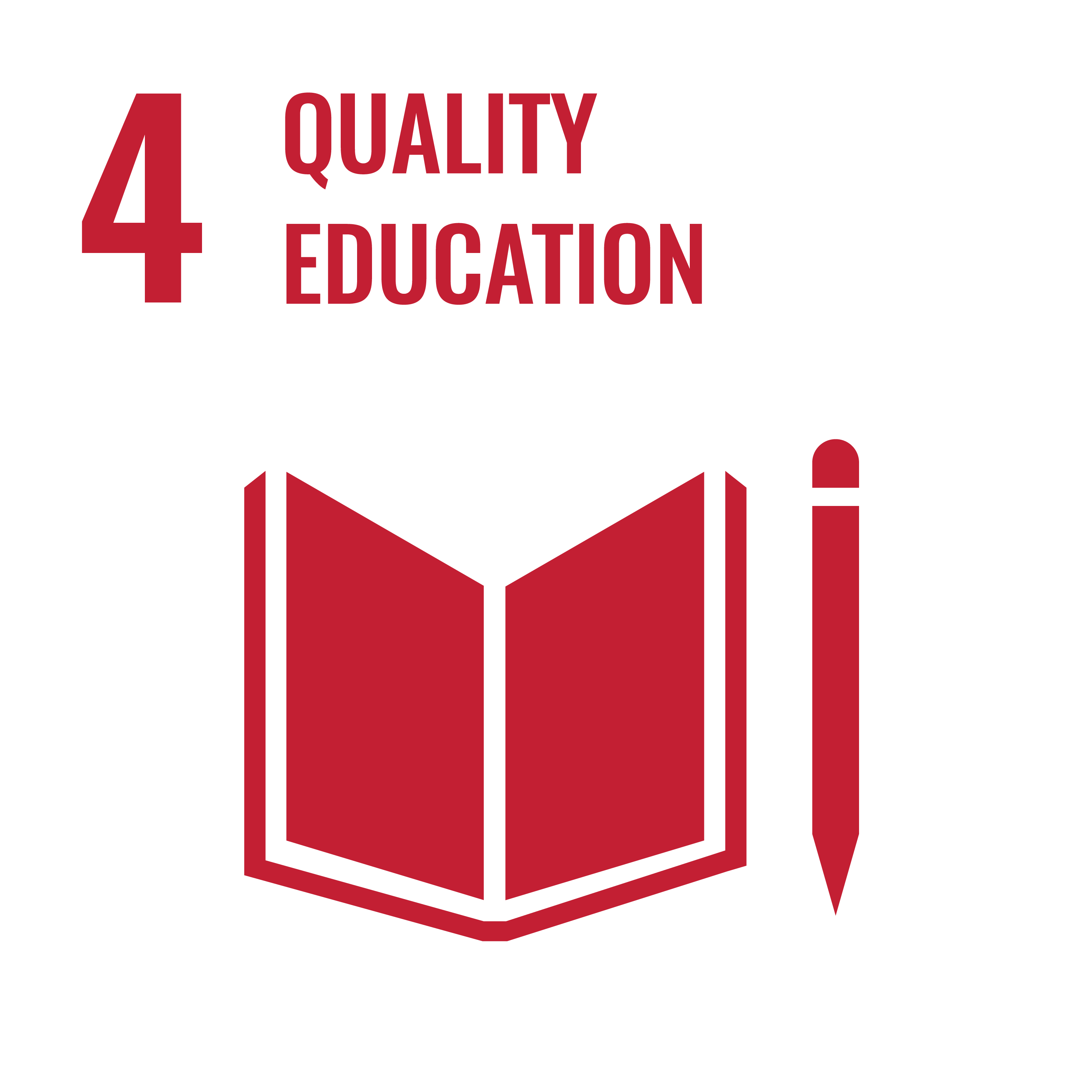}}} & \parbox[t]{\SDGright\textwidth}{\textbf{4 Quality education:\ Ensure inclusive and equitable quality education and promote lifelong learning opportunities for all}
\vspace{\recskip}
\begin{itemize}[leftmargin=20pt]
\setlength{\itemsep}{\recskip}
\item Research develops and uses scientific methods to establish a general body of knowledge that can be passed on in educational settings.
\item Researchers are often teachers for their respective fields and have an effect on the teaching culture.
\item Researchers and institutes, through their conduct and integrity, have an impact on the credibility of science in society.
\item Transparent reporting on efforts towards more sustainable research has a positive impact on the credibility of scientists, and helps avoid greenwashing.
\end{itemize}}\\

\parbox[t]{\SDGleft\textwidth}{\raisebox{\iconskip}{\includegraphics[scale=\SDGscale]{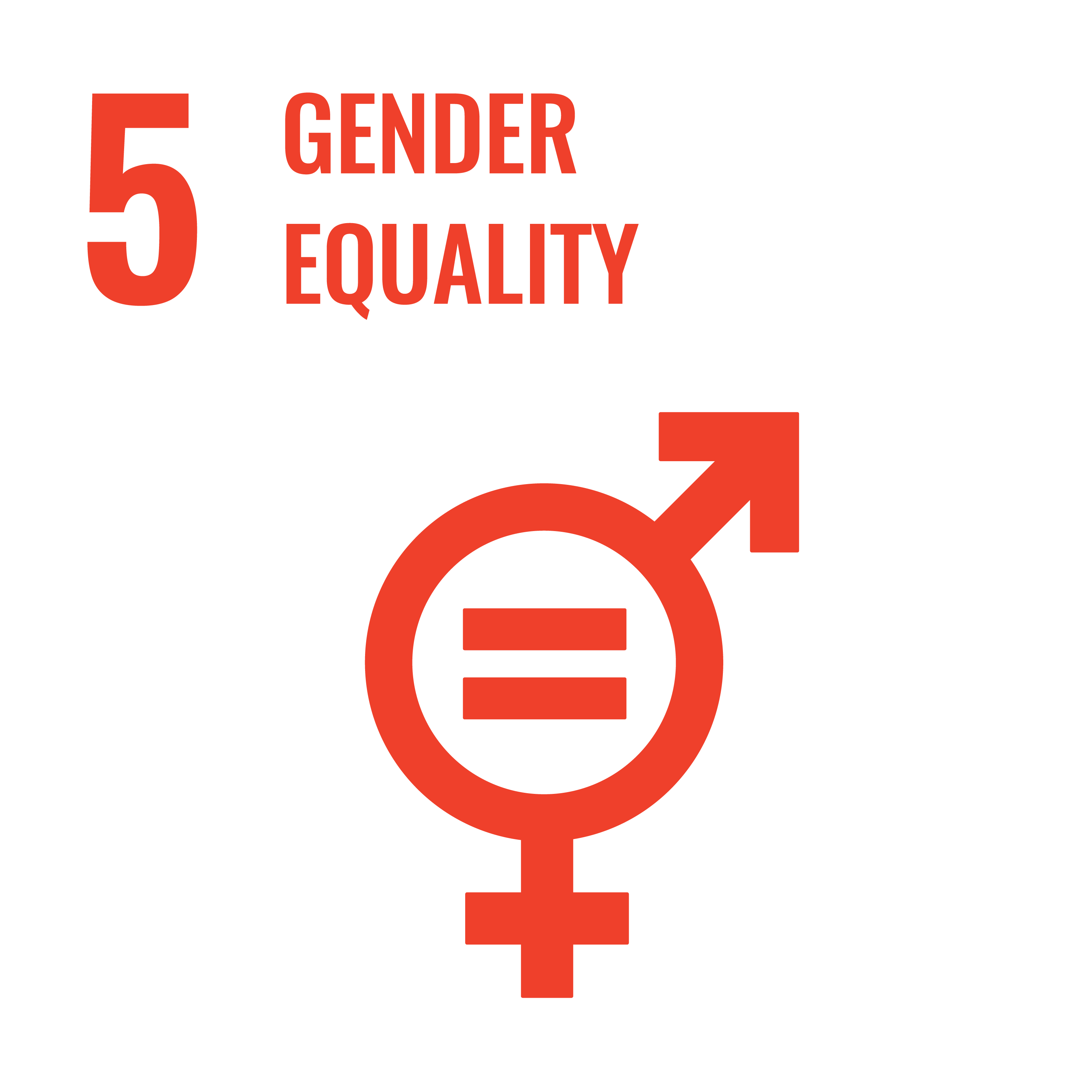}}} & \parbox[t]{\SDGright\textwidth}{\textbf{5 Gender equality:\ Achieve gender equality and empower all women and girls}
\vspace{\recskip}
\begin{itemize}[leftmargin=20pt]
\setlength{\itemsep}{\recskip}
\item As an historically male-dominated field, \ACR\ should strive for a workplace culture that counters discrimination in science and acts for the visibility, acceptance and representative participation of all genders.
\end{itemize}}\\

\parbox[t]{\SDGleft\textwidth}{\raisebox{\iconskip}{\includegraphics[scale=\SDGscale]{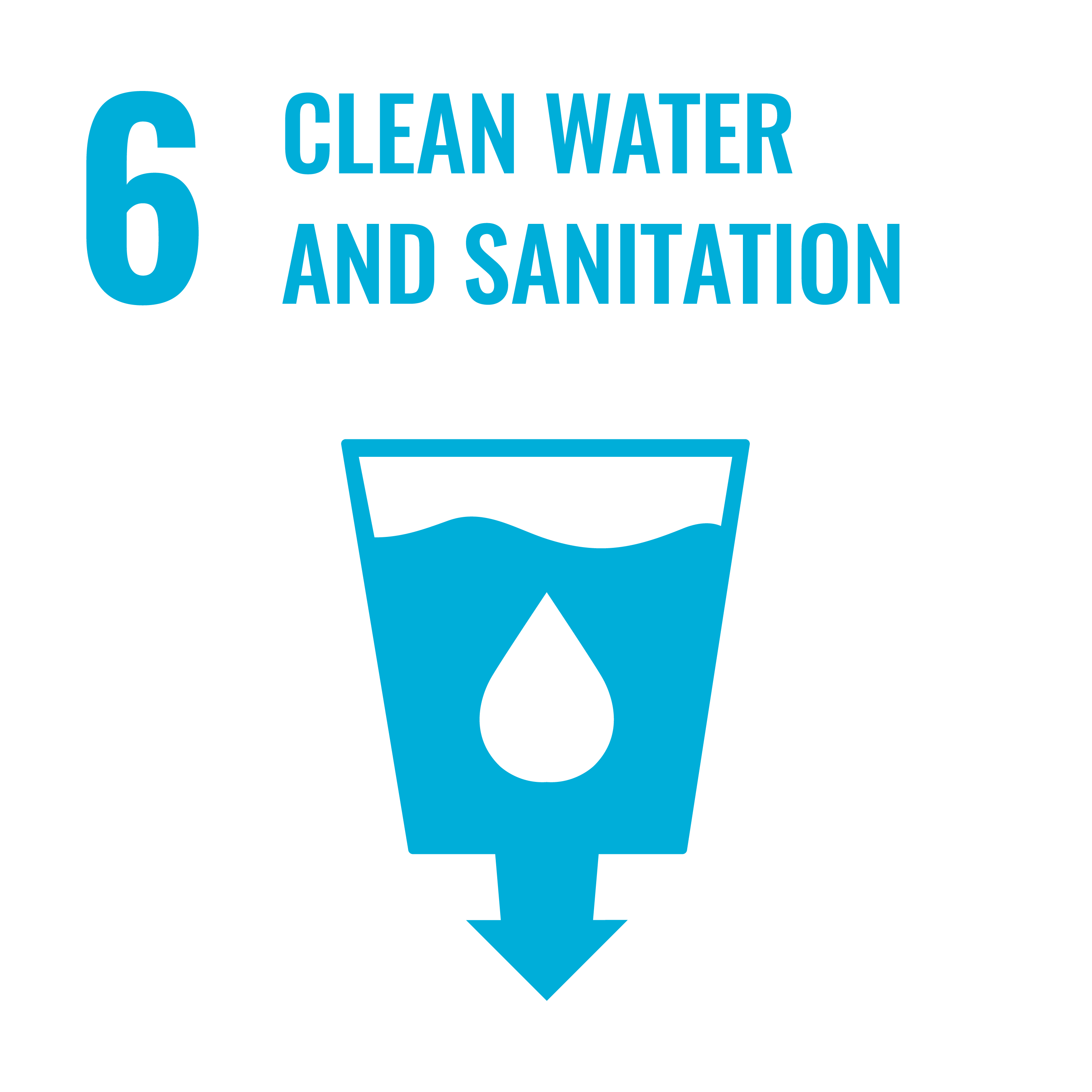}}} & \parbox[t]{\SDGright\textwidth}{\textbf{6 Clean water and sanitation:\ Ensure availability and sustainable management of water and sanitation for all}
\vspace{\recskip}
\begin{itemize}[leftmargin=20pt]
\setlength{\itemsep}{\recskip}
\item Our research requires the use of water for various purposes  (heating, cooling, cleaning, sanitation, food production and preparation, etc.). Its sources are affected by our needs and behaviour.
\item \ACR\ research creates waste water. The treatment of this has an impact on the water quality in the linked aquatic ecosystems.
\item The behaviour and lifestyle choices of our community in professional and private life have an impact on the water needs in the surrounding and indirectly linked area.
\end{itemize}}\\

\parbox[t]{\SDGleft\textwidth}{\raisebox{\iconskip}{\includegraphics[scale=\SDGscale]{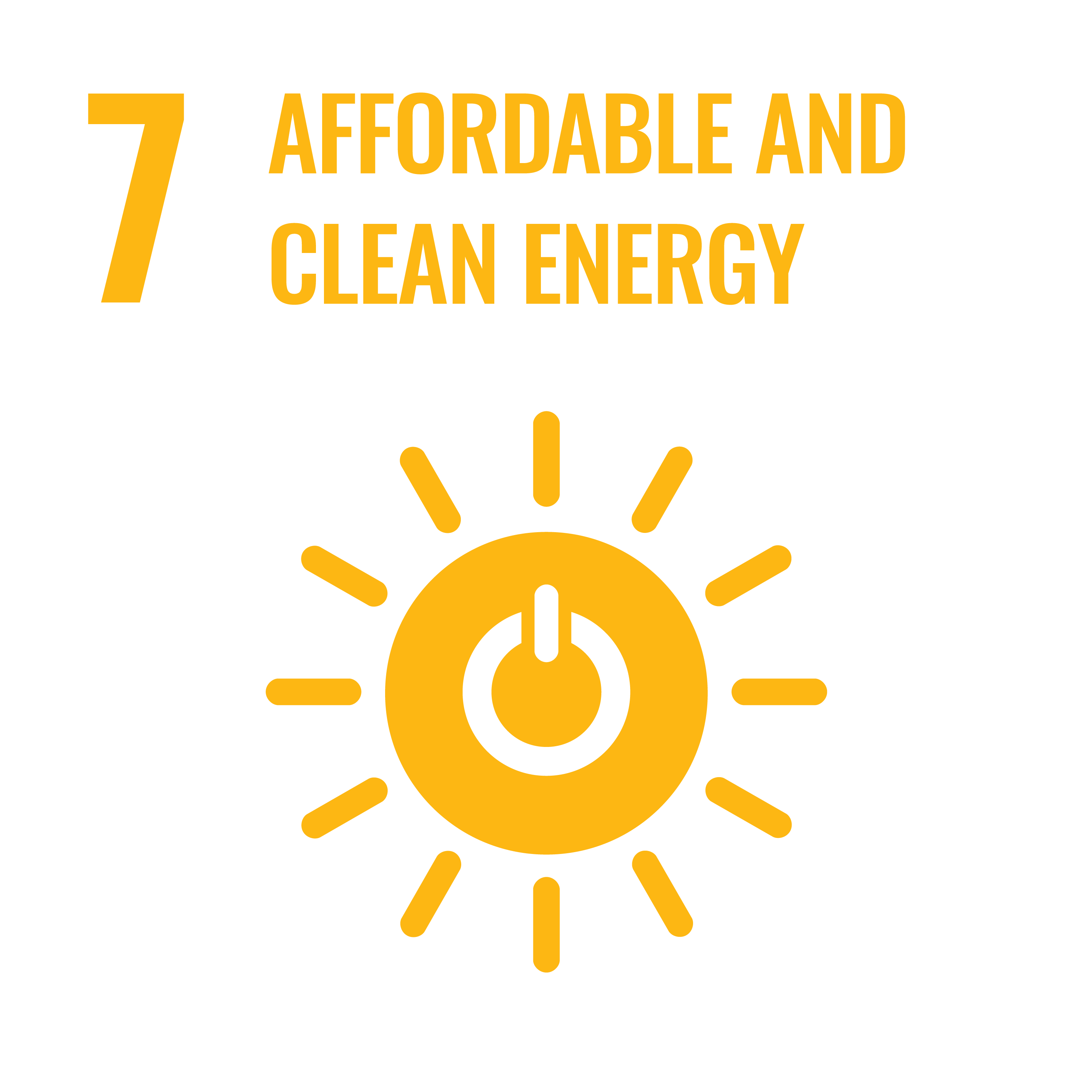}}} & \parbox[t]{\SDGright\textwidth}{\textbf{7 Affordable and clean energy:\ Ensure access to affordable, reliable, sustainable and modern energy for all}
\vspace{\recskip}
\begin{itemize}[leftmargin=20pt]
\setlength{\itemsep}{\recskip}
\item The sources of energy planned and used for institutes, accelerators and experiments have an environmental impact on a global level.
\item The high consumption and the resulting financial impact of research facilities have an impact on the energy market.
\end{itemize}}\\

\parbox[t]{\SDGleft\textwidth}{\raisebox{\iconskip}{\includegraphics[scale=\SDGscale]{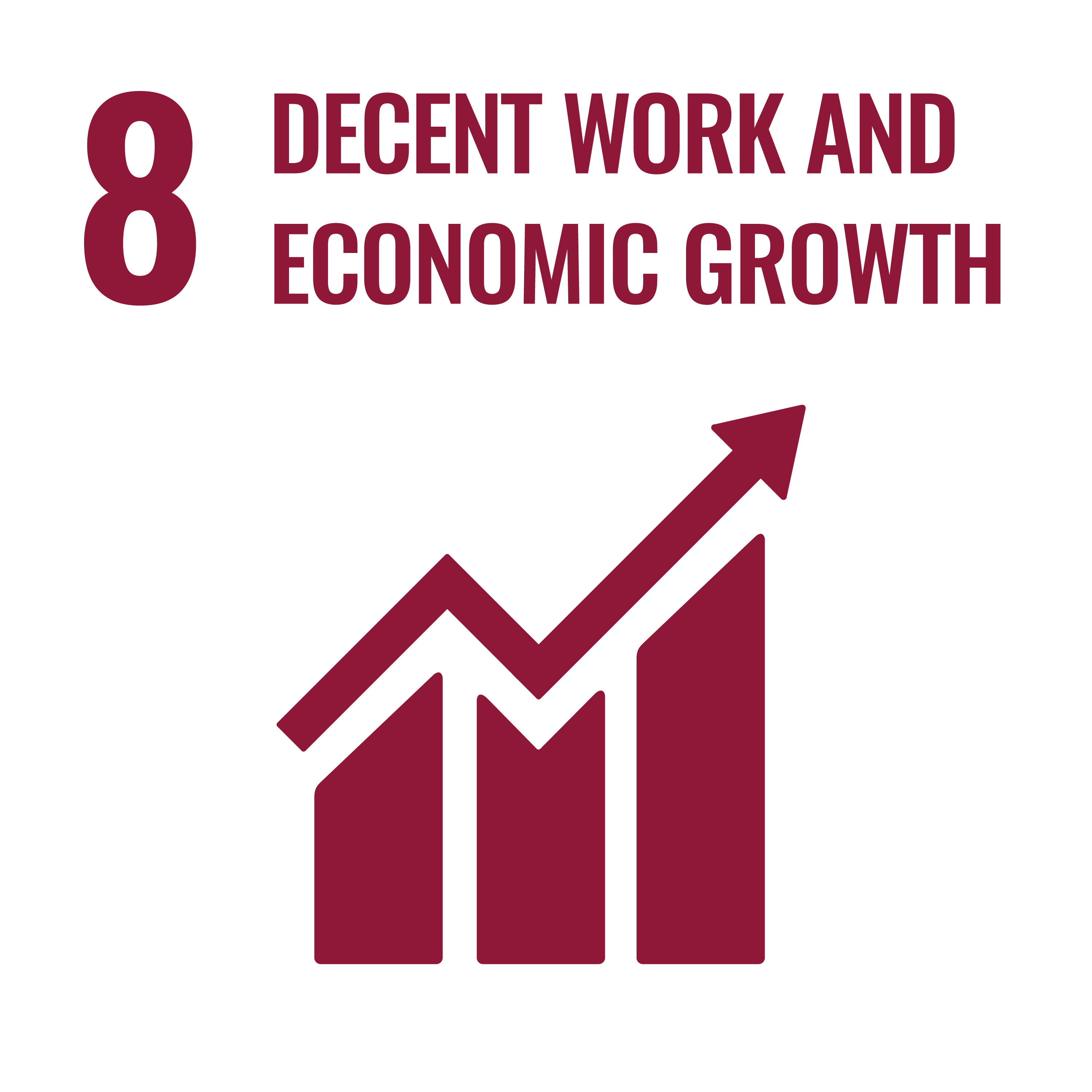}}} & \parbox[t]{\SDGright\textwidth}{\textbf{8 Decent work and economic growth:\ Promote sustained, inclusive and sustainable economic growth, full and productive employment and decent work for all}
\vspace{\recskip}
\begin{itemize}[leftmargin=20pt]
\setlength{\itemsep}{\recskip}
\item The terms of employment contracts and working culture in \ACR\ research influence employees' living conditions.
\end{itemize}}\\

\parbox[t]{\SDGleft\textwidth}{\raisebox{\iconskip}{\includegraphics[scale=\SDGscale]{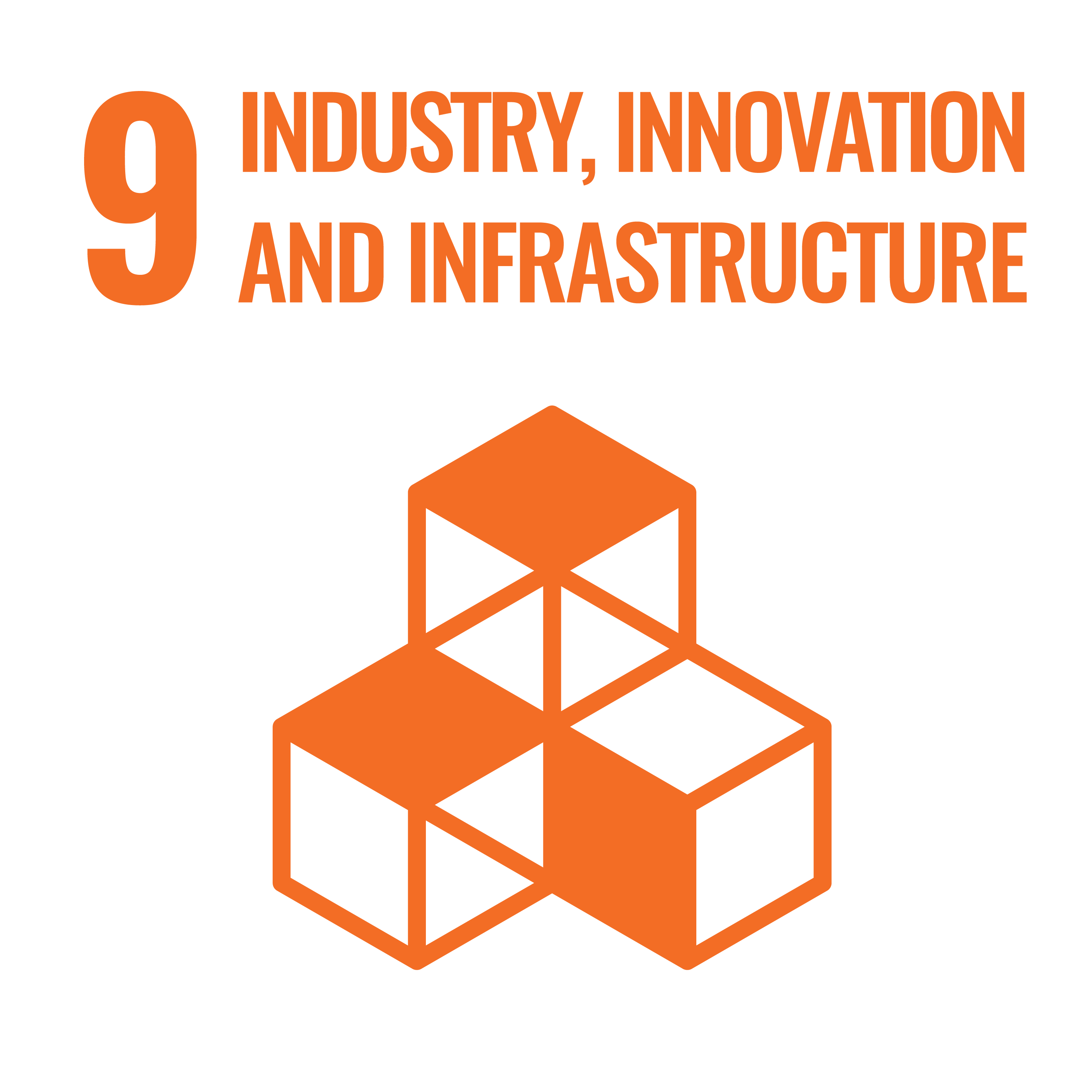}}} & \parbox[t]{\SDGright\textwidth}{\textbf{9 Industry, innovation and infrastructure:\ Build resilient infrastructure, promote inclusive and sustainable industrialization and foster innovation}
\vspace{\recskip}
\begin{itemize}[leftmargin=20pt]
\setlength{\itemsep}{\recskip}
\item Innovation is at the core of \ACR\ research.
\item Institutes influence the local infrastructures on which they rely and construct infrastructure for research.
\item Industry and \ACR\ research are linked as knowledge and products are transferred. This transfer can be shaped actively.
\end{itemize}}\\

\parbox[t]{\SDGleft\textwidth}{\raisebox{\iconskip}{\includegraphics[scale=\SDGscale]{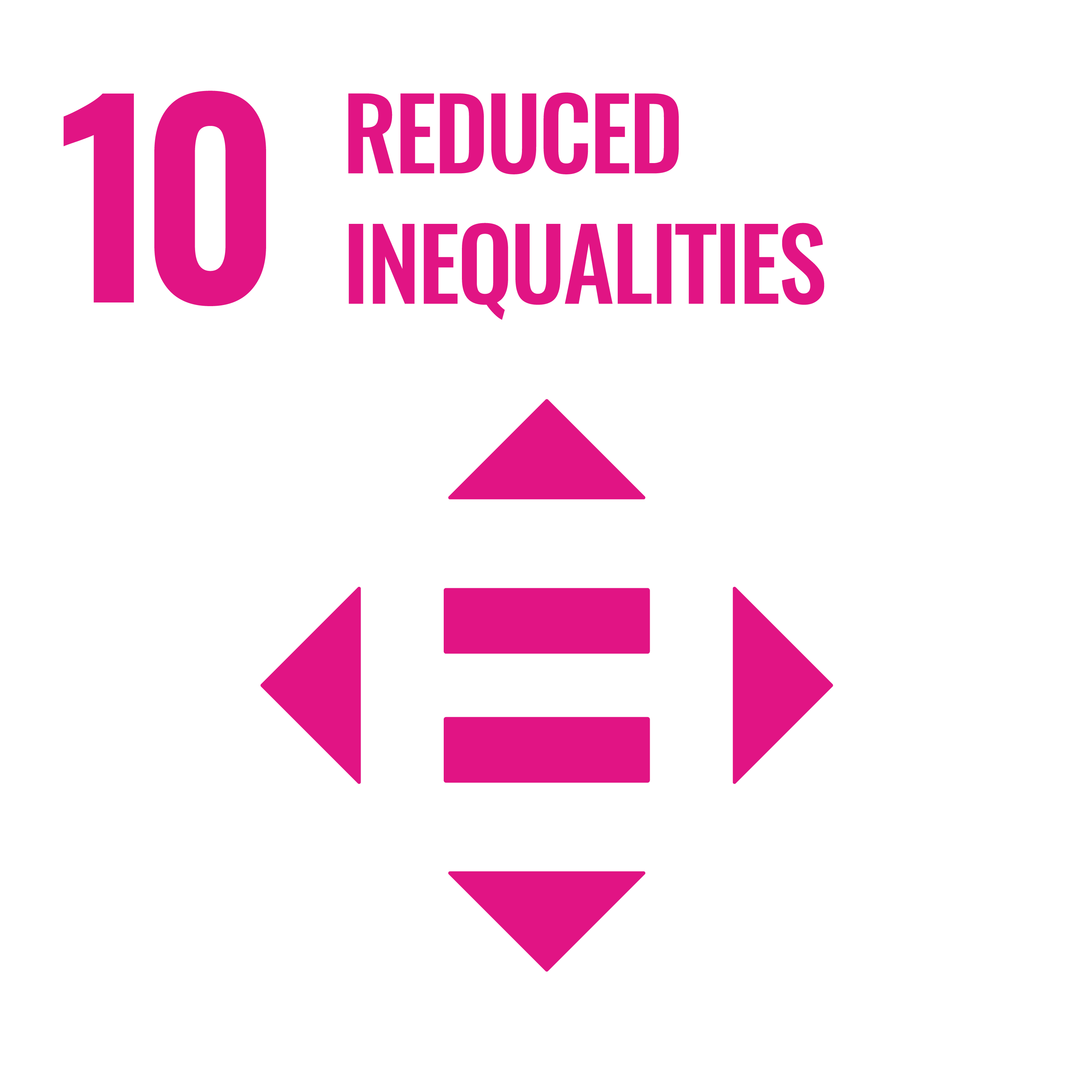}}} & \parbox[t]{\SDGright\textwidth}{\textbf{10 Reduced inequalities:\ Reduce inequality within and among countries}
\vspace{\recskip}
\begin{itemize}[leftmargin=20pt]
\setlength{\itemsep}{\recskip}
\item International research facilities have an opportunity to set an example by redressing historical inequalities between states through the judicious use of science policy and funding, with a particular focus on restoring scientific capital and resources where they have been depleted (see \bpref{BP:sesame} for an example).
\end{itemize}}\\

\parbox[t]{\SDGleft\textwidth}{\raisebox{\iconskip}{\includegraphics[scale=\SDGscale]{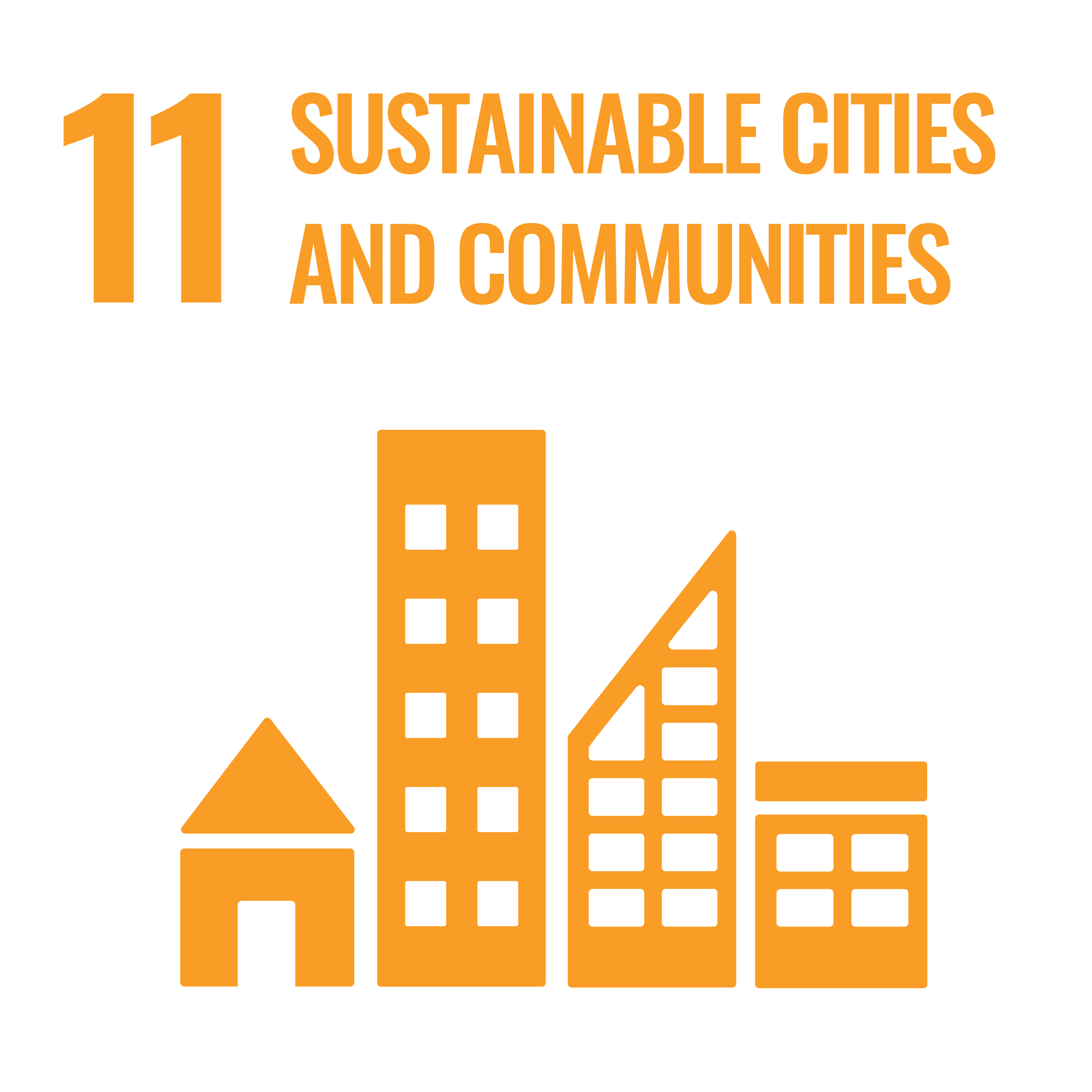}}} & \parbox[t]{\SDGright\textwidth}{\textbf{11 Sustainable cities and communities:\ Make cities and human settlements inclusive, safe, resilient, and sustainable}
\vspace{\recskip}
\begin{itemize}[leftmargin=20pt]
\setlength{\itemsep}{\recskip}
\item The campuses of research facilities have an impact on the cities and neighbourhoods in which they are built.
\item The behaviour and lifestyle choices of our community in professional and private life have an impact on our local communities.
\end{itemize}}\\

\parbox[t]{\SDGleft\textwidth}{\raisebox{\iconskip}{\includegraphics[scale=\SDGscale]{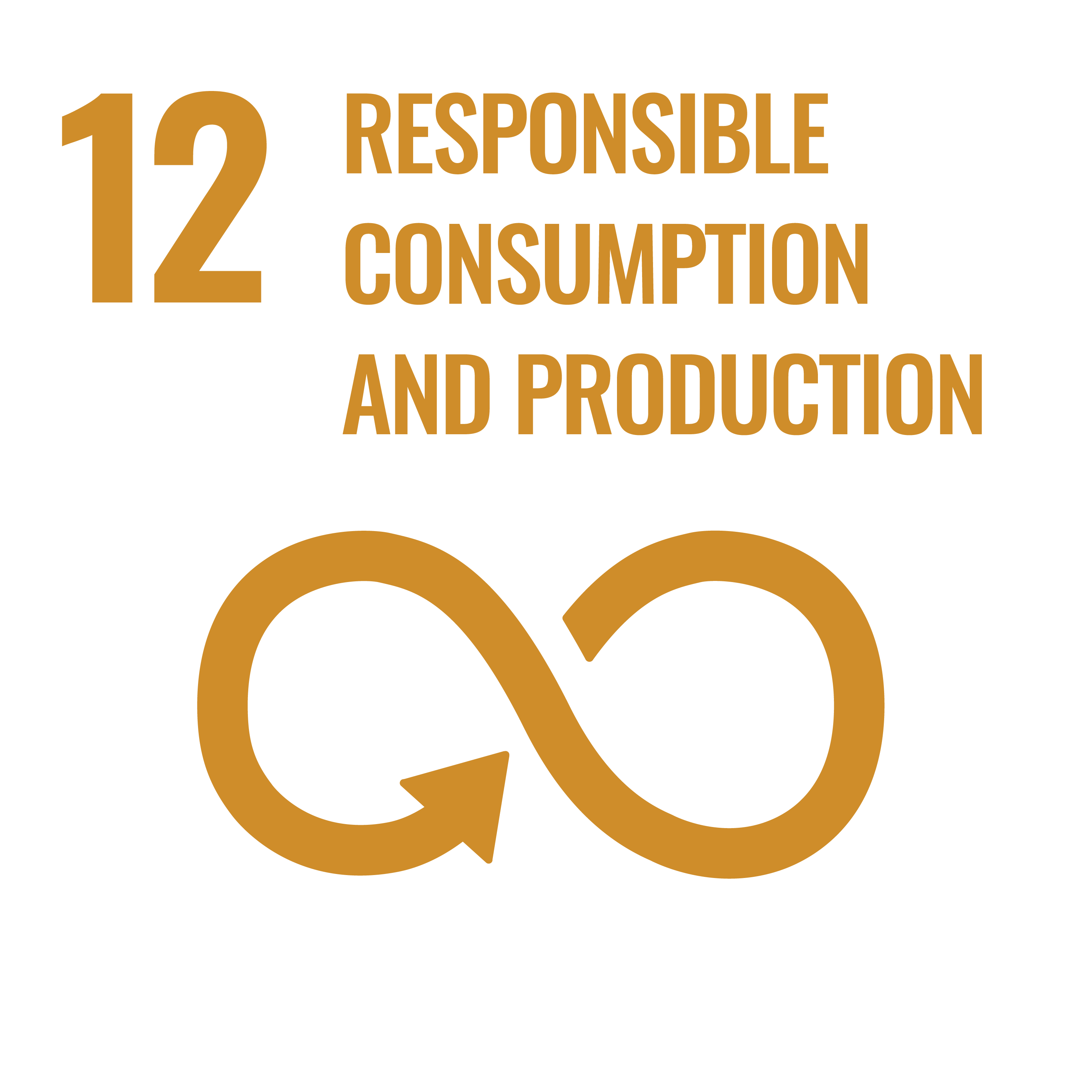}}} & \parbox[t]{\SDGright\textwidth}{\textbf{12 Responsible consumption and production:\ Ensure sustainable consumption and production patterns}
\vspace{\recskip}
\begin{itemize}[leftmargin=20pt]
\setlength{\itemsep}{\recskip}
\item The facilities, accelerators, machines, and experiments we build use up resources and energy in their design, construction, overall lifetime (\eg maintenance) and disposal.
\item The disposal of obsolete equipment and other waste generated by the work we do has an impact on our environment.
\item Our daily choices on consumption have a wider effect on the systems which produce them, \eg food and travel.
\end{itemize}}\\

\parbox[t]{\SDGleft\textwidth}{\raisebox{\iconskip}{\includegraphics[scale=\SDGscale]{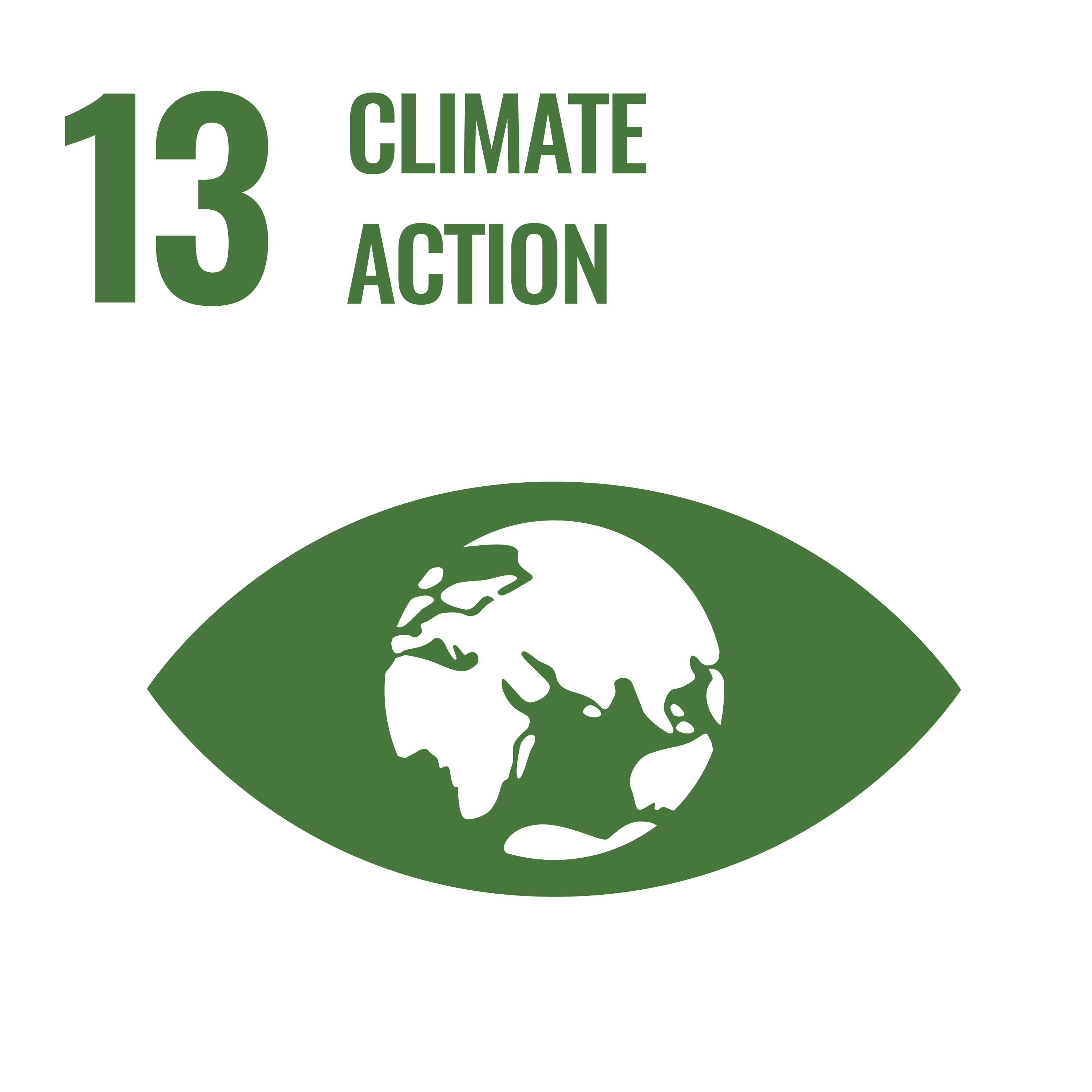}}} & \savenotes\parbox[t]{\SDGright\textwidth}{\textbf{13 Climate action:\ Take urgent action to combat climate change and its impacts*}\footnote{Footnote by the UN: *Acknowledging that the United Nations Framework Convention on Climate Change is the primary international, intergovernmental forum for negotiating the global response to climate change.}
\vspace{\recskip}
\begin{itemize}[leftmargin=20pt]
\setlength{\itemsep}{\recskip}
\item The emission of various gases by \ACR\ research has an impact on the Earth’s climate.
\item The sources of the electrical and thermal energy used by \ACR\ facilities impact the global climate.
\item The behaviour and lifestyle choices (eating, travel, product consumption) of our community in professional and private life have an impact on the global climate.
\end{itemize}}\spewnotes\\

\parbox[t]{\SDGleft\textwidth}{\raisebox{\iconskip}{\includegraphics[scale=\SDGscale]{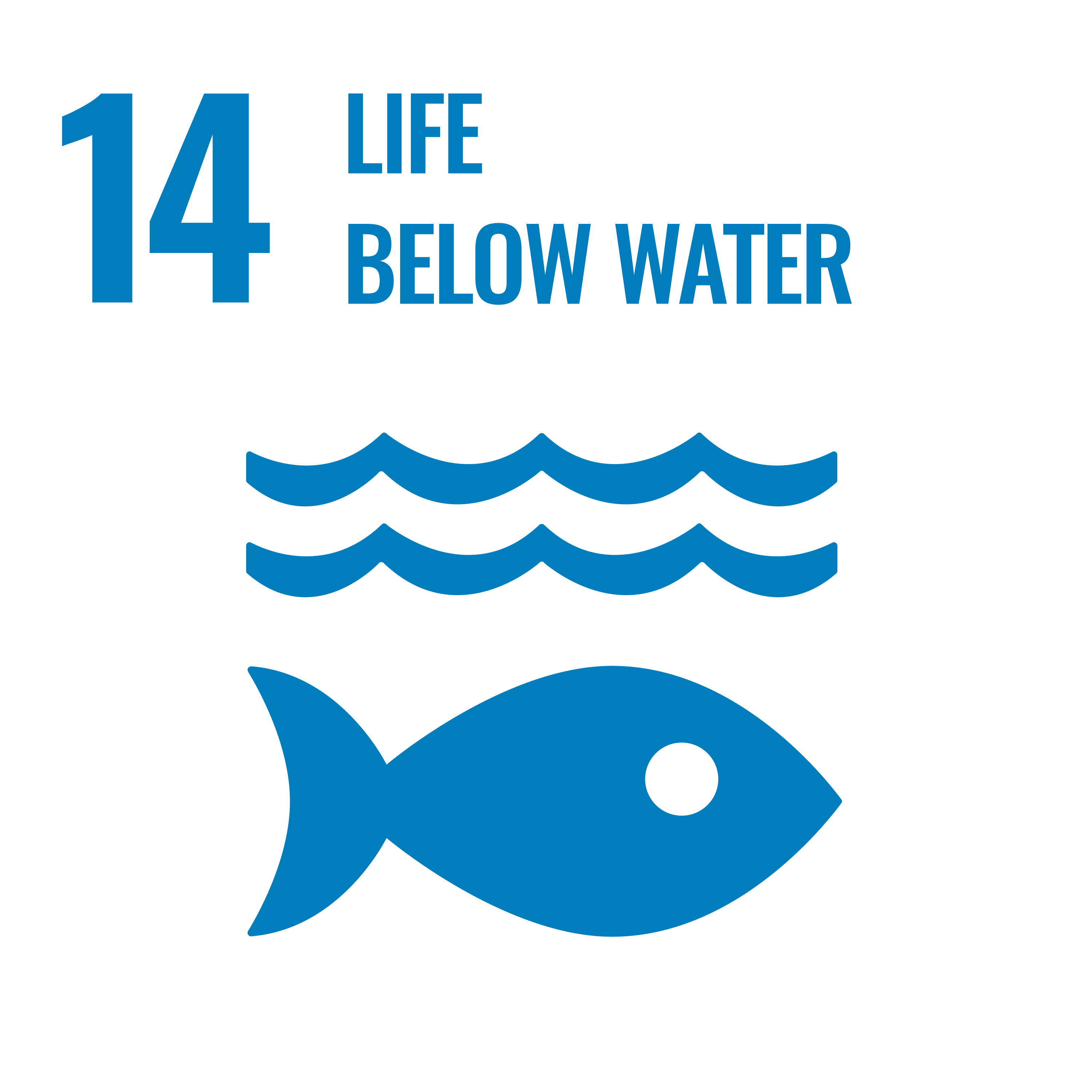}}} & \parbox[t]{\SDGright\textwidth}{\textbf{14 Life below water:\ Conserve and sustainably use the oceans, seas and marine resources for sustainable development}
\vspace{\recskip}
\begin{itemize}[leftmargin=20pt]
\setlength{\itemsep}{\recskip}
\item Some of the \ACR\ experiments and facilities are built within or close to aquatic ecosystems, \eg Antarctica, and therefore affect these both directly and indirectly.
\item Many goods, products and experiments used in research are travelling the oceans prior to use.
\item The industries that produce the goods that we consume use water and produce waste products, some of which ends up in the ocean.
\item The behaviour and lifestyle choices of our community in professional and private life have an impact on the oceans, through the demand for clean water, and the production of waste water and residues, including microplastics.
\end{itemize}}\\

\parbox[t]{\SDGleft\textwidth}{\raisebox{\iconskip}{\includegraphics[scale=\SDGscale]{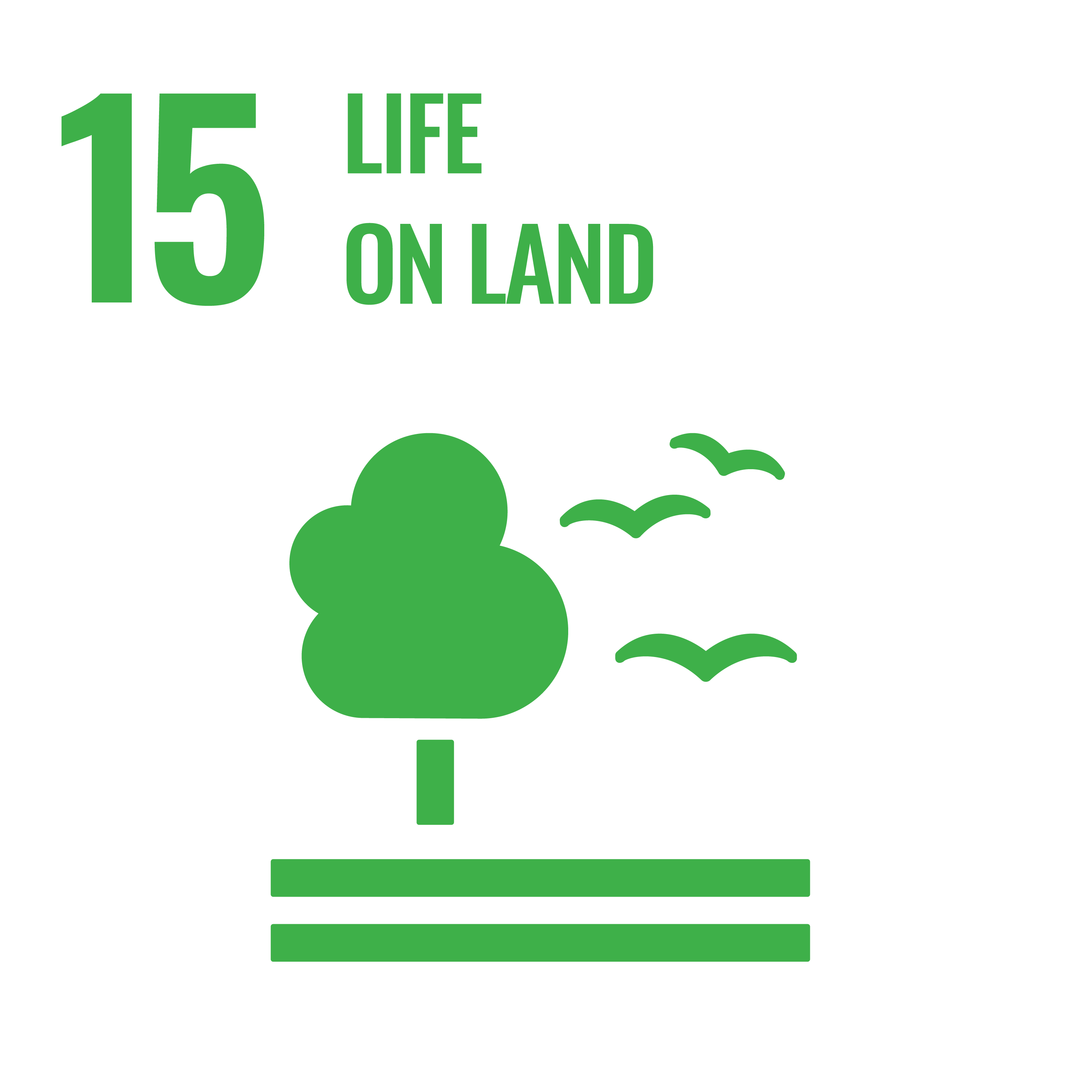}}} & \parbox[t]{\SDGright\textwidth}{\textbf{15 Life on land:\ Protect, restore and promote sustainable use of terrestrial ecosystems, sustainably manage forests, combat desertification, and halt and reverse land degradation and halt biodiversity loss}
\vspace{\recskip}
\begin{itemize}[leftmargin=20pt]
\setlength{\itemsep}{\recskip}
\item Campuses are ecosystems.
\item Expanding the campuses of research institutes can have an impact on surrounding ecosystems.
\item Our consumption has direct (\eg deforestation for agriculture and construction) and indirect (\eg our emissions give rise to more frequent extreme weather events) effects on land use, damaging ecosystems.
\item The behaviour and lifestyle choices of our community in professional and private life have an impact on the land and its ecosystems, because of the extraction of resources and the production of waste or residues.
\end{itemize}}\\

\parbox[t]{\SDGleft\textwidth}{\raisebox{\iconskip}{\includegraphics[scale=\SDGscale]{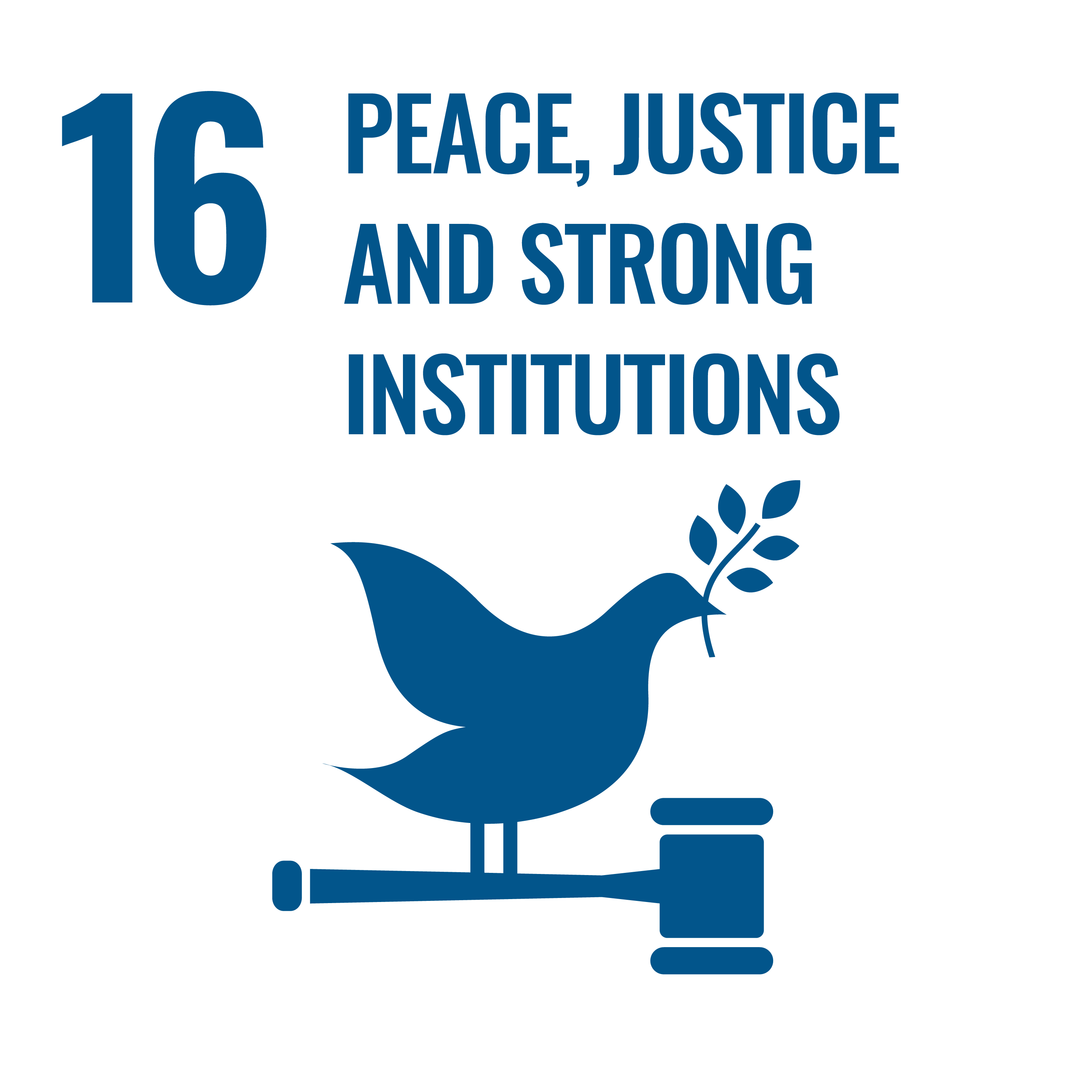}}} & \parbox[t]{\SDGright\textwidth}{\textbf{16 Peace, justice and strong institutions:\ Promote peaceful and inclusive societies for sustainable development, provide access to justice for all and build effective, accountable and inclusive institutions at all levels}
\vspace{\recskip}
\begin{itemize}[leftmargin=20pt]
\setlength{\itemsep}{\recskip}
\item \ACR\ is an international field demonstrating harmonious partnership in working towards common goals, and can serve as a model for peaceful international collaboration.
\item \ACR\ is part of society, and it is composed of institutions that can help shape the societies and politics within which they are embedded.
\item Large-scale \ACR\ projects can have a positive impact on industrial and political partnerships.
\item Transparent reporting on efforts towards more sustainable research has a positive impact on the credibility of scientists, and helps avoid greenwashing.
\end{itemize}}\\

\parbox[t]{\SDGleft\textwidth}{\raisebox{\iconskip}{\includegraphics[scale=\SDGscale]{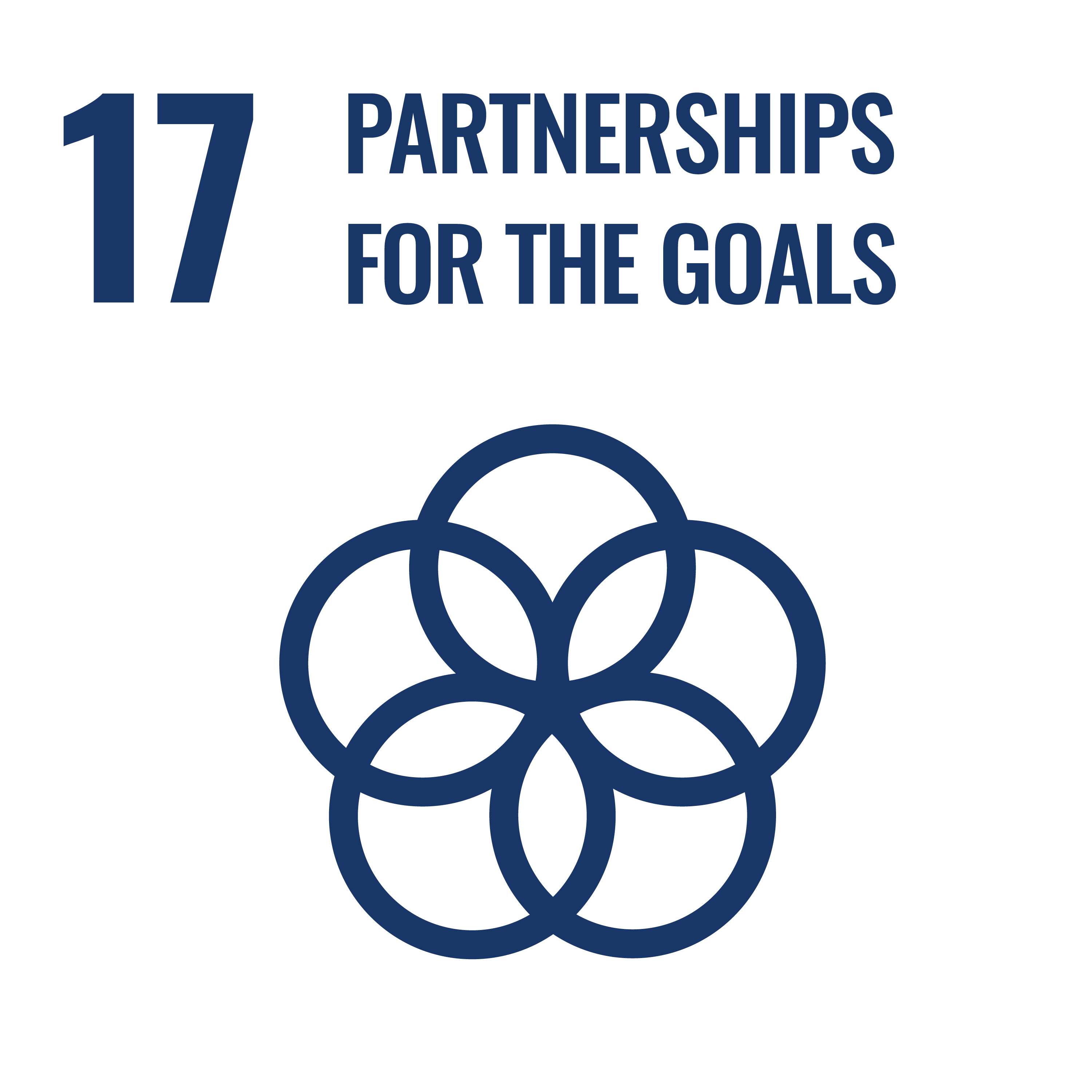}}} & \parbox[t]{\SDGright\textwidth}{\textbf{17 Partnership for the Goals:\ Strengthen the means of implementation and revitalize the global partnership for sustainable development}
\vspace{\recskip}
\begin{itemize}[leftmargin=20pt]
\setlength{\itemsep}{\recskip}
\item As an international community and a driver of research and innovation, we can influence our partners and work together to strengthen a sustainable society around the globe.
\end{itemize}}

\end{longtable*}
\renewcommand*{\arraystretch}{1}

\RaggedRight
\sloppy
\newpage


\section{Computing}
\label{sec:Computing}


\begin{center}
\includegraphics[width=\SDGsize]{Sections/Figs/Common/SDG_7_CleanEnergy.png}~%
\includegraphics[width=\SDGsize]{Sections/Figs/Common/SDG_9_IndustryInnovation.png}~%
\includegraphics[width=\SDGsize]{Sections/Figs/Common/SDG_12_ResponsibleConsumption.png}~%
\includegraphics[width=\SDGsize]{Sections/Figs/Common/SDG_13_ClimateAction.png}
\end{center}


\exSum

\noindent Computing is an integral part of basic research, being used for theoretical modelling, simulation (including lattice simulation), and data analysis. With increasing data sets and demands for accuracy, computing resource consumption is expected to rise. 
This poses concerns in the context of climate sustainability. 
Within \ACR, \eg the High-Luminosity phase of the Large Hadron Collider (\acrshort{hl-lhc}), expected to be operational from the end of this decade, will require 5 to 10 times the computing capacity needed for the Large Hadron Collider (LHC), with data storage needs reaching about ten exabytes~\cite{ATLAS_Computing_HLLHC_Roadmap}. 
At the same time, some lattice quantum chromodynamics (QCD) calculations, applied, \eg to studying heavy quark decays and anomalous magnetic moments, can be too expensive to pursue, even if approximately 10\% of open-science super-computing in the United States is devoted to such studies~\cite{Shanahan}.
Up to 88\% of the electricity consumption of an astronomy researcher at MPIA, shown in \fref{fig:Intro-ComparativeEmissions}, is due to (super)computing~\cite{Jahnke2020}, and CERN's (now defunct) data centre in Hungary is responsible for a third of its electricity emissions when the LHC is not running~\cite{Environment:2737239}.\footnote{The total emissions due to electricity use of a CERN researcher during LHC shutdown is about a third that of a Fermilab researcher (see Appendix~\ref{sec:DataforFig1.4} for details). CERN, however, uses mainly the French grid with its low-carbon energy, reducing its computing carbon footprint. For further discussion, see \sref{sec:Energy}.}

\ACR\ research infrastructure ranges from local and portable computing, to high-performance computing (HPC) and high-throughput computing (HTC)\footnote{Generally, synchronisation requirements of large parallel HPC applications place substantial constraints on runtime scheduling choices and use of power-saving functionalities.  HTC applications, on the other hand, can be naturally run in parallel, but are constrained by memory consumption, and data access and transfer.} in centralised computing centres that {---} depending on the application {---} deal with large volumes of experimental data. As an international community, we also rely on communication technologies and the ability to move these large volumes of data around the globe. The infrastructure we use to do so, comprising hardware, the data centres within which the hardware is housed, and cloud computing resources used for data storage, contributes to our community's energy consumption and the waste that our research generates.  Furthermore, the energy efficiency of hardware is ultimately limited by the efficiency of the computer programmes that run on this hardware, making the \acrshort{ghg} emissions of \ACR\ researchers dependent upon the choice of software architecture.

This chapter covers sustainability in procurement, and  extending and optimising the life cycle of computing equipment in \sref{subsec:hardware}, choice and optimisation of software in \sref{subsec:software}, and energy savings in data centres in \sref{subsec:infrastructure}.  For a full discussion of sustainable sourcing in a broader context, as well as information on E-waste and its impact, see \sref{sec:Waste}. A brief explanation of the life-cycle analysis used to estimate the cradle-to-grave environmental impact of infrastructure and technology can be found in \sref{sec:Technology}.  For other aspects of energy use, see \sref{sec:Energy}.

\newpage
\begin{reco2}{\currentname}
{
\begin{itemize}[leftmargin=6 mm]
\setlength{\itemsep}{\recskip}
\item Make sustainable personal computing choices by considering the necessity of hardware upgrades, the repurposing of hardware, and the environmental credentials of suppliers and their products. 

\item Assess and improve the efficiency and portability of codes by considering, \eg the required resolutions and accuracy.

\item Assess and optimise data transmission and storage needs.

\item Follow best practice in open-access data publishing, prioritising reproducibility and limiting repeat processing.

\item Read the discussions on E-waste, right to repair and sustainable procurement in Section~\ref{sec:Waste}.

\end{itemize}
}
{
\begin{itemize}[leftmargin=6 mm]
\setlength{\itemsep}{\recskip}
\item Right-size IT requirements and optimise hardware life cycles.
\item Schedule queueing systems with environmental sustainability in mind, so as to maximise the use of renewables, accounting for the geographical location of servers/data centres.
\end{itemize}
}
{
\begin{itemize}[leftmargin=6 mm]
\setlength{\itemsep}{\recskip}
\item Ensure that environmental sustainability is a core consideration when designing and choosing sites for large computing infrastructure, such as data centres, including, e.g., the availability of renewables, the efficiency of cooling systems and the reuse of waste heat.

\item Proceduralise the repair, upgrade and repurposing of existing computing, the de-inventorising of personal equipment for leaving personnel or for donation, and the responsible recycling of retired hardware.

\item Select cloud computing services for their carbon emission mitigation policies.

\end{itemize}
}
\footnotetext{Some of the above recommendations are based on those made by Jan Rybizki~\cite{Rybizki}.}
\end{reco2}


\newpage

\subsection{Hardware}
\label{subsec:hardware}

When considering the future of sustainability in \ACR, the hardware aspect of computing is of great concern. Hardware is both energy- and resource-consuming. The manufacture, transport, energy consumption, and disposal of each piece of hardware contribute substantially to the environmental footprint of the HPC that \ACR\ relies on to analyse large swathes of data. 

Manufacture is the largest source of hardware GHG emissions, with primarily fossil-fuel-powered manufacturing chains contributing as much as 80--85\% of lifetime emissions of a personal computing device~\cite{Greenpeace_Oeko, OxfordLCA}.  Moreover, production is notoriously resource-intensive~\cite{Greenpeace_Oeko}, with the mining of the necessary metals and `conflict' minerals responsible for a number of negative environmental and social effects.  Improper disposal of substances found in computing equipment is also linked to environmental hazards and a variety of other risks.    For an in-depth discussion, see \sref{sec:Waste}.

One way to mitigate the impact due to production is by purchasing modular equipment, which allows for easy upgrades and repurposing of hardware.  In fact, the extension of hardware lifetime has been increasingly demonstrated to have major benefits over upgrading to more efficient technology.  A study by the University of Edinburgh Department for Social Responsibility and Sustainability~\cite{UniEd} found that simply using 174 computer monitors for six years instead of four saved 33 \acrshort{tco2e}, which, when incorporated into standard practice, would not only reduce purchasing costs, but would result in annual GHG savings of 380 \tCdOe.  It is also crucial that institutions be able and willing to support repairs. This applies in particular to personal equipment, \eg laptops,  which come with additional peripherals such as display, keyboard, and housing, as compared with HPC units in data centres. 

Furthermore, prioritising suppliers that implement sustainable sourcing, including recovery of secondary materials, and manufacturing methods would partially mitigate the resource burden, as would enabling circularity and appropriate E-waste recycling.  As one example, TCO certification \cite{TCO_Certified} is the world-leading sustainability certification for IT products. TCO-certified compliance is independently verified both pre- and post-certification. TCO certification also covers data centre products, which could be given preference over uncertified ones for cluster computing. For more information on sustainable procurement, including some hallmarks of sustainability in raw materials supply chains, see \sref{subsec:Resources}.  For further discussion of E-waste, see \sref{subsec:Waste}.

A secondary source of hardware emissions is energy consumption during its use~\cite{Greenpeace_Oeko}, with the majority coming from processors, memory, and runtime of jobs.  Processor upgrades and the optimisation of memory type can greatly reduce energy consumption. See \csref{case:LHCb} for details of energy-efficient hardware purchase at the LHCb experiment at CERN.

It is important to ensure `energy proportionality' in hardware use, \ie that energy consumption is proportional to computing performance over the full range of applications \cite{energy-prop-computing}.    Often, hardware designed to be most efficient at maximum performance load in practice spends most of its time idle, or performing less intensive computations. This can be addressed by, \eg running jobs at high utilisation rate on as few servers as possible.  

Implementing parallelisation within processors can also reduce the number of processors needed, and by replacing central processing units (CPU) with graphics processing units (GPU), the energy usage can be reduced.  For certain tasks relevant to \ACR\ applications, other even more specialised processors are available, such as Google's tensor processing unit (TPU)~\cite{Tensor}.  This consumes less power than its predecessors, although it suffers from poor energy proportionality: at 10\% load, it consumes almost 90\% of the power it would consume at 100\% load \cite{TensorPerformance}.

However, it should always be tested whether parallelisation does reduce the overall energy usage of a task, as an increase in energy consumption per second could counteract the benefits of reduced runtime.  Another aspect to take into consideration when implementing parallelisation is the particular application, and its requirements in terms of memory, scalability, and data access. Reference~\cite{Abdurachmanov:2014xka} discusses these issues in the context of the Worldwide Large Hadron Collider Computing Grid (WLCG).  It also gives suggestions for power-aware software applications and scheduling that could reduce power consumption.  Some of the advocated changes are software specific and are further detailed in \sref{subsec:software}. The Green500 list~\cite{GreenList} ranks the most energy efficient high-performance computing systems. 
The GHG emissions of the computer centres that house them, however, depend critically on their infrastructure. This aspect is further discussed in~\sref{subsec:infrastructure}.


\subsection{Software}\label{subsec:software}

Software is integral to the work of \ACR. It underpins how the global \ACR\ community communicates, shares data, produces papers and graphics, and acquires, manages, processes and analyses huge amounts of data from experiments, observatories and simulations.

It is therefore pivotal that the software developed and used by the \ACR\ community is efficient in order to minimise CPU hours, and to facilitate data sharing and long-term reproducibility. This requires a balance to be struck between portability and optimisation for particular architectures.  While not directly linked to environmental sustainability, initiatives focused on software sustainability in \ACR, such as the Institution for Research and Innovation in Software for High Energy Physics (IRIS-HEP)~\cite{IRISHEP} and the HEP Software Foundation~\cite{HSF}, may provide an important platform for accelerating the inclusion of  environmental considerations in software development. Doing so is compatible with the FAIR principles~\cite{FAIR} for scientific data management, that software (and data) should be Findable, Accessible, Interoperable and Reusable.

Much of the code used in \ACR\ computing relies on libraries and public codes. Experiments use general frameworks and software infrastructure provided by experts in the experiments. They can have a tremendous impact on the energy efficiency of the employed code and, in some cases, work to meet strict requirements posed by the computing environment. Decisions on the computing language employed can be crucial, with Fortran and C/C++ specifically suited for numerical calculations, while others prioritise convenience or readability over performance.  Changes in processor architecture have been utilised through dedicated and collaborative  efforts, leading to a factor of 2 improvement in the performance (and energy efficiency) of the reconstruction code of the ATLAS experiment at CERN~\cite{ATL-SOFT-PUB-2021-002}. Other examples of software improvement are recent changes to the software framework and architecture at LHCb (see \csref{case:LHCb}) and improvements in a Monte Carlo (MC) generator core code, having led to an improvement in speed of a factor of 50 (see \bpref{CS:SoftwareOptimisation}).  In the case of cosmological analyses, it has been suggested that the Likelihood Inference Neural Network Accelerator (LINNA) can lead to efficiency increases that would save  \$300,000 in energy costs and around 2,200 \acrshort{tco2} in first-year Rubin Observatory's Legacy Survey of Space and Time (LSST) analyses~\cite{To:2022ubu}.

Sustainable use of software can also be encouraged at an individual level. The energy used in a job directly correlates with the memory assigned/available for a job, so mitigation by individuals can be easily implemented through assigning the correct memory used and by optimising code~\cite{Karayakin}. Further examples of conscientious use of software include limiting resolution or precision to that which is necessary, effective testing to avoid wasted CPU hours, good practice in data retention to avoid data loss and the need to rerun analysis or simulations, and scheduling CPU hours when a higher percentage of the local energy mix is from renewables.

\begin{bestpractice}[Optimisation of software\label{CS:SoftwareOptimisation}]{Optimization of software}%
A targeted effort enabled by the UK-based SoftWare and InFrastructure Technology for High Energy Physics (SWIFT-HEP)~\cite{SWIFTHEP} project recently brought together experimentalists and Monte Carlo (MC) developers to greatly improve the computational efficiency of higher-order perturbative calculations by focussing on two major components of general purpose MC event generators: The evaluation of parton-distribution functions along with the generation of perturbative matrix elements. A dedicated CPU profiling illustrated that for the cost-driving event samples employed by the ATLAS experiment at CERN to model irreducible Standard Model backgrounds, these components dominate the overall run time by up to 80\%. Improved interpolation and caching strategies in the main evaluation tool for parton-distribution functions used by the experiments~\cite{Buckley:2014ana}, along with the introduction of a simplified pilot run in the MC generator Sherpa~\cite{Sherpa} achieves a reduction of the computing footprint by factors of around 50, while maintaining the formal accuracy~\cite{Bothmann:2022thx}. The speed-up translates into a direct CPU (and hence energy) saving, paving the way towards affordable and sustainable state-of-the-art event simulation in the \acrshort{hl-lhc} era.

\end{bestpractice}


\subsection{Infrastructure}\label{subsec:infrastructure}

Even the most energy-efficient data centres are not environmentally sustainable if they are powered by carbon-based fuels~\cite{Bloom:2022gux}.  
However, provided energy from renewable sources is available, this can be easily addressed. There are many advantages to doing so, owing to the flexibility of HTC.
Inherent fluctuations in supply of electricity from renewables can be managed using a smart queueing system that runs jobs at times where electricity has a large renewable component,
 or directs them to data centres where this is the case.  
Moreover, a carefully managed HTC system can even help stabilise fully renewable power grids in response to local imbalances in supply and demand: 
an instantaneous reduction in the CPU clock frequency by up to 60\% ensures per-second grid stabilisation~\cite{bruers2023resourceawareresearchuniversematter},
and a similar technique can be employed on longer time scales, \eg hourly, in response to changes in the carbon intensity of electricity (or equivalently, market price). 
For longer periods, with higher latency, this can also involve powering down nodes. 
The reduced work can be compensated by operating older hardware longer, but only when the electricity price is low.  
See \sref{subsec:grid} for further discussion of renewables-based grid infrastructure.  

Another source of GHG emissions associated with computing is the construction and operation of the large data centres within which IT equipment is housed.  Although emissions due to construction can be significant, particularly if concrete is used, our focus in the remainder of the section will be on cooling the facilities and equipment, which is responsible, on average, for almost one third of facility power use. A judicious choice of location for the centre can minimise these energy costs, by provision of a cooler external environment, or other means to cool efficiently.  Proximity to a large body of water, \eg could make water cooling an attractive option.  Care must be taken, however, to ensure minimal disruption to the natural environment.  Waste heat from the data centre can also be reused to heat nearby infrastructure. For examples of best practice in data centre design and construction, see \bpref{BP:CSCS}, \bpref{BP:Prevessin}, and \bpref{BP:GreenITCube}.  For more information on energy-efficient LHCb computing infrastructure, see \csref{case:LHCb}. 


\begin{bestpractice}[\label{BP:CSCS}Cooling in Swiss National Supercomputing Centre\\
{\footnotesize \noindent Information taken from CSCS fact sheets \cite{CSCSLakeWater,CSCSInnovativeBuilding} and vetted by the organisation.}]{Cooling in Swiss National Supercomputing Centre (CSCS)}%
    The Swiss National Supercomputing Centre (CSCS) is a three-floor concrete building in Lugano that houses the “Piz Daint” supercomputer and the system used by MeteoSwiss for weather predictions, among other things.  It currently operates at a Power Usage Effectiveness (\acrshort{pue}) rating of 1.20 at 25\% of full load, with a design PUE of 1.25. At CSCS, high-efficiency cooling is achieved with a state-of-the art cooling system using the water from Lake Lugano, extracted at a depth of 45 m and a temperature of 6\degree C. (Lake Lugano is a 288 m deep glacial lake.) 420 litres of this water per second are pumped to the facility over a distance of 2.8 km into large heat exchangers, where it meets and cools the water in the internal cooling circuit for the supercomputers. The resulting warmer water is then sent to a heat exchanger in a second cooling circuit, which cools the components with a lower thermal sensitivity, as well as the building itself in the summer, before being returned to the lake. The return flow of water falling back into the lake is used to produce electricity via a microturbine in the pumping station further reducing the power consumption of the pumps by 30\%.  Due to modular cooling and room concepts, the different parts of the facility are equipped only as necessary.  Not only does this reduce the initial budgetary outlay, but it also results in increased flexibility to react to future hardware needs, while keeping the PUE close to its final design value from the outset.
\end{bestpractice}

\begin{bestpractice}[\label{BP:Prevessin}Sustainable design for Pr\'evessin Computing Centre (PCC), CERN\\{\footnotesize \noindent Edited contribution from Wayne Salter, IT Project Manager for the PCC.}]{Sustainable design for Pr\'evessin Computing Centre, CERN}%
CERN has, for some time, expressed the desire to build a second Data Centre (DC) on its Pr\'{e}vessin site (named the PCC) to augment the capacity being provided by its Meyrin DC, in particular in light of the increased demands from the LHC experiments in the HL-LHC era.
In 2019, a project was approved to build a turn-key DC with an initial capacity for computing of 4 MW, but with the possibility to upgrade the IT capacity in two steps to 8 MW and finally to 12 MW. 
A call for tender was initiated at the end of 2019 for the design, construction and 10-year operation and
maintenance of a new DC, and the result of the tender was
adjudicated at the CERN Finance Committee in December 2020. A contract was signed with the winning
consortium in July 2021 and construction began at the beginning of 2022.
The DC is expected to be operational in the final quarter of 2023. An
important aspect included in the thinking for the new DC was
sustainability and, in particular, energy efficiency. As such, the
specification required a target \acrshort{pue} of 1.1,
but contractually allows for a PUE of no worse than 1.15, for energy
recuperation of at least 25\% of the heat generated by the IT equipment
and for a roof with vegetation.

When considering the increased energy efficiency compared with CERN's
existing Meyrin DC, which now has a PUE of around 1.5 after
many years of efforts to bring this down, this equates to significant
energy savings. Assuming the PCC running at full first-phase capacity of
4 MW with a PUE of 1.1, cf.~1.5 for the current CERN Data Centre, then
the annual saving in terms of electricity would be 14 GWh. Obviously,
should the PCC eventually be upgraded and used at its full final
capacity then the savings could be tripled, cf.~with running a similar
capacity with the PUE of the current Meyrin DC. It should be
noted that the PUE of the current data centre is the result of many
years of efforts to improve the energy efficiency, which have
substantially reduced its PUE, but that further improvements would now
be complex and costly.

In addition to aiming for high energy efficiency, the design of the PCC
also allows for the heat produced by the IT equipment to be recuperated
and used to help power a new building heating plant that will soon be
built close to the PCC to replace an existing ageing and inefficient
heating plant. 
The specification for the PCC required the possibility
to recover a minimum of 25\% of the generated heat per phase, implying
1.3 MW per phase leading to a total of 3 MW once the full 12 MW
configuration would be operational. However, during the design phase, it
has been decided to request 3 MW already during the first phase. In the
second phase, the heat recuperation will be increased to 4 MW.

During hot weather, water is sprayed on the heat exchanger elements of
the dry coolers to improve their efficiency. In the original design, this
water was lost, resulting in a significant water consumption over
the year. However, with sustainability and environmental protection
considerations in mind, it was decided to make efforts to reduce the
water consumption as far as possible without impacting the efficiency of
the cooling solution. As such, it was agreed with the contractor to
change the design to include water re-circulation at the level of the dry
coolers and hence substantially reduce the water consumption. In the
first phase, the annual water consumption is estimated to be reduced by almost 60\% from
21,455 ${\rm m}^3$ to 8,645 ${\rm m}^3$, based on the average meteorological data for the area.

To further improve sustainability and to make the building more
ecologically friendly, it was decided to request that vegetation be
planted on the roof of the building, which is effectively in two halves.
The first half contains the IT rooms (two per floor for three floors) and
the second half is for all the technical rooms. The roof is similarly
split in two. The first half is used for the dry coolers and associated
technical infrastructure and hence cannot be used for vegetation, but
the second half will be planted with grass covering an area of
approximately 1,250 ${\rm m}^2$.
\end{bestpractice}

\begin{bestpractice}[\label{BP:GreenITCube}The Green-IT Cube at GSI/ FAIR~\cite{GreenITCube}\\{\noindent\footnotesize Edited contribution from Tetyana Galatyuk on behalf of KHuK (Komitee f\"{u}r Hadronen- und Kernphysik).}]{Green-IT Cube}%
\noindent At  GSI Helmholtzzentrum für Schwerionenforschung in Darmstadt, the Green-IT Cube~\cite{GreenITCube} was constructed in 2014 to host the computing systems of the FAIR particle accelerator facility under construction close to GSI, as well as numerous other scientific computing systems. It has a total capacity of 12 MW and 768 racks, distributed over 6 floors. The partial \acrshort{pue} of the installation, that is the PUE across some part of the data centre,\footnote{For a detailed explanation of partial PUE, see, \eg Sec.~VII of Ref.~\cite{GreenGridPUE}.} reaches 1.07 at a load of less than 25\%, which meets the design value. In acceptance testing at higher loads an even better partial PUE has been observed.
 
This became possible due to the award-winning~\cite{GreenITCubeAward} innovative design of the Green-IT Cube, which was developed at the Frankfurt Institute of Advanced Studies. The innovative design based on water cooled back-door heat exchangers allowed not only for a low PUE, but also for an advanced 3D building design, which reduced the ground print of the compact data center. At the same time, it reduced the building material needed, further reducing the environmental impact. Parts of the excess heat are used to heat office buildings on the GSI campus.
 
The patented design has received many innovation and data center awards and was successfully transferred into industry.
\end{bestpractice}


\RaggedRight
\sloppy

\newpage
\section{Energy}
\label{sec:Energy}

\exSum


\begin{center}
\includegraphics[width=\SDGsize]{Sections/Figs/Common/SDG_7_CleanEnergy.png}~%
\includegraphics[width=\SDGsize]{Sections/Figs/Common/SDG_9_IndustryInnovation.png}~%
\includegraphics[width=\SDGsize]{Sections/Figs/Common/SDG_11_SustainableCities.png}~%
\includegraphics[width=\SDGsize]{Sections/Figs/Common/SDG_12_ResponsibleConsumption.png}~%
\includegraphics[width=\SDGsize]{Sections/Figs/Common/SDG_13_ClimateAction.png}~%
\includegraphics[width=\SDGsize]{Sections/Figs/Common/SDG_16_PeaceJusticeStrongInstitutions.png}~%
\includegraphics[width=\SDGsize]{Sections/Figs/Common/SDG_17_PartnershipForGoals.png}

\end{center}


\noindent
The operation of experimental equipment and computing facilities at large-scale research centres has a significant energy footprint. 
In addition, energy is required for the construction and disassembly of infrastructure, for heating and cooling buildings, and for transport of goods and people. To comply with the Paris Agreement, future facilities must be effectively climate-neutral, and this presents a significant challenge for \ACR.

Particle collider experiments are particularly power-hungry, with the LHC at CERN being a prominent example. With its particle accelerators, detectors and extensive infrastructure, CERN consumes up to 1.2~TWh of electricity annually, of which 55\% is due to LHC operations~\cite{Environment:2737239,CERNenergymanagement, CERNEnvironment2023}. CERN plans to significantly increase the scale of its installations in a push towards higher energies and intensities. Doing so responsibly will require a concerted effort to minimise power consumption and increase the energy efficiency of the infrastructure, and a careful analysis of how to source the remaining energy needs sustainably.

CERN receives most of its electricity from the French grid, which is currently characterised by a high share of low-carbon nuclear power, suppressing its electricity-related \CdO\ emissions in comparison with other facilities  (see \fref{fig:Intro-ComparativeEmissions}).\footnote{CERN's annual electricity emissions range from 9,000 \acrshort{tco2e} (LHC in shutdown) to 15,000 \tCdOe\ (LHC in operation)~\cite{Environment:2737239}.} However, taking into consideration the decreasing share of nuclear power in the French grid over the last 15 years, as well as the wider common European electricity market, where fossil fuels account for 37\% of electricity production on average~\cite{BP2022}, the outlook is more worrying.\footnote{For a live visualisation of the carbon emissions of electricity by country, see Ref.~\cite{ElectricityMap}.}

\begin{figure}[t]
     \centering
    \includegraphics[width=0.9\linewidth]{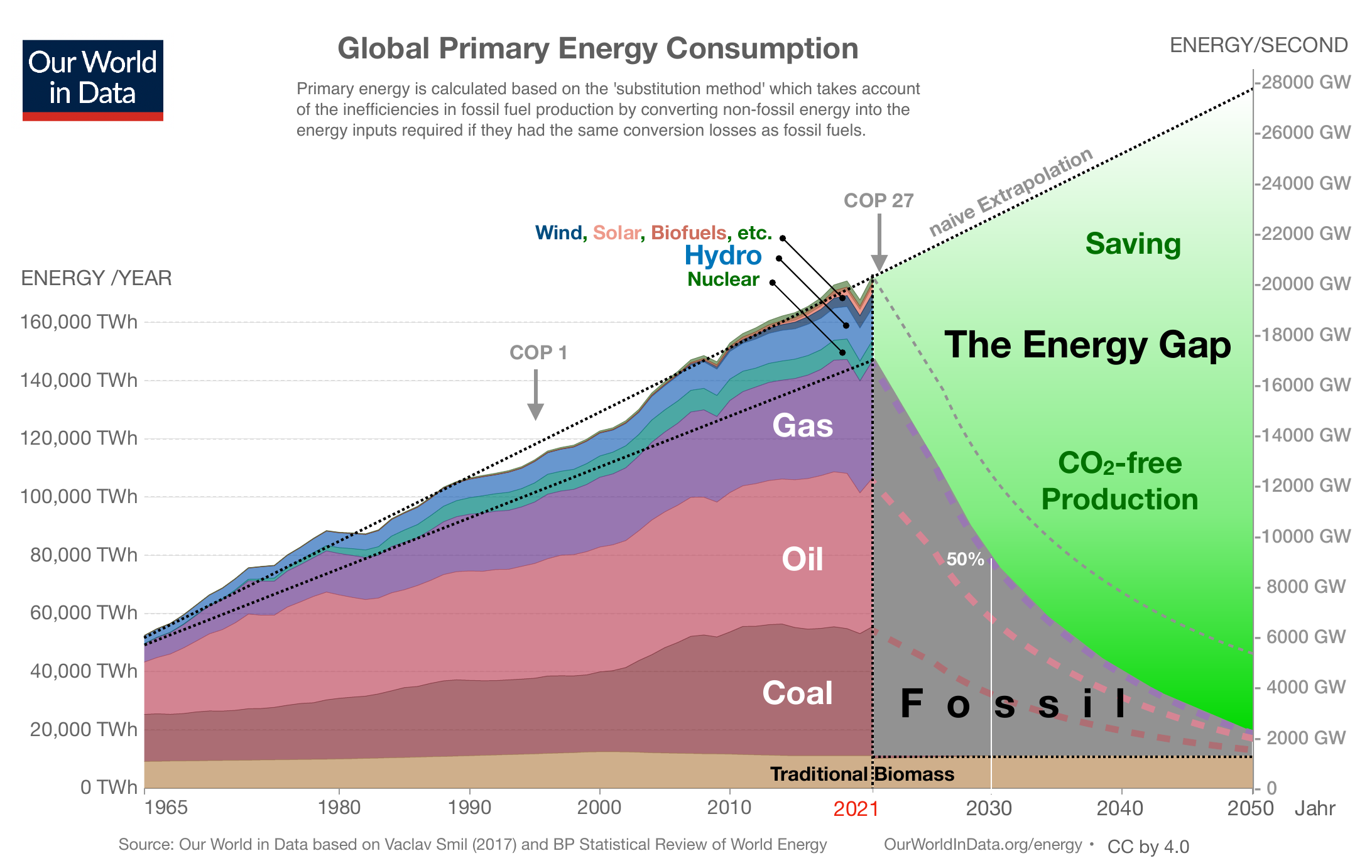}     
    \caption[Primary energy consumption is dominated by fossil fuels]%
        {Global primary energy consumption is dominated by fossil fuels, the use of which has been increasing steadily despite repeated warnings from the climate change conferences of the United Nations (COP) dating as far back as 1995. Decreasing emissions by 50\%, as recommended by the IPCC to avoid irreversible tipping points~\cite{OECDTippingPoints} (see blue line in \fref{fig:ene-netco2}) creates a large energy gap that must be filled by additional climate-neutral power generation, or by energy savings and recuperation. Consumption was extrapolated linearly from 1965--2021 to account for additional demand from emerging countries. Left part of figure taken from Ref.~\cite{OWDgap}, based on data from Refs.~\cite{Smil,BP2022}, reused and adapted under the terms of the \href{https://creativecommons.org/licenses/by/4.0/}{Creative Commons Attribution 4.0 International (CC BY 4.0) license}.\label{fig:ene-gap}}
 \end{figure}

It is important to place the energy needs of \ACR\ research infrastructure within the context of the world's necessary and rapid transition to zero-carbon energy sources. Global primary energy consumption in 2019 was approximately 160,000~TWh (equivalent to an average power consumption of 18,000~GW), around 80\% of which comes from \CdO-emitting fossil fuels~\cite{OWDfuels}. Moreover, demand is rising, primarily due to the growth and industrialisation of emerging economies.  A 50\% reduction of \CdO\ emissions by 2030, as stipulated in the Paris Agreement in combination with the latest IPCC scenarios (see~\fref{fig:ene-netco2}), will create a huge global energy gap, as shown in \fref{fig:ene-gap}~\cite{physikkonkret}. 


Many experimental technologies such as \CdO\ capture and storage (CCS) will not be viable for large-scale implementation within this short time frame~\cite{IPCCMitigationReport}. Filling the energy gap with solar, wind, and nuclear power requires upscaling existing facilities by more than an order of magnitude within seven years, a not inconsiderable task. Therefore, substantial increases in energy efficiency will be indispensable. Even tripling the output of existing solar, wind and nuclear installations by 2030 (which may itself be unrealistic), bridging the fossil energy gap would still require a 40\% global efficiency increase compared to today. Substantial energy savings requires systemic changes in technology and behaviour, such as transitioning from combustion engines to electric motors, from gas heating to heat pumps and from cars to rail. The global situation is therefore likely to result in energy becoming scarce and expensive in the coming decades, with the potential to directly limit our capabilities to conduct energy-intensive experiments and data analysis in basic research.

Section~\ref{sec:Ene-Transition} elaborates on the wider context of global production of low-carbon energy and focuses on potential sources of sustainable energy for \ACR\ research infrastructure, as well as energy savings and recuperation in \sref{sec:Ene-Saving}. Saving energy through structural and organisational changes are described elsewhere:\ see \sref{sec:Computing} for computing, \sref{sec:Travel} for mobility, and \sref{sec:Waste} for procurement.


\clearpage

\begin{reco2}{\currentname}
{
\begin{itemize}[leftmargin=6 mm]
\setlength{\itemsep}{\recskip}
\item Save energy in all ways practicable, e.g., by avoiding unnecessary heating or cooling of workspace, and by turning off electrical items when not in use.

\item Read the sections about computing (\sref{sec:Computing}) and mobility (\sref{sec:Travel}).

\end{itemize}
}
{
\begin{itemize}[leftmargin=6 mm]
\setlength{\itemsep}{\recskip}
\item Ensure that energy efficiency is a major focus in experimental design, and prioritise technologies that minimise consumption and maximise energy recovery.

\item Monitor, report, and assess energy usage with the aim of reducing consumption and resulting emissions.

\item Read the section on research infrastructure and technology (\sref{sec:Technology}).
\end{itemize}
}
{
\begin{itemize}[leftmargin=6 mm]
\setlength{\itemsep}{\recskip}
\item Ensure that energy efficiency is a major factor in the renovation of existing estates and the design and construction of new infrastructure.

\item Prioritise moving to sustainable and renewable energy sources via both local generation, and energy import and export.

\item Collate and publish energy usage and emissions statistics, stratifying by source, e.g., heating, experimental infrastructure, computing, transportation, and procurement.

\item Advocate for environmentally sustainable energy policy.
\end{itemize}}

\end{reco2}


\newpage

\subsection{Low-Carbon Energy}
\label{sec:Ene-Transition}

\begin{figure}[!h]
     \centering
     \includegraphics[width=\linewidth]{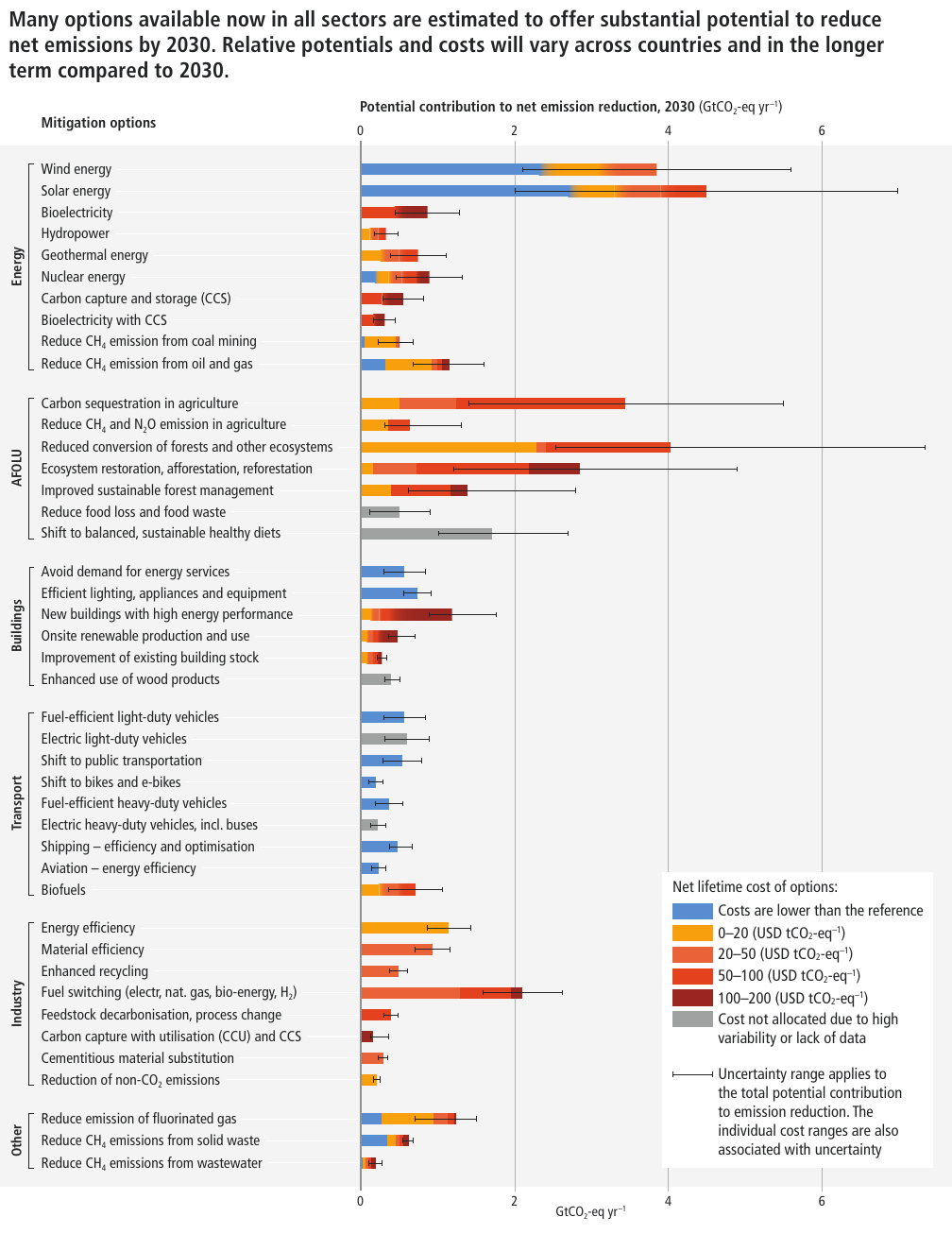}
     \caption[IPCC mitigation potential for alternative sources of energy]{Figure taken from the IPCC Sixth Assessment Report~\cite{IPCCMitigationReport}. [Original caption: Costs shown are net lifetime costs of avoided greenhouse gas emissions. Costs are calculated relative to a reference technology. The assessments per sector were carried out using a common methodology, including definition of potentials, target year, reference scenarios, and cost definitions. The mitigation potential (shown in the horizontal axis) is the quantity of net greenhouse gas emission reductions that can be achieved by a given mitigation option relative to a specified emission baseline. Net greenhouse gas emission reductions are the sum of reduced emissions and/or enhanced sinks.  \emph{Continues overleaf.)}
\label{fig:ene-mitigation}
         }
\end{figure}

Transitioning the energy demands of \ACR\ research to \CdO-neutral sources will likely require a mix of sources: solar, wind, hydro, geothermal, and nuclear power, many of which will be strongly location dependent. Despite their relatively low cost (see \fref{fig:ene-mitigation}), these geographical limitations, combined with the challenges of rapidly scaling up grid renewables and nuclear power, 
(see the paragraph on nuclear power for details),
will make the transition to carbon-neutral energy difficult for \ACR. (See, however, the Green ILC initiative, \csref{cs:GreenILC}, for ongoing efforts to design a particle accelerator facility powered entirely by renewable sources.)  A recent report by the Carbon Tracker Initiative~\cite{CarbonTracker21} suggests large-scale transmission or import of sustainable energy as a solution for ``stretched'' regions, where energy demand outstrips local renewable supply by over an order of magnitude.  For further discussion, see \sref{subsec:import}, and \csref{case:desert} for an example use case. Alternatively, efforts could be made to site future facilities near abundant sources of renewable energy, which could have the additional benefit of contributing to developing economies over a longer period. 
The Synchrotron-Light for Experimental Science and Applications in the Middle East (SESAME) project ~\cite{Sesame} is one example (see \bpref{BP:sesame}).  International organisations like CERN,  with their history of cooperation across political and ideological boundaries, are uniquely positioned to spearhead such initiatives.

\begin{bestpractice}
[SESAME: The world’s first large accelerator complex fully powered by renewable energy \label{BP:sesame}]{SESAME: A solar-powered light source}%
\noindent SESAME, the accelerator complex providing "Synchrotron-light for Experimental Science and Applications in the Middle East", is an international research centre near Jordan's capital Amman~\cite{SesameWebpage2,IRBeamline}. SESAME was developed in the late 1990s with the intention to foster scientific cooperation in the Middle East under the auspices of UNESCO after the CERN model~\cite{SESAMEScienceDiplomacy}. SESAME currently operates five beamlines~\cite{IRBeamline,XAFSBeamline,XRPowerBeamline,HESEBbeamline,BEATBeamline} and aims to operate eight beamlines during its first phase~\cite{SesameWebpage}. The beamlines provide different synchrotron light wave lengths to enable a wide range of investigations related to chemistry, physics, materials science, medicine, and cultural heritage.

The SESAME storage ring is filled using two subsequent pre-acceleration stages, a microtron (providing electrons with energies of 22.5 MeV) and a booster synchrotron, accelerating electrons up to 800 MeV. The main storage ring is operated at 2.5 GeV, making SESAME a competitive third-generation light source~\cite{SESAMEScienceDiplomacy}.

The complex is remarkable in two aspects: It recycles the BESSY I microtron and uses also parts of the BESSY I in its booster synchrotron and is nominally fully powered by solar power. The replacement of the BESSY I located in West Berlin with a new facility located in the eastern part of the city after the German reunification triggered the idea to donate the infrastructure and eventually the founding of SESAME.

The European Union has invested significant funds towards the building of the main storage ring, and to ensure the success of the SESAME facility.  It also provided grants to the Jordanian government, which enabled the building of an on-grid solar plant about 30 km from the SESAME site. The USD 7.05 million plant inaugurated in 2019 has a 6.48 MW power capacity, which is enough to nominally power SESAME for the next few years of its operation. This is the first contribution of the EU towards the running costs of SESAME, out of which electricity consumption amounts to approximately USD 1 million per year (about 30–40\% of its annual budget)~\cite{SESAMEScienceDiplomacy}. As the power fed into the Jordan power grid is credited to SESAME, it will enable the running of the accelerator for free once the investment costs have been amortised. It should be noted, however, that this mode of operation only nominally satisfies the energy needs, but does not address issues with energy storage for running at night
\end{bestpractice}

\begin{casestudy}
    [Green ILC\label{cs:GreenILC}]{Green ILC}%
    The Green ILC is a site-specific project designing the ILC with sustainability considerations~\cite{GreenILC}.
    Producing and managing stable and reliable renewable energy for a high-precision machine, such as a particle accelerator, is a difficult task. However, the Green ILC project maintains that 100\% of the billion kWh of power needed per year to run the ILC could be sourced successfully from local renewable energy sources~\cite{GreenILC2}. 
    The project focuses on saving energy through maximizing scientific output per kW of energy consumed and through optimizing the power efficiency of machine components.
    In addition, the project aims to recycle energy, particularly through the storage and re-use of heat, including considering new waste-heat storage technologies.
    Finally, other ideas such as alternative sustainable building materials are in consideration.
\end{casestudy}

\newpage

\newpage

\begin{figure}[!h]
     \centering
     \captionsetup{labelformat=empty}
     \captionsetup{singlelinecheck=off}
     \caption[IPCC mitigation potential for alternative sources of energy]{

    \emph{(Figure~\ref{fig:ene-mitigation} Contd.)} The baseline used consists of current policy (2019) reference scenarios from the AR6 scenarios database (25/75 percentile values). The assessment relies on approximately 175 underlying sources, that together give a fair representation of emission reduction potentials across all regions. The mitigation potentials are assessed independently for each option and are not necessarily additive.\medskip
    
    The length of the solid bars represents the mitigation potential of an option. The error bars display the full ranges of the estimates for the total mitigation potentials. Sources of uncertainty for the cost estimates include assumptions on the rate of technological advancement, regional differences, and economies of scale, among others. Those uncertainties are not displayed in the figure.\medskip
    
    Potentials are broken down into cost categories, indicated by different colours (see legend). Only discounted lifetime monetary costs are considered. Where a gradual colour transition is shown, the breakdown of the potential into cost categories is not well known or depends heavily on factors such as geographical location, resource availability, and regional circumstances, and the colours indicate the range of estimates. Costs were taken directly from the underlying studies (mostly in the period 2015-2020) or recent datasets. No correction for inflation was applied, given the wide cost ranges used. The cost of the reference technologies were also taken from the underlying studies and recent datasets. Cost reductions through technological learning are taken into account (FOOTNOTE 70).\medskip 
    
    When interpreting this figure, the following should be taken into account:
    \begin{itemize}
    \item The mitigation potential is uncertain, as it will depend on the reference technology (and emissions) being displaced, the rate of new technology adoption, and several other factors. Cost and mitigation potential estimates were extrapolated from available sectoral studies. Actual costs and potentials would vary by place, context and time.
    \item Beyond 2030, the relative importance of the assessed mitigation options is expected to change, in particular while pursuing long-term mitigation goals, recognising also that the emphasis for particular options will vary across regions (for specific mitigation options see sections C4.1, C5.2, C7.3, C8.3 and C9.1 of Ref.~\cite{IPCCMitigationReport}).
    \item Different options have different feasibilities beyond the cost aspects, which are not reflected in the figure
    (cf. Section E.1 of Ref.~\cite{IPCCMitigationReport}). 
    \item The potentials in the cost range 100 to 200 USD per \acrshort{tco2e}  may be underestimated for some options.
    \item Costs for accommodating the integration of variable renewable energy sources in electricity systems are expected to be modest until 2030, and are not included because of complexities in attributing such costs to individual technology options.
    \item Cost range categories are ordered from low to high. This order does not imply any sequence of implementation.
    \item Externalities are not taken into account.
    \end{itemize}\medskip
    
    FOOTNOTE 70: For nuclear energy, modelled costs for long-term storage of radio-active waste are included.]  \medskip
    
    The 'reference technology' used as a cost benchmark is electricity production for coal and gas-fired power generation in China, India, Europe and North America, as calculated by the International Energy Agency (IEA); potentials were assessed with respect to the `Current Policies' scenario in the IEA report, World Energy Outlook 2019~\cite{IEA2019}.}
\end{figure}

\paragraph{Solar}

Solar energy is abundant, and ubiquitous.  Its intensity depends on latitude, with the highest efficiencies in the deserts of the sun belts north and south of the equator. According to Ref.~\cite{CarbonTracker21}, populating an area of 450,000 km$^2$ with solar panels would be sufficient to satisfy global energy demands.  This corresponds to an area two-thirds the size of  Texas, or 4\% of European landmass.\footnote{Taking into account geographical differences in PV cell operating efficiencies, there is no more than a factor of two difference between the solar PV power output achievable in the highest-output (Namibia) and lowest-output (Ireland) countries, of all countries lying between the 60th parallels, assuming solar panel installation in technically feasible areas.  For more details, see Ref.~\cite{ESMAP2020}}  According to the IPCC~\cite{IPCC2022reportSPM}, ``The global technical potential of direct solar energy far exceeds that of any other renewable energy resource and is well beyond the total amount of energy needed to support ambitious mitigation over the current century.''  However, large-scale adoption of photovoltaic (PV) panels poses concerns due to resource use, particularly the energy-intensive production of silicon used to produce the panels (see \bpref{bp:SiWafer} for a partial life cycle analysis of a monocrystalline silicon wafer), and end-of-life waste generation (see \sref{sec:Waste} for more information). These impacts can be partially mitigated by material recovery in PV cell recycling, and their reuse~\cite{DANIELAABIGAIL2022103539}.

The price of solar panels has dropped by almost 90\% in the last two decades~\cite{owid-cheap-renewables-growth}, and they can be easily retrofitted onto existing infrastructure (sometimes at significant cost). Unfortunately, solar power can be unavailable when it is needed most:\ at night and during winter (in countries at higher latitudes), leading to a need to increase its efficiency and storage capacity.  An overview of energy storage technologies can be found in \sref{subsec:grid}. See \csref{case:solar} for a study of the implementation of in-house solar power at CERN.

\paragraph{Wind}

By comparison with solar energy, wind energy is more sensitive to local conditions. In Europe, competitive locations for wind energy, with costs below \EUR{0.06} per kWh, are primarily offshore, and are concentrated along the coasts of the North and Baltic Seas, the Bay of Biscay, and the English Channel~\cite{EEAWindEnergy}.
Landlocked countries, such as Switzerland or Austria, are generally less suited for production of energy through wind turbines. Producing 25\% of, \eg Swiss energy demand from wind power would require populating 100\% of its agricultural farmland with wind turbines (although this does not preclude growing crops beneath the windmills).  By contrast, fulfilling a quarter of Danish or Estonian energy demand would require less than 4\% of the farmland~\cite{EEAWindEnergy}. 

\paragraph{Hydroelectric}

Water power is even more reliant on local conditions, such as high flows or water volumes and large altitude difference, which naturally limits its applicability.  However, the energy output of hydroelectric plants is steady, and can be adjusted to demand very quickly, making it a good complement to other renewable sources. It can also be used for energy storage, see \sref{subsec:grid}. The largest hydroelectric capacity is in China, which produces almost 30\% of the global hydroelectric power~\cite{IHA2021}, thanks to its large projects in the Yangtze River valleys.

Mega-dams, however, constitute a large intervention on the natural environment, and consequently come with associated risks, such as landslides, earthquakes, and destruction of habitats, and can themselves be a source of the potent \acrshort{ghg} methane when flooded flora rots. The Three Gorges dam in particular has been controversial both domestically and abroad~\cite{ThreeGorges}. In arid areas, or during periods of drought, which are expected to become more prevalent due to climate change, hydroelectric power may be in competition with agricultural needs, and climate change may jeopardise future yields of existing dams.

While the potential for marine power generation from ocean currents, tides, waves, and gradients in salt and temperature (collectively known as Ocean Energy Technologies) is huge, there is no technology currently mature enough to produce marine power at large scale~\cite{OET}.

\paragraph{Geothermal}

Geothermal energy is a stable source of renewable energy. It consists of residual heat from the time when the Earth was formed and of heat newly produced inside the Earth due to radioactive decay of hot elements in the mantle, by tidal forces due to the Moon and Sun, or by friction along tectonic plate boundaries. Although it has low intensity compared with solar energy, its technical potential of about $1.4 \times10^6$ TW-years is around three times total global energy consumption~\cite{Britannica}.

The most easily exploitable is the `shallow' geothermal energy stored in the upper few metres of the Earth's surface.  This can be employed to provide space heating or cooling for buildings and urban areas using buried pipes containing a circulating fluid as a heat exchanger, and a geothermal heat pump~\cite{Narsilio2018}. The low thermal conductivity of the ground limits the total amount of geothermal heat that can be exploited and depends strongly on rain and ground water. Modern geothermal heat pumps use the ground as heat storage and not so much as heat source. The ground heat that is extracted in winter can be regenerated in summer by using the heat pump for cooling the building.  

Geothermal power generation, however, requires higher temperatures.  Easily accessed only in areas of volcanic activity, and along plate boundaries, much geothermal power-generating potential is locked up below common drilling depths, where the rock is less porous.  Enhanced Geothermal Systems (EGS)~\cite{IRENA2017} induce porosity by fracturing deeper, hotter rock using high-pressure water injection, to allow for fluid circulation.  This hydrothermal `fracking' has attracted significant controversy for its environmental and geological impacts.  In addition to polluting nearby groundwater sources by injection of toxic chemicals into the Earth, it brings a risk of induced seismic activity if unwittingly carried out near a `locked' dormant fault, and was thought to be responsible for triggering a magnitude-5.4 earthquake in the South Korean city of Pohang in November 2017~\cite{Kim2018,Grigoli2018}, among others. 

\paragraph{Nuclear}
Nuclear power is a source of low-carbon electricity.
Nevertheless, an energy source is only termed sustainable if it does not carry any significant long-term risk for future generations. This definition of sustainability based on the Brundtland Report~\cite{Brundtland} has also been adopted by the International Atomic Energy Agency (IAEA)~\cite{IAEA}, which argues for the `weak sustainability' of nuclear power. 

The share of primary energy from nuclear power has been decreasing on all continents except Asia over the last two decades, and has fallen below 4\% of global primary energy production~\cite{OWDnuclear}. According to the IAEA, nuclear reactors have a median construction time of 93 months~\cite{IAEANuclear}, not including planning and permissions.  See~\csref{cs:nuclear_gap} for an estimate of how many nuclear power stations must be constructed to cover global energy needs. 

Safety, security and climate resilience of the reactors, and availability of fuel, as well as storage of spent fuel, are important challenges.  The exact form these take is crucially dependent on future technological developments, to which \ACR\ research contributes directly.  Today, several new reactor types are being developed, which promise to have additional safety features, an efficient use of more abundant isotopes and less long-lasting nuclear waste.  \ACR\ research can also contribute to non-proliferation efforts, see \eg Ref.~\cite{DUNE:2022jhf}.  Bringing developing technologies to maturity and commercial viability takes time, and in the near term the IPCC does not assess favourably the mitigation potential for nuclear energy, see Figure~\ref{fig:ene-mitigation}. 


\begin{casestudy}[Filling the energy gap with nuclear reactors\label{cs:nuclear_gap}]{Filling the energy gap with nuclear reactors\label{case:nuclear}}%
    A typical nuclear reactor produces on the order of 1~GW$_{\rm el}$ (i.e., electrical power actually generated).
    Based on the ``substitution method'' used in \fref{fig:ene-gap}, this corresponds to 2.5~GW primary energy.\footnote{The substitution method accounts in a simplified way for the inefficiencies in energy usage and conversions of different primary energy sources, and assumes that electricity is 2.5 times as useful as fossil fuels of the same energy content.  The factor 2.5 comes from the 40\% efficiency in fossil power plants~\cite{BP2022} and is consistent with comparing the numbers in Ref.~\cite{OWDprod}.} 
    Filling the entire global energy gap using nuclear power would require $\sim$8,800 additional nuclear power plants within 18 years (at 1.3--1.5 GW electricity output per power plant), which corresponds to building and commissioning an average of 9 new nuclear power plants every week in that period. 
    A community like \ACR\, with experience in planning and implementing large projects, knows that such a huge technological conversion in such a short time represents a significant challenge, especially in the absence of a global road map for such a transition.
\end{casestudy}


\begin{casestudy}[Local solar power at CERN\label{case:solar}]{Local solar power at CERN}%
    In-house solar power production is not sufficient to cover the full needs of a huge laboratory such as CERN. Nevertheless, it can make up an important contribution to foster a fast transition to renewables.\footnote{For a discussion on potential future energy system configurations for Switzerland, see Ref.~\cite{Zuettel}.} Research centres are often characterised by the many flat rooftops which make excellent locations for installing solar photovoltaic (PV) panels. However, PV infrastructure can be heavy and may not be sufficiently supported by large flat roofs.\\

        \begin{center}
    \vspace{1.5em}
    \captionsetup{type=figure}
    \includegraphics[width=0.8\linewidth]{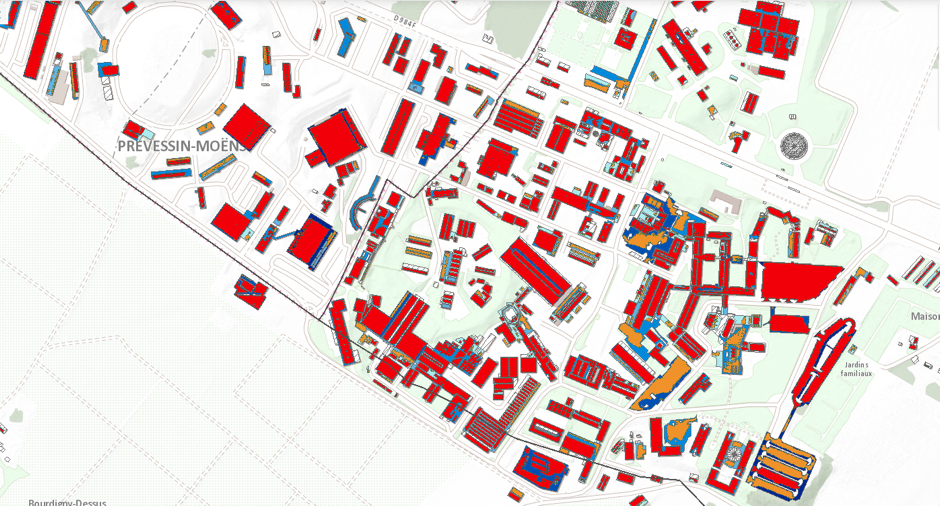}
    \caption[CERN roofs suitable for solar cells]%
        {Map of CERN buildings. Rooftops that are suitable for PV installation in respect to their received solar irradiation are shown in red (very suitable) and yellow (suitable). In addition other areas like \eg parking lots could also be covered by PV-panelled roofs. Reproduced with permission from Ref.~\cite{SIG}.\label{fig:ene-cernroof}}
    \end{center}

    Using publicly available tools provided by the Canton of Geneva~\cite{SIG} and the Swiss Federal Department of Energy~\cite{BFE}, it is possible to estimate the solar potential of these rooftops. Similar public tools are now available for most countries, provided by local governments or Non-Governmental Organizations (NGOs). Figure~\ref{fig:ene-cernroof} shows part of the main CERN site as taken from the Geneva solar cadastre~\cite{SIG}. Buildings in red are classified as “optimal” for their orientation towards the sun. The large rectangular building in the middle is assembly hall 157. The cadastre lists an estimate of 392 MWh per annum of electricity generation for the south-west half of this 2,055 m$^2$ roof, with the other part capable of producing an additional 335 MWh per annum. CERN has 653 buildings with a total roof area of 421,000~m$^2$,\footnote{This number does not include areas that are otherwise assigned, \eg parking spaces for personal vehicles, which can also be roofed with PV panels.} which amounts to approximately 80~GWh annual electricity generation potential. A comparison with the electricity consumption in 2019 of 428~GWh~\cite{Environment:2737239}, when the LHC was not in operation, shows that around 18\% of CERN’s basic (non-LHC) electricity demand could be produced locally with solar power.

    Using the cadastre, the cost for electricity from rooftop PV for CERN can be estimated to be fixed around at \EUR{50}/MWh for the next 30 years. This cost is well below current wholesale market spot prices in France (>\EUR{120}/MWh), but also below the average price over summer 2021 (> \EUR{70}/MWh)~\cite{FranceMix}.
\end{casestudy}


\subsubsection{Renewable grid infrastructure}
\label{subsec:grid}
For global power grids to rely more heavily on intermittent sources of renewable energy, such as wind and solar photovoltaic, assessed as having the highest climate mitigation potential to 2030 by the IPCC (see \fref{fig:ene-mitigation}), much of the existing grid infrastructure and controls will need to be updated to smooth out fluctuations in supply and demand~\cite{PowerGridInertia}.  Grid inertia, which acts as a short-term buffer in fossil-fuel grids in periods where electricity demand outstrips supply, is significantly lower for intermittent sources, compromising the stability of the grid. `Smart grid' infrastructure must provide peak-shifting capabilities and fast frequency response to stabilise electricity supply despite the lower intrinsic inertia of the grid.  It will also need to draw upon a novel and expanded energy storage capacity to bridge longer-term gaps between supply and demand~\cite{LDESEnergyStorage} (for further discussion, see below), as well as the capability to regulate bi-directional flow of electricity, well-suited to the distributed generation of intermittents.  Inverter-based resources with low inertia are ideally suited to near-instantaneous response~\cite{PowerGridInertia}.  Electronic control of this response, coupled with developing `grid-forming' technologies, which allow inverters to emulate a traditional grid's stable frequency, as well as automated demand-side response to voluntarily disconnect non-critical loads momentarily, have the potential to transform our existing networks~\cite{PowerGridInertia}.  

Existing solutions have been utilised successfully in several fully renewable island microgrids, and on a larger scale in the Electric Reliability Council of Texas (ERCOT), the smallest of the three power grids in the United States, which achieved 58\% instantaneous wind penetration in 2019 \cite{PowerGridInertia}.  Scaling up these solutions requires further research, although several highly cited studies argue for the feasibility of 100\% renewable-based grids world-wide at low cost, eschewing any fossil fuel or nuclear energy component (for a comprehensive review, see Ref.~\cite{HEARD20171122} and also Ref.~\cite{BROWN2018834}).  

\paragraph{Energy storage}
The feasibility of pure renewable-based grids is crucially dependent on an increased energy storage capacity to smooth out fluctuations (on timescales ranging from diurnal to seasonal) in the supply of intermittent renewables and demand~\cite{LDESEnergyStorage}.  While the cost of Li-ion batteries have plummeted 40-fold in the past 35 years~\cite{Ziegler_2020}, and notwithstanding the implications of resource extraction (see, e.g., \sref{sec:Waste}), their development has focused on short-duration portable energy storage, as driven by needs of the electric vehicle industry.  Projections show that in order to minimise the costs of a net-zero energy system, storage capacity must increase by almost an order of magnitude, from 160 GW (and 9 TWh total capacity) today, to 1.5--2.5 TW (85-140 TWh) globally  by 2040~\cite{LDESEnergyStorage}.  

Most existing and planned storage capacity is in pumped storage hydropower, a mechanical form of storage where water pumped into a reservoir at high elevation turns a turbine as it flows to one at lower elevation~\cite{LDESEnergyStorage}.  As well as being geographically limited, however, these open-air reservoirs suffer the same environmental problems as other hydropower projects (see discussion above), and are similarly subject to the vagaries of the climate. 
A new promising approach to pump storage is the use of undersea bowls that are evacuated to store energy which is restored when they are filled up with water again. An even simpler approach is to build a large ring wall inside of a deep lake, and empty and refill the internal area using pump-turbines. This way, pumped-storage hydroelectricity does not require a separate upper and lower lake, and a single lake is sufficient.
Defunct open pit mines, large natural lakes and even the sea allow for new opportunities to install large-scale storage devices with less environmental impact~\cite{Dueren,pumpstorage}.  

Interesting alternatives include storing energy as compressed air; or as latent heat in, \eg aluminium alloy~\cite{LDESEnergyStorage}.
However, there is need for significant investment in the energy storage sector to bring new ideas, including novel mechanical, thermal, electrochemical and chemical storage methods to commercial viability~\cite{LDESEnergyStorage}.    For more details on capacity and market-readiness of promising long-duration energy storage methods, see Ref.~\cite{LDESEnergyStorage}.

\subsubsection{Energy import and export}
\label{subsec:import}
The uneven geographical distribution of sources of renewable energy leads to the question of whether large-scale import and export of renewable energy could be a cost-effective way of closing the energy gap. For Europe, detailed studies have shown that energy import by cable, as well as by chemical energy carriers, have comparable or lower costs compared to local energy harvesting~\cite{10.1371/journal.pone.0281380}.

Technical options to transport electricity over long distances have improved significantly in the last decades. In South America and China, projects to transport electricity over more than 2,000~km by Ultra High Voltage Direct Current (UHVDC) lines are already operational~\cite{Champion}. The Viking Link UK-Denmark project provides another example~\cite{VikingLink}, and a summary of all current and planned European transmission projects is provided in~\fref{fig:ene-cernhvdc}. These include projects to import renewable energy from North Africa to Europe, such as the Xlinks Morocco-UK Power Project~\cite{Xlinks}, which aims to connect a solar and wind energy facility in Morocco's Guelmim Oued Noun region to the UK energy grid by 3,800 km HVDC sub-sea cables by 2030.


\begin{center}
\vspace{1.5em}
\captionsetup{type=figure}
     \includegraphics[width=0.8\linewidth]{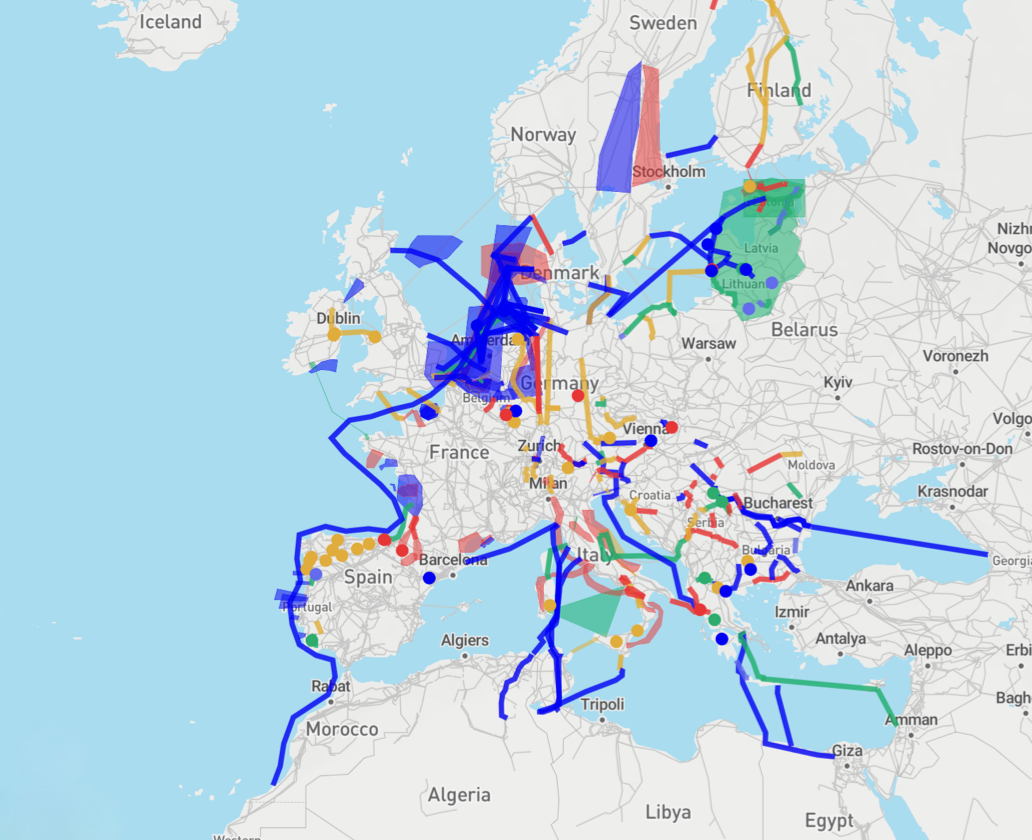}
    \caption[Map of transmission projects]{Transmission projects within and to Europe, reused with permission from the TYNDP 2024 Project Sheets~\cite{TYNDP2024}. Projects under construction are shown in green, planned and in permitting are in yellow, planned and not yet in permitting are in red, and projects under consideration are in blue. Dots indicate storage projects.}
    \label{fig:ene-cernhvdc}
\end{center}


While the import and export of energy is a promising solution on the technical and economic level, constructing wind or solar farms, \eg in the sun belts of Africa, to then export the power to Europe involves geopolitical and social considerations. Resource and person-power extraction from Africa to the benefit of Europe and America has a long, reprehensible colonial history. Notwithstanding the potential for electricity generation by solar and wind power, large unused tracts of land, and existing energy trade partnerships with North African countries for fossil fuels, it is of the utmost importance to make fair power trade agreements between the continents that ensure strong integration into local communities and include the local population in the planning and implementation of such projects and related infrastructure. In this way, a win-win situation for all stakeholders should be ensured, and well-planned cooperation has the potential to act in a geopolitically stabilizing way in line with the 16th and 17th UN \acrshort{sdg}s ("peace, justice and strong institutions" and "partnership for the goals").  If achievable, equitable import and export of renewable energy could be considered as part of a catalogue of solutions to cover future global energy needs.

The \ACR\ communities have a record of successful collaboration between nations, and they could therefore be important players in proving the viability of such partnerships.  One possible scenario~\cite{Dueren,Dueren+2011+263+275} for the transmission of solar energy is detailed in \csref{case:desert}.


\begin{casestudy}[CERN-LINK --- Clean power from the desert\label{case:desert}]{CERN-LINK --- Clean power from the desert}%
The \ACR\ community, and CERN in particular, has a long history of effective cooperation across geographical and socio-political boundaries, in the pursuit of science.  
CERN brought scientists from East and West together during the Cold War, and Arabic and Israeli people together for the SESAME project (see \bpref{BP:sesame}), the first accelerator laboratory powered by solar energy from the desert~\cite{Sesame}. This makes CERN ideally placed to spearhead a project to import energy from countries rich in renewable energy sources, and transport it across international boundaries.  This type of spin-off could help cover CERN's energy needs, while also reinforcing the idea that fundamental research has the potential to solve problems outside its immediate purview in new and innovative ways.

The technology for long-distance energy transmission, in the form of High-Voltage Direct Current (HVDC) lines, is commercially available, and is being increasingly employed globally (see \fref{fig:ene-cernhvdc} for current and planned transmission projects in Europe). Prominent examples include the Viking Link UK-Denmark project, which broke ground in July 2020~\cite{VikingLink}, as well as the planned XLinks Morocco-UK power project~\cite{Xlinks}.

Even so, it is of paramount importance to consider the social implications of any such projects and ensure that fair power trade agreements are put in place for the mutual benefit of all stakeholders, including the local communities hosting the energy-harvesting infrastructure. It is also important to acknowledge that additional environmental considerations are required when planning and implementing a project of this type, in terms of minimising the impacts on local ecosystems, as well as the marine environments across which the underwater cables would be installed. 


\begin{center}
\vspace{1.5em}
\captionsetup{type=figure}
     \includegraphics[width=0.8\linewidth]{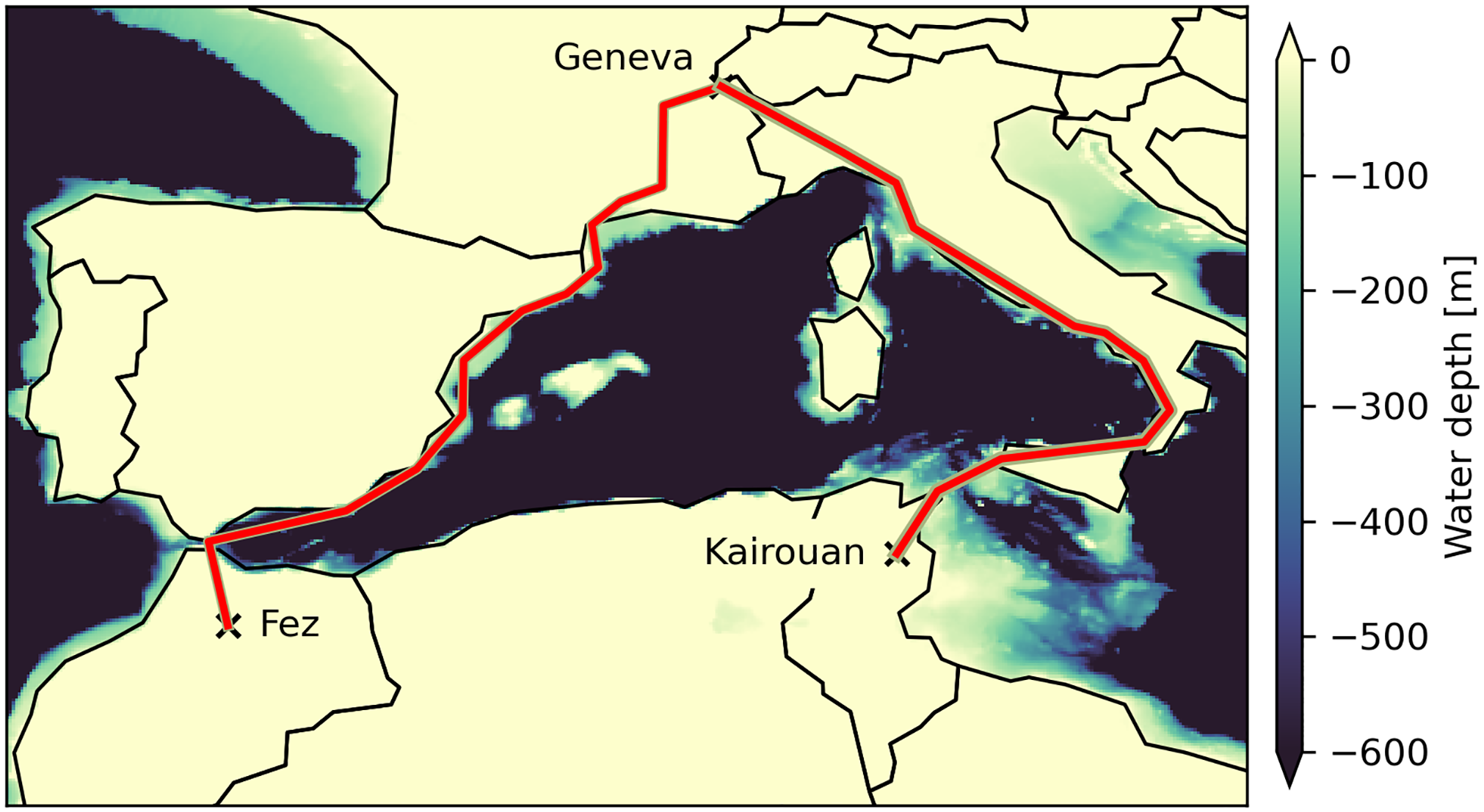}
     \caption[Map for CERN-LINK]{Potential CERN-LINK cables (in red) connecting North African solar power plants in Morocco or Tunisia with CERN and the central European electricity grid.  The cables from Fez would run just offshore, whereas those from Kairouan would need to be built onshore, the sea depth making underwater development challenging.
     Figure taken from Ref.~\cite{HamppPLOS}, reused under the terms of the \href{https://creativecommons.org/licenses/by/4.0/}{Creative Commons Attribution 4.0 International (CC BY 4.0) license}.}
     \label{fig:ene-cernlink}
\end{center}


A scenario for connecting, \eg Morocco, Algeria or Tunisia to Southern France, Spain or Italy by sub-sea cable is plausible from a technological point of view (for a detailed feasibility study, see Ref.~\cite{MoroccoHVDV}), and could be employed for \ACR\ applications (see~\fref{fig:ene-cernlink}). Costs are estimated to be around \EUR{0.06--0.08}/kWh for a year-round power supply of 3.6~GW day and night, depending on the place of generation and the demand-side flexibilities~\cite{Dueren, Dueren+2011+263+275, Hampp, HamppPLOS}. This estimate includes infrastructure costs for generating the electricity, buffer storage and transmission line costs, assuming the use of concentrated solar power stations with thermal energy storage. Substituting this with H$_2$ for storage and transport purposes would increase costs by a factor of 2--3, to \EUR{0.16--0.20}/kWh~\cite{HamppPLOS}.  Feasibility and cost estimates agree well with those for previously proposed commercial projects~\cite{Xlinks}. These estimates can be compared with the current price of electricity at CERN, which has 80\% of its power supplied via the ARENH system~\cite{ARENH_website} at a fixed rate of \EUR{0.042}/kWh to December 2025 (to be renegotiated subsequently).  The remaining 20\% is supplied at market rate, which is subject to large geopolitically induced fluctuations, particularly in recent years.

Electricity import on this scale, corresponding to the output of 3 nuclear power plants, would exceed by far the power needs of CERN.  The surplus power could be returned to the European electricity grid to power other research institutions and universities that join the initiative. However, the feasibility of any such project would rely on the capacity of international grid interconnector infrastructure and agreements on inter-grid energy transfer.

\end{casestudy}


\subsection{Energy Saving and Recuperation}
\label{sec:Ene-Saving}
A first step in reducing energy usage is energy monitoring, which will allow assessment of where improvements are needed. The best energy saving measures will be individual to each location and facility, making it hard to recommend specific actions here, although insulating buildings and ensuring that the heating/cooling systems are maximally efficient are universally applicable measures. For an example of energy recuperation in the context of basic research, see \bpref{BP:DESY_recycling}.

Most of the energy budget for many high energy experiments is due to the accelerators and detectors. Initiatives to reduce their energy use are many and varied. Relevant references for detectors are collected in \sref{sec:Technology}, see also the discussion on energy saving in the LHCb experiment at CERN in \csref{case:LHCb}.  A particularly impressive example of energy-efficient accelerator design is the Energy Recovery Linac Test Accelerator (CBETA)~\cite{Bartnik:2020pos}, based in Cornell. This accelerator saves energy, both by recovering the energy of the bunched particles to accelerate the next batch, and by using permanent magnets to guide the particle beam.  See \bpref{BP:EnergyRecoveryAccelerator} for more details and \csref{case:wakefield} for energy savings using plasma wakefield acceleration technology.

\begin{bestpractice}[Recycling energy at DESY\label{BP:DESY_recycling}]{Recycling energy at DESY}%
   For existing experiments, where minimising energy usage was not a factor in the design process, it is still possible to save energy retroactively through recycling of energy/heat. The Deutsches Elektronen-Synchrotron (DESY) in Hamburg, Germany's largest accelerator centre, is currently using the waste heat, which is generated by condensation of the helium that is used to cool the accelerator, to heat their buildings. This saves 7.5 GWh a year, which is approximately a third of the heat energy used on campus~\cite{DESY}. Together with the University for Applied Sciences in Hamburg, they are also investigating the potential for recycling waste heat from other sources, \eg the many magnets used in the accelerator. First results suggest that it should be sufficient to heat all buildings on campus in this way~\cite{DESYsustainableReport2022}.
\end{bestpractice}

\RaggedRight
\sloppy
\newpage


\section{Food}
\label{sec:Food}

\begin{center}
\includegraphics[width=\SDGsize]{Sections/Figs/Common/SDG_2_ZeroHunger.png}~%
\includegraphics[width=\SDGsize]{Sections/Figs/Common/SDG_3_GoodHealth.png}~%
\includegraphics[width=\SDGsize]{Sections/Figs/Common/SDG_6_CleanWater.png}~%
\includegraphics[width=\SDGsize]{Sections/Figs/Common/SDG_12_ResponsibleConsumption.png}~%
\includegraphics[width=\SDGsize]{Sections/Figs/Common/SDG_13_ClimateAction.png}~%
\includegraphics[width=\SDGsize]{Sections/Figs/Common/SDG_14_LifeBelowWater.png}~%
\includegraphics[width=\SDGsize]{Sections/Figs/Common/SDG_15_LifeOnLand.png}
\end{center}


\exSum

\noindent Over a quarter of global \acrshort{ghg} emissions comes from food production~\cite{USEPA}, and a quarter of this food is lost in the supply chain, or discarded by consumers~\cite{Searchinger2018}.  Plant and animal agriculture also has extensive and profound negative impacts on the environment through land use, freshwater use and pollution, and terrestrial acidification. 

A recent article in \textit{Science} argued that limiting warming to 1.5\degree C will not be possible without ``ambitious changes to food systems'' even if fossil fuel emissions are immediately halted~\cite{Clark370}.  The most impactful change reported in the study is a global switch to the healthy, plant-rich diet recommended by the EAT-Lancet commission \cite{WILLETT2019447}, which can be implemented immediately, on an individual level.  Supplementing this with measures such as reducing food waste and increasing efficiencies in food production could result in a net carbon-neutral food system by 2100 \cite{Clark370}.
While sourcing sustainably grown food and `eating local' can have a positive impact on food-related emissions (of which transportation is responsible for 6\%, see \fref{fig:food_impact}), the largest impact can be achieved by reducing the consumption of high-methane emitters, such as beef, lamb and dairy products~\cite{PooreNemecek2018,OWID-Sustainable,OWID-Local, Scarborough2023}.  References~\cite{PooreNemecek2018,Scarborough2023} argue that this is the case despite significant differences in carbon intensity of food production due to differing geography and agricultural practices.

 However, choices related to the food that we eat are deeply personal. Diet is often loaded with cultural and social significance, and can readily be politicised. As such, it is important to acknowledge that changes to food systems will be a gradual process and will not have a `one size fits all' solution. Their equitable implementation will require cross-disciplinary analysis of the implications of such changes for all stakeholders. This includes producers and the local populations to which they belong, and steps must be taken to ensure communities are empowered and resilient to multiple overlapping pressures, from climate change and markets~\cite{WalshDilley2020}. The devastating impact of the quinoa boom and bust on pastoral communities in Bolivia provides one well-documented example~\cite{WalshDilley2020, Rodas2021}. Moreover, such changes must account for global disparities in wealth, and the variations in availability and access to food sources, to avoid further cementing geographic inequalities in diet. Notwithstanding the care that these factors necessitate, a significant proportion of the \ACR\ community is in the privileged position to be able to reduce their consumption of animal-derived food products and minimise food waste.


\newpage
\begin{reco2}{\currentname}
{
\begin{itemize}[leftmargin=6 mm]
\setlength{\itemsep}{\recskip}
\item Reduce consumption of animal products, especially those that result in the highest emissions, \eg ruminant meat, and dairy, where alternative sources of nutrition are equitably available.

\item Minimise food waste.

\end{itemize}
}
{
\begin{itemize}[leftmargin=6 mm]
\setlength{\itemsep}{\recskip}
\item Prioritise plant-based options in conference catering, and optimise service method to reduce food waste.
\end{itemize}
}
{
\begin{itemize}[leftmargin=6 mm]
\setlength{\itemsep}{\recskip}
\item Undertake comprehensive and transparent local audits of the sustainability of food service.

\item Incentivise the consumption of plant-based products at on-site restaurants by increasing their variety and quality,  and subsidising their cost.

\item Highlight the environmental impact of food choices through service layout and labelling.

\item Strive for zero food waste by, e.g., providing multiple portion sizes and donating unused food.

\item Read section on waste (Section~\ref{sec:Waste}) and limit food-service waste \eg through industrial composting of biodegradable food containers.
\end{itemize}}
\end{reco2}

\newpage
\subsection{Food Production\label{sec:agriculture}}

\begin{figure}
    \centering
    \includegraphics[width=\textwidth]{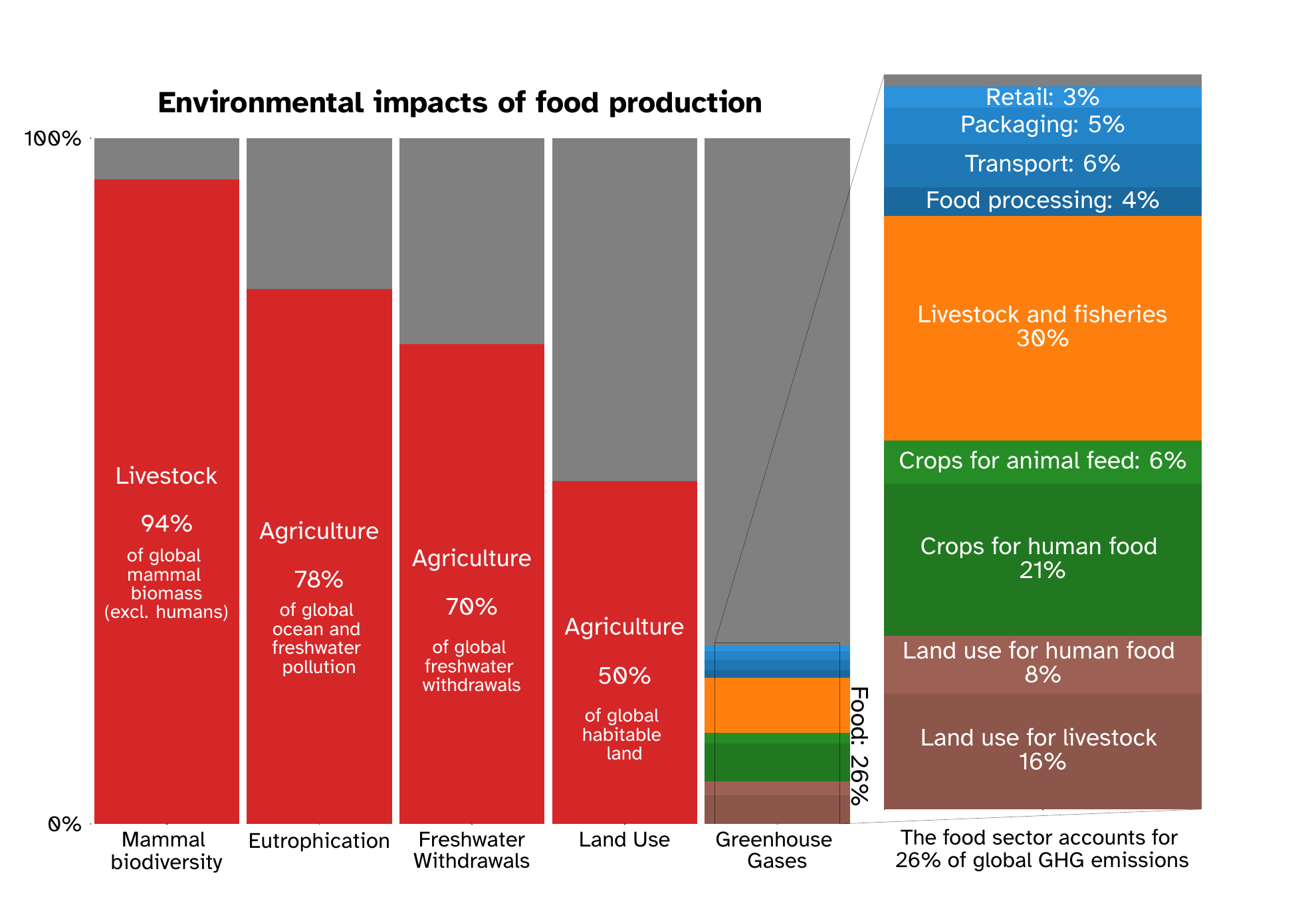}
    \caption[Environmental impact of food production]{ Environmental impact of food production, with fine-grained partitioning of GHG emissions by food sector. Figure modified from Ref.~\cite{OWID-Food} under the terms of the \href{https://creativecommons.org/licenses/by/4.0/}{Creative Commons Attribution 4.0 International (CC BY 4.0) license}, based on data from Refs.~\cite{PooreNemecek2018} and~\cite{Bar-On6506}.
    \label{fig:food_impact}}
\end{figure}

\fref{fig:food_impact} reveals the overall environmental impact of food production.  The agriculture sector uses 70\% of the world's fresh water reserves and has caused eutrophication\footnote{Excessive fertiliser runoff to freshwater environments causing algal blooms, oxygen depletion, and fish die-offs.} of most of the world's oceans and freshwater.  It is responsible for large-scale deforestation and habitat loss~\cite{OWID-Food, PooreNemecek2018, Xu2021}, resulting in an historic low in mammalian biodiversity, with total mammal biomass dominated by humans and their livestock \cite{Bar-On6506}. 

Animal agriculture is responsible for just over half  of GHG emissions from the food sector, due to direct emissions from livestock and fisheries, land use, and production of crops for animal consumption.\footnote{Organic animal-derived foods often have higher yields of GHG emissions, partly because of the animals' lower productivity~\cite{Pieper2020}.}  It accounts for three-quarters of global agricultural land use, while providing just a fifth of the world's calories, and under 40\% of its protein supply~\cite{OWID-Food, PooreNemecek2018, Xu2021}.  Consumption of animal-source protein has a large global variance, ranging from, \eg 995 kcal/capita per day in the USA to 164 kcal/capita per day in, \eg Zambia~\cite{WILLETT2019447}.

The over-use of antibiotics in animal agriculture is partially responsible for the development of antimicrobial resistance in "superbugs"~\cite{AMBR2005}, and may be a risk factor for the emergence of new zoonotic diseases~\cite{Jones8399, Espinosa, Morand}.  Furthermore, there is substantial evidence linking high intake of red meat to an increased rate of heart disease~\cite{meat-health}.

\begin{figure}[tb]
    \centering
    \includegraphics[width=0.9\textwidth]{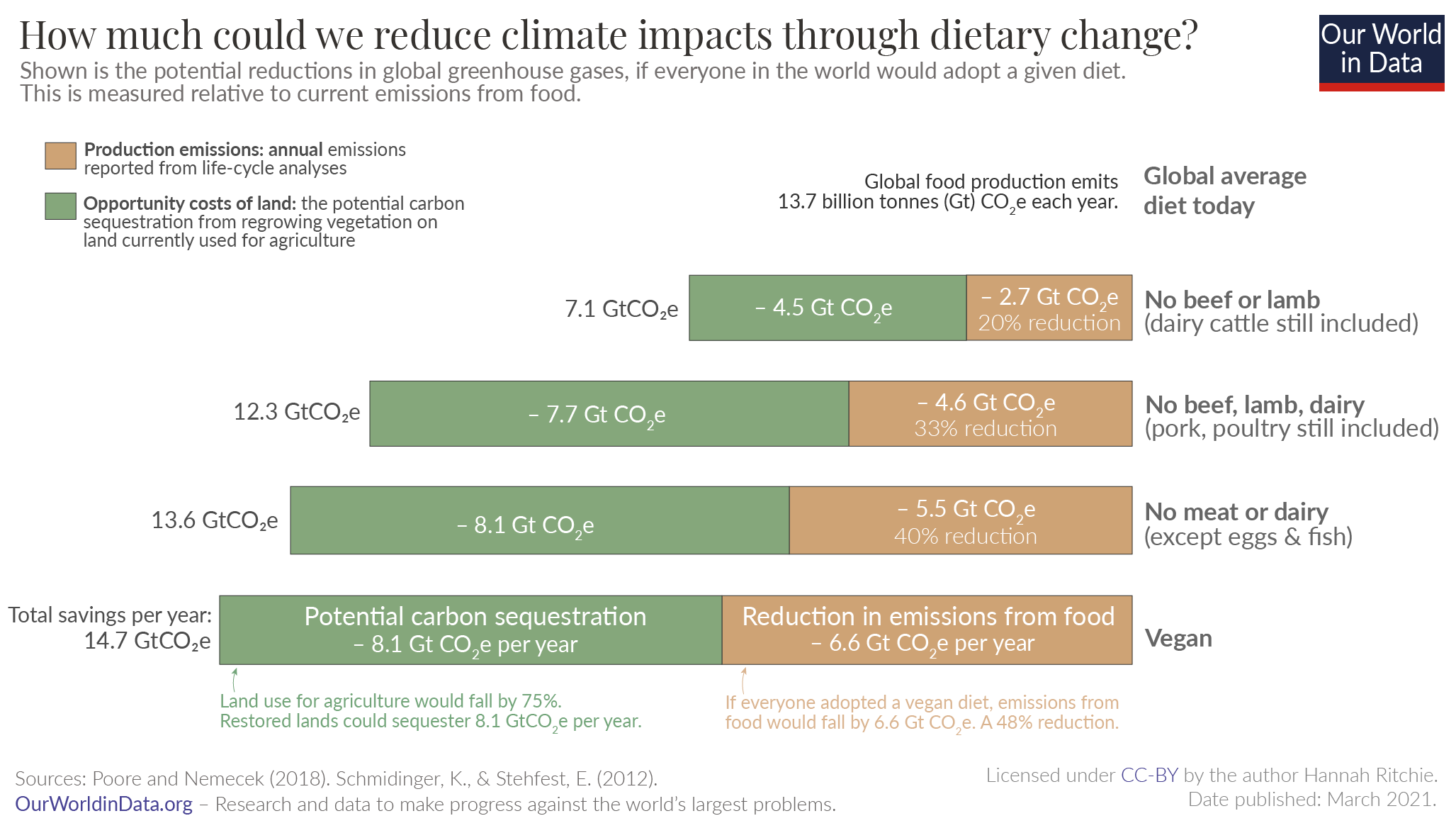}
    \caption[Annual carbon opportunity due to changes in diet]{Potential reduction in \acrshort{ghg} emissions due to changes in diet, relative to current emissions from food.  Figure reused from Ref.~\cite{OWID-CarbonOpportunity} under the terms of the \href{https://creativecommons.org/licenses/by/4.0/}{Creative Commons Attribution 4.0 International (CC BY 4.0) license}, based on data from Refs.~\cite{PooreNemecek2018,Schmidinger}.
    \label{fig:foodchanges}}
\end{figure}

Shifting consumption away from animal products to a more plant-based diet would significantly reduce both the environmental and healthcare costs of food systems. The potential annual reduction in GHG emissions from eliminating different food groups from our diet is shown in \fref{fig:foodchanges}.  
Beef, lamb, and dairy, responsible for the largest cumulative global emissions, are also among the highest emitters per 100 grams of protein, see \fref{fig:ghg-per-protein}.  In addition, animal products are generally more expensive than plant products~\cite{Springmann2021}, as well as being less inclusive of people with dietary restrictions or preferences, due to religion, lifestyle choices, food allergies or intolerances.

\subsection{Food Service}

Rigorous and transparent institutional audits of food service sustainability can readily effect positive changes to reduce the consumption of carbon-intensive foods and move towards zero waste, while accounting for the wider implications of sustainability within the context of local food systems.

Several universities and other institutes for higher education have implemented measures to limit or eliminate consumption of animal-derived foods and reduce food waste, including eliminating red meat from their cafeterias, increasing the quality and variety of plant-based options, changing the cafeteria layout, and modifying default meal options and food labelling~\cite{Berlin, EPFL, Cambridge, Goldsmiths}.    By way of illustration, we quantify the GHG savings due to replacing beef with alternative sources of protein in the weekly menu of CERN Restaurant 1 in \csref{case:CERNR1}.

At conferences and workshops, the primary purpose of any food served is to create additional opportunities for attendees to mingle and discuss.  As such, conference organisers enjoy more leeway to make sustainable food choices the default option for these short-term, small-scale events.  \bpref{catering_bp} contains two examples of successful physics conferences with plant-based catering. They highlight, among other things, the importance of institutional partnerships with plant-friendly caterers, and organisers should push for these if they do not already exist.  For further discussion on sustainability at conferences, see Section~\ref{sec:ConferenceWaste}.

\begin{casestudy}[Sustainable catering at CERN Restaurant 1\label{case:CERNR1}]{Sustainable catering at CERN Restaurant 1}%
    CERN Restaurant 1 (R1) serves an average of 2,000 meals per day~\cite{CERNACCUMeeting}. It offers five hot meal options daily, and has recently overhauled its menu options to include a larger variety of vegetarian and plant-based options, including at least one plant-based main course.  We assume each of the five mains are chosen with equal likelihood, and neglect cold food options, such as salads and sandwiches.

    \begin{center}
        \captionsetup{type=figure}
        \includegraphics[width=0.85\textwidth]{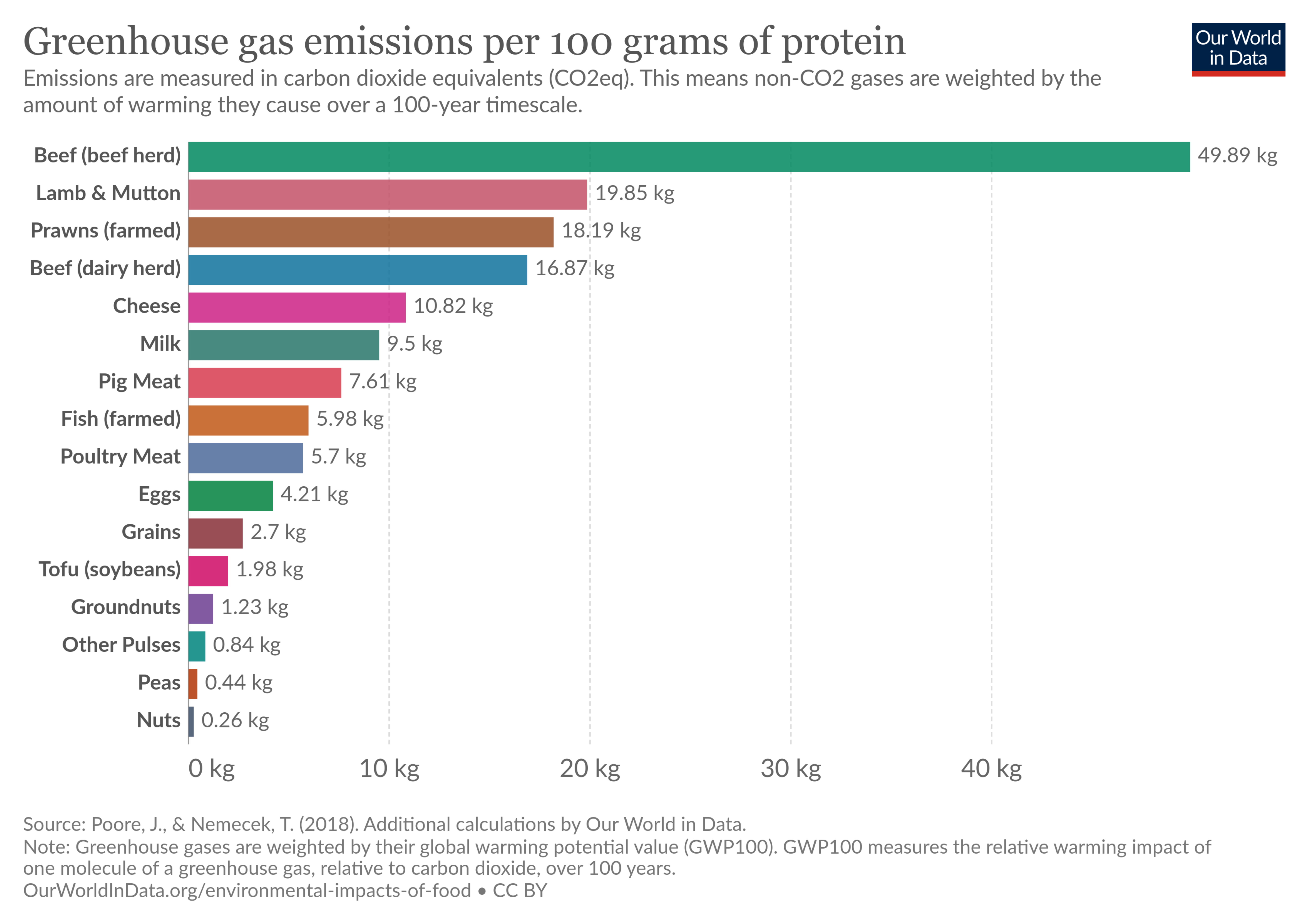}
        \caption[GHG emissions per 100 g of protein]{\acrshort{ghg} emissions in \acrshort{co2e} per 100 g of protein. Figure reused from Ref.~\cite{OWID-Food} under the terms of the \href{https://creativecommons.org/licenses/by/4.0/}{Creative Commons Attribution 4.0 International (CC BY 4.0) license}, based on data from Ref.~\cite{PooreNemecek2018}.}
        \label{fig:ghg-per-protein}
    \end{center}

    In, \eg the week beginning 27 June 2022, beef, fish and seafood were each served three times as the primary component of the meal, veal once, poultry five times, and fish twice.  We assume that these distributions are representative of a typical weekly menu at R1 and that the beef originated from beef-herd cows. The \acrshort{ghg} emissions of the various forms of protein are shown in \fref{fig:ghg-per-protein}.

    Substituting each gram of protein from beef with a gram of protein from chicken or fish reduces emissions by 440~g \acrshort{co2e}.  
    Assuming a serving contains 20~g of protein\footnote{Recommended intake is around 15--30~g of protein per meal~\cite{MAYO}.}, substituting all beef meals at R1 with farmed fish or chicken would result in a reduction of its annual carbon footprint by 528 \acrshort{tco2e}.\footnote{Substituting 1,200 beef meals weekly over 50 weeks, each meal consisting of 20~g protein, with chicken or fish leads to a reduction of $1,200\times 50\times 20\times 0.440$ kg \CdOe \; in emissions. Note however that farmed chicken and fish give rise to significant environmental impacts in sectors other than GHG emissions~\cite{KUEMPEL2023}.  The estimated emissions overshoots CERN's reported 2019 beef-related emissions by a factor of two \cite{HSENote}.  The reason for this discrepancy is unclear, since details of the calculation from Ref.~\cite{HSENote} were not shared.}
    This corresponds to approximately  260 return flights between London and New York. 
    Both the emissions savings and the overall environmental impact would be even greater if plant-based substitutions were made.

    Instituting one weekly meat-free day (taking as a benchmark a day where one beef, one fish, and one poultry meal were served, and replacing them with one tofu-based meal and two pulse-based meals) would result in a reduction of 735 \tCdOe\ annually, where the bulk of the savings comes from the beef replacement.
\end{casestudy}

\begin{bestpractice}[Plant-based catering at conferences and workshops\label{catering_bp}\\
{\footnotesize\noindent We thank Hannah Wakeling (for WIPC 2019) and Stefan Fredenhagen (for YRISW 2019) for sharing their experience as part of the respective organisational teams.}]{Plant-based catering at conferences and workshops}%
\noindent The Women in Physics Canada Conference 2019~\cite{WIPC} at McGill University in Montréal was designed as a `sustainable' conference. Ecologically friendly choices were made by offering purely plant-based catering, sustainable goodie bags, and use of reusable tableware (see also Section~\ref{sec:CateringTableware}). Most of the feedback regarding these measures was positive. An important point for the organisers was to advertise the catering as sustainable, and not only as vegan, since according to their experience this ``helped the way the catering was received'' by the participants. The organisers mentioned that it can be difficult to find a vegan caterer if the only choices are partners of the university hosting the conference, but it was nevertheless possible in their case.\\

   The `Young Researchers Integrability School and Workshop (YRISW) 2019: A modern primer for 2D CFT'~\cite{YRISW} in Vienna offered only plant-based catering. The organisers of the school selected this option as the ``most inclusive approach'', where people are not separated according to their eating habits. They wanted to advertise plant-based food to the participants, and ``reduce the environmental impact of the event''. The limited food options also reduced the total cost of the catering. The organisers received positive feedback, not only for the food itself but also for the ``effort to reduce the ecological impact of the school''. The organisers emphasised the importance of finding a specialist plant-based caterer to ensure the quality and flavour of the food.
\end{bestpractice}

\RaggedRight
\sloppy
\newpage


\section{Mobility}
\label{sec:Travel}


\begin{center}
\includegraphics[width=\SDGsize]{Sections/Figs/Common/SDG_3_GoodHealth.png}~%
\includegraphics[width=\SDGsize]{Sections/Figs/Common/SDG_9_IndustryInnovation.png}~%
\includegraphics[width=\SDGsize]{Sections/Figs/Common/SDG_11_SustainableCities.png}~%
\includegraphics[width=\SDGsize]{Sections/Figs/Common/SDG_12_ResponsibleConsumption.png}~%
\includegraphics[width=\SDGsize]{Sections/Figs/Common/SDG_13_ClimateAction.png}
\end{center}


\exSum

\noindent Mobility constitutes a significant portion of the emissions of a \ACR\ researcher.  This includes short daily commutes between the home and the workplace, and longer-distance business travel (see \fref{fig:Intro-ComparativeEmissions}).

Transport accounted for almost a fifth of total global emissions in 2016~\cite{OWIDsector}, and is the sector that saw the highest growth in pre-COVID years~\cite{SLOCAT}.  Demand for car, rail and air transport is expected to continue to increase over time with increasing global population and income levels.

Unsurprisingly, self-powered mobility, such as walking and cycling, are the most carbon-efficient means of transportation, with train travel being the next best.  A quantitative comparison between these and other options requires further details to be specified, such as the distance travelled, the fuel efficiency of the vehicle used and the number of passengers carried, and the underlying electricity mix for the country of travel.  In the UK, for instance, driving alone in a medium-sized petrol-fuelled car yields smaller \acrshort{ghg} emissions than air travel for distances shorter than 1,000 km, whereas flying in economy class beats driving over longer distances~\cite{OWIDtravel} (data taken from Ref.~\cite{BEIS}).\footnote{\label{radiative_forcing_foot}These estimates include a ``radiative forcing'' factor of 1.9 for air travel, which accounts for a larger warming effect due to aeroplanes emitting GHGs high in the atmosphere.} 
For a detailed comparison of emissions due to various forms of transport within France, see \fref{fig:emiMobility}.

When and how we travel, however, are not always free choices, being constrained by existing transport infrastructure, local geography, our research, finances, and caring responsibilities.  
Universities and \ACR\ institutions, with their large and progressive workforce, can help tip the balance in favour of the more environmentally sustainable option with a judicious combination of policy, incentives, on-site infrastructure and advocacy.

Our current societal infrastructure is set up to facilitate individual travel by car. Universities have the potential to act as instigators of change in this. Making public transport and cycling the preferred options when possible will increase demand for these more sustainable forms of transport and thus encourage cities to improve the infrastructure for them.

Nevertheless, efforts to limit emissions resulting from travel must be balanced against legitimate needs for mobility: the establishing and maintenance of close collaborative relationships, sharing of research outputs, individual exposure and career development, and travel to research facilities. 
Changes to our travel culture and policies must be implemented so as to benefit and, at the very least, not to worsen barriers to inclusion, by avoiding the disenfranchisement of members of our community such as early career researchers, members of our community from the Global South or those otherwise geographically isolated. 

\newpage
\begin{reco2}{\currentname}
{
\begin{itemize}[leftmargin=6 mm]
\setlength{\itemsep}{4pt}
\item Re-assess business travel needs, using remote technologies wherever practicable.
\item Choose environmentally sustainable means of transport for daily commutes as well as unavoidable business travel, amalgamating long-distance trips where possible.
\end{itemize}
}
{
\begin{itemize}[leftmargin=6 mm]
\setlength{\itemsep}{4pt}
\item Define mobility requirements and travel policies that minimise emissions, while accounting for the differing needs of particular groups, such as early career researchers or those who are geographically isolated.

\item Re-assess needs for in-person meetings, and prioritise formats that minimise travel emissions and diversify participation by making use of hybrid, virtual or local hub participation, and optimising the meeting location(s).

\end{itemize}
}
{
\begin{itemize}[leftmargin=6 mm]
\setlength{\itemsep}{4pt}
\item Support environmentally sustainable commuting by improving on-site bicycle infrastructure, subsidising public transport and providing shuttle services.

\item Disincentivise car travel where viable alternatives exist, facilitate car pooling, and provide on-site charging stations.

\item Incentivise the reduction of business travel, e.g., by implementing carbon budgets with appropriate concessions.

\item Ensure unavoidable travel is made via environmentally sustainable means through flexible travel policies and budgets, and the use of travel agents that offer multi-modal itineraries.  Employ carbon offsetting only as a last resort.

\item Remove any requirement on past mobility as an indication of quality in hiring decisions. 

\item Advocate for improved and environmentally sustainable local and regional transport infrastructure.

\end{itemize}
}
\end{reco2}
\subsection{Commuting}

Changes in commuting patterns are typically affected by life circumstances, including changes in education, employment and residence~\cite{BEIGE2017179}.  The viability of environmentally sustainable mobility, like walking, cycling and taking the train, depends crucially on characteristics of the home and workplace locations, including the distance between them, and their local environment. These properties are seen to influence the relative importance of commuting and business emissions for different \ACR\ institutions. 

For example, CERN, Fermilab, and the Department of Physics at ETH Zurich have wildly different \CdO\ emission profiles due to personal transportation. While emissions due to commuting were roughly equal to those for business travel at Fermilab, commuting outweighed business travel for CERN, and conversely, business travel swamped commuting emissions for ETH Zurich. This reflects the unique environment and characteristics of each these research centres: 

\begin{itemize}
\item ETH Zurich is located in an urban centre and is well connected to the local public transport. In 2008, only 1,700 \acrshort{tco2e} were recorded for commuting, with 7.5 to 10 times larger emissions attributable to business travel (using numbers from 2006--2012). 
The emissions per capita for all staff (including researchers) are significantly smaller than those for Fermilab or CERN. 

\item Fermilab and CERN have more rural settings, with a 77\% majority of CERN employees commuting by car from France. Fermilab's commuter emissions~\cite{FermilabEnvReport} of about 6,000 \tCdOe\ are approximately on par with business travel emissions,\footnote{In a typical year, Fermilab’s approximately 1,900 staff members commute an average distance of 15.6 miles each way mostly by car. This translates into 5,987 \tCdO\ when assuming 250 working days per years and using 404~g of \CdO\ per mile as per the US~Environmental Protection Agency~\cite{USEPA}. This is only 5\% less than Fermilab's emissions from air travel, calculated from 8.2 million (or 42\%) fewer miles flown in 2020 using 200~g per air km~\cite{RefAGU}.} while CERN quotes 5,836 \tCdOe\ of commuter emissions compared to 3,330 \tCdOe\ business travel emissions for its approximate 4,000 staff members. The small amount of travel emissions compared to emissions from commuting reflects to some extent the status of CERN as a scientific centre, to which other members of the community are expected to travel and where travel is easier to avoid, also because the experiments are located at CERN.
\end{itemize}

While ETH Zurich, Fermilab and CERN face different boundary conditions, all three of them, and \ACR\ institutions in general, should aim to reduce emissions from commuting, even if these contribute to their overall budget to a different degree.

\begin{figure}
    \centering
    \caption[Mobility emissions]{GHG emissions
for different means of transport (in g\CdOe\ per km).  Emissions from electricity and vehicle production as well as fuel combustion are included.  All data is for 2019--2020, and comes from the database of Ref.~\cite{labos1p5} --- see, in particular, Ref.~\cite{labos1p5emi} --- and assumes the electricity comes from the French grid, which is a factor of 10 less carbon-intensive than other countries~\cite{OWDintensity}.  For a comparison with the UK, see Ref.~\cite{OWIDtravel}. Note that emissions from personal transport do not scale linearly with number of passengers.
\label{fig:emiMobility}}
\includegraphics[width=\textwidth]{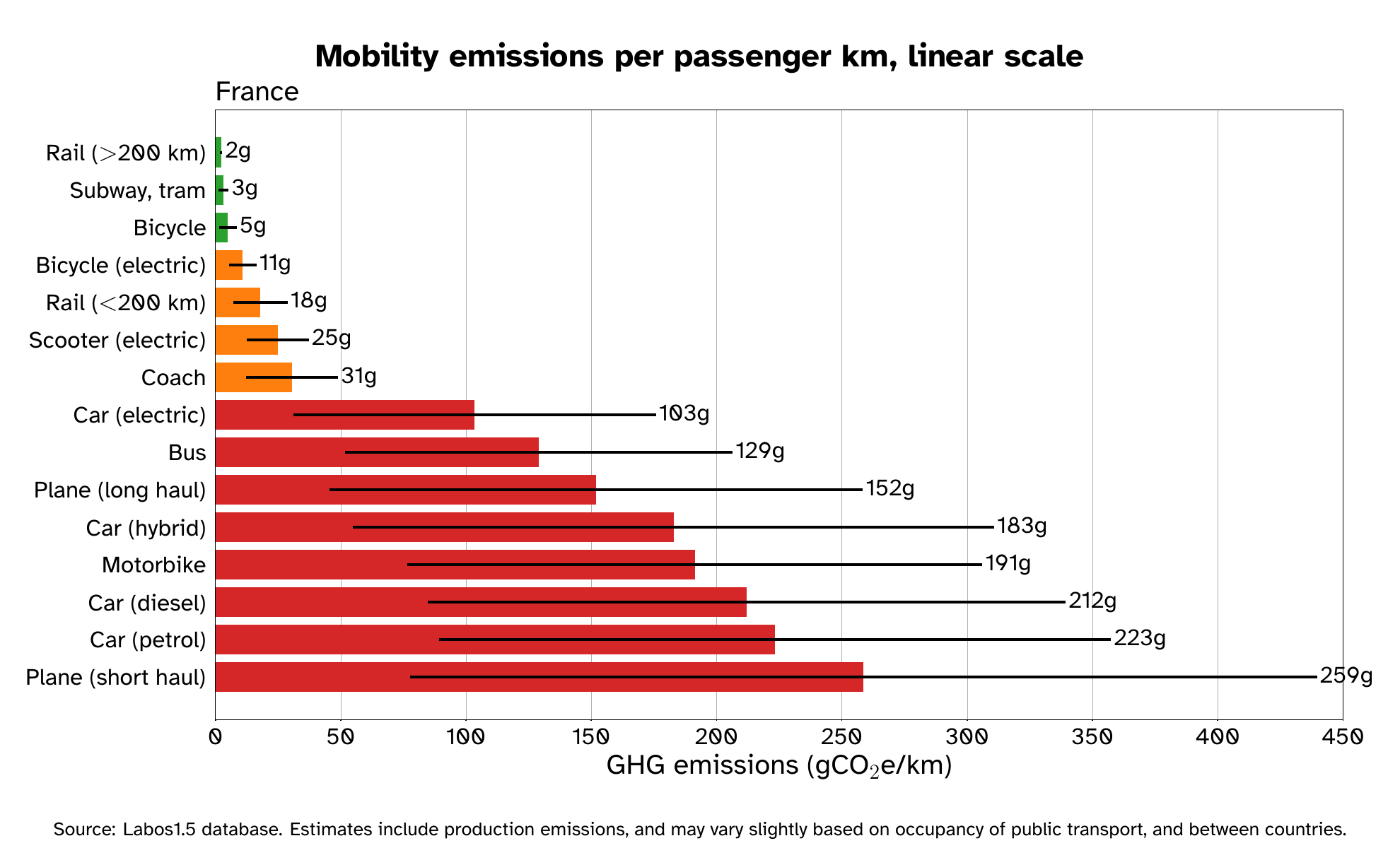}
\end{figure}

This reduction requires an interplay of institutional and individual actions:\ while institutions cannot force employees to choose more environmentally sustainable commuting habits, they can incentivise them through various measures, from the availability of bicycle-friendly infrastructure, such as showers and secured/covered parking, to financial incentives for greener transportation. They can also allow employees to avoid long commutes by formalising telecommuting options, which have become more normalised since the start of the COVID-19 pandemic, and use their standing to push local authorities towards better public transit/cycling/carpool infrastructure. Individuals and groups can, on the other hand, push for these actions at the institutional level. Table~\ref{tab:subCommute} collects some means by which academic and \ACR\ institutes promote `green' transport. An estimate of the emissions per distance
of different forms of transport in France is presented in \fref{fig:emiMobility}.


\begin{table}
\ra{1.05}
\centering
{\scriptsize
\caption[Measures and subsidies for greener commuting]{Institutional/country-wide measures to encourage sustainable commuting amongst employees. This is a non-exhaustive list; similar initiatives are also offered by other employers.
\label{tab:subCommute}}

\begin{tabular}{m{0.1\textwidth}m{0.22\textwidth}m{0.58\textwidth}}
   \toprule
    Institute &
    Initiative&
    Comments\\ 
    
DESY&
    Reduced-price ticket for public transport for all employees&
    The non-transferable ticket, with a 30\% subsidy for employees, is also usable outside working hours, and allows free network-wide travel for an additional adult and up to 3 children (age 14 and under) on weekends and holidays.
    This requires a subscription of more than 6 months. Once
    suspended, a cooling-off period of 9 months is required to be eligible for re-subscription. This is problematic if the employee is posted abroad for a few months. With the implementation of the "Deutschlandticket" in Germany, the terms have slightly changed~\cite{DESYsustainableReport2022,hvvJobticket1,hvvJobticket2}.\\
    \midrule
FNAL&
    Shuttle service
    to and from Chicago Metra trains for all employees &
    There are two scheduled connections in the morning and three in the afternoon, on demand at other times, from 06:30 to 18:50. A ticket costs \$2.25 (cash only), payable to the bus driver \cite{FNALPace}. The regular shuttle does not connect to the Metra station serving the fast train to Chicago (Route 59); FNAL do not offer pre-tax public transport ticket purchase.\\
    \midrule
France &
    Public transit subsidy \cite{transitFR} or \EUR{300}/year for all employees who cycle or carpool \cite{mobdouceFR} &
    Honours system for the \EUR{300}/year. 
    The adoption of the roughly 50\% reimbursement on public transport subscription depends on
    how well connected each institute is.\\
    \midrule
Germany & 
    General tax reimbursement for commuting depending on distance & 
    For each km travelled to work, \EUR{0.30} is deducted from the taxable income (\EUR{0.38} per km above 21 km per one way starting from 2022)~\cite{GermanyTax}.  While this was originally thought to cover the expenses of private cars, it now applies to all means of transport, so that also cyclists or pedestrians obtain the same financial advantage even if they have no direct costs.\\
    \midrule
University of Sheffield  &
    Bike to work scheme for all employees~\cite{shefbike} & There is the possibility to borrow an e-bike (or a bike) for free for 2 months in order to test commuting by bike, to rent bikes throughout the semester and to buy reconditioned bikes. Over 1,400 cycle parking spaces are available throughout campus and at the residences.
    Services are provided for free bike checks and at-cost servicing and repairs for staff and students funded by the university.
    (All UK universities.) Financial help is available to buy an e-bike. (However, this is based on reducing the university's financial contribution to the pension scheme over a set amount of time.)
    \\
   \bottomrule
   
\end{tabular}}
\end{table}


\clearpage

\subsection{Business Travel}

A global scientific endeavour such as \ACR\ will mandate some amount of long-distance travel, e.g., to experimental sites, or to build close working relationships.  However, the current academic culture, which rewards hyper-mobility, is neither environmentally sustainable, nor equitable to all scientists.  Visa rules and prohibitive long-haul travel costs can make participation in conferences extremely challenging, especially for researchers from the Global South. Moreover, the freedom to travel can be heavily restricted for people with disabilities, health impairments or caring responsibilities. For example, the burden of childcare is still unequally distributed, and this burden falls predominately on female shoulders~\cite{McCarthy}.

Emissions from commercial aviation is a long-recognised problem, contributing 2.5\% of \CdO\ emissions and 3.5\% of ‘effective radiative forcing’ (a closer measure of aviation's impact on warming as explained in footnote \ref{radiative_forcing_foot})~\cite{Rit20} in 2018.  Note that the majority of these emissions derive from the one-tenth of the world’s population that can afford air travel.  Almost all \ACR\ scientists belong to the 4\% of the population taking international flights, and many fall within the 1\% classified as the most frequent flyers~\cite{GOSSLING2020102194}.  These statistics highlight the inequalities inherent in travel emissions.

More troubling is that global aviation statistics belie the significance of business travel emissions for many \ACR\ researchers, which are comparable to, and in some cases even dwarf, their direct and indirect emissions (see \fref{fig:Intro-ComparativeEmissions}).  This is clearly in tension with the push to net-zero emissions, particularly given that we do not expect the aviation sector to decarbonise at the same rate as the rest of the transport sector~\cite{Gota2019,Science18}.

Emissions related to conference travel have been studied in detail and dominate conference-related emissions~\cite{Spinellis, nature_better_confs}, contributing annual emissions 30\% larger than the total annual transportation emissions for Geneva (720 kilotonnes \CdO~\cite{Geneve}).  The \CdO\ emissions for a single conference trip amount to about 7\% of an average individual’s total \CdO\ emissions~\cite{Spinellis}. This might be even worse for \ACR\ researchers, for whom frequent trips to experimental sites and meeting venues to undertake international collaborations are common.  See also emissions estimates for business travel of members of the LHCb experiment at CERN in \csref{case:LHCb}.

Discussions about reducing business travel are highly charged, as active engagement with other members of the scientific community is integral to scientific practice. Any changes that we make to \ACR\ travel culture have to be considered in the context of other aspects of our working practices, such as hiring decisions, where any curbs on travel may, \eg disproportionately impact early career researchers. At the same time, the reprioritisation of business travel and a move toward a greater share of virtual/hybrid formats can have a positive impact both on the climate and on inclusivity.

For necessary travel, sustainable alternatives to air travel should be prioritised where possible, keeping in mind that the increased travel time and costs of sustainable travel as compared with air travel could make this choice difficult for researchers with caregiving responsibilities, or limited travel budgets.  In \csref{case:traveltoCERN}, we compare emissions, travel time and cost of different modes of travel to CERN, from various starting points within Europe, for CMS Week in January 2022.  \ACR\ institutions and funding bodies are beginning to implement more sustainable travel policies, including travel top-ups for green travel; we highlight two examples in \bpref{bp:Erasmus} and \bpref{bp:DESY_travel}.

If the community is to rethink this travel culture and move toward more hybrid/virtual modes of engagement, we must recognise that these require additional planning to maximise engagement, which amounts to much more than streaming the in-person event format (see~\csref{cs:CfH}). It is also important to appreciate that virtual participation requires an internet-ready device and stable connection, and devices with which to connect, which may not be universally available in lower-income countries. A possible remedy for this might be the concept of hub conferences, where the conference has several locations spread globally (see, \eg Ref.~\cite{Parncutt}).  In \csref{case:ICHEP}, we study travel emissions and participation in the context of the last 5 instances of the International
Conference for High Energy Physics (ICHEP) conference, and assess the reduction in emissions from optimising the conference location, moving to a hub model, or hybrid/virtual forms of attendance.

The operation of large detector equipment requires the presence of personnel that can take action in case of problems, and call experts if needed. This 24-hours-a-day attendance is called a “shift” in high-energy physics jargon and is considered to be part of the training as a particle physicist. Many collaborations set requirements on the minimal number of shifts that qualify to be listed as an author of scientific publications. These requirements have a climate impact as they require physicists to regularly travel to the experimental site. There is thus an increasing tendency to have remote shifts. The CLEO collaboration in Cornell pioneered them by having professional local shifters and allowing physicists to have shifts remotely~\cite{Wilksen:2005zz}. In the case of the future FCC accelerator, the experimental site may be as far as 50 km away from the CERN main site, adding additional commute travel.


\begin{casestudy}[Sustainable travel to CERN\label{case:traveltoCERN}]{Sustainable travel to CERN}%
The itineraries in \tref{tab:CERNtravel} were found for travel to CERN for, e.g., CMS Week, 24$^\text{th}$--28$^\text{th}$ January 2022, as found on 30$^\text{th}$ November 2021. Although emissions were significantly smaller for rail travel as compared with travel by car or air, as expected, this must be weighed against the increased travel time, and in many cases, cost of rail travel.  Note that the air travel times are underestimated as they do not include travel to the airports, which are usually distant from the city centres, or the usual buffer time required for check-in and security formalities.  For itineraries that include sleeper trains, the additional cost of the train could offset a night's hotel accommodation at origin or destination.

\paragraph{Methodology}

Prices and carbon footprint were rounded to nearest whole number.  For prices not given in euros, currency conversions were made using Google currency converter.  Carbon footprints for one-way travel were calculated using Ref.~\cite{ecopassenger} and then doubled, using all default assumptions, except for toggling on the climate factor for flights.  Precise departure and arrival information was not used for calculation of the flight footprint.  Since some airports are not included as possible destinations, the footprint was calculated from the central train station in the origin city to the central train station in the destination city, and the footprint of travel to the airport is assumed negligible (in comparison to the flight).  Train fares quoted are for the most convenient train journeys from the central station in the origin city to the central station at the destination.  For longer journeys, preference was given to itineraries with overnight trains to maximise efficiency per euro spent, assuming savings on an additional night in a hotel.  For all overnight trains, quoted prices include reservation in shared sleeper cabin. Female-only occupancy can be specified. Note that in many cases there may be a limited number of `super saver' tickets that are available for purchase ahead of time.  Air fares were for the ‘best’ option available on Skyscanner~\cite{skyscanner}, with the inbound flight arriving in time for an assumed midday start of meetings at CERN, and the outbound flight departing after 15:00 hours on Friday.  Flight prices were taken directly from the airline where possible, and include a standard-sized cabin bag, but not necessarily a checked bag.  Durations include flight time only and do not include airport check-in times.
\end{casestudy}


\begin{landscape}
{\tiny
\centering
\ra{1.05}
\extrarowheight=\aboverulesep
\addtolength{\extrarowheight}{\belowrulesep}
\aboverulesep=0pt
\belowrulesep=0pt
\captionsetup{type=table}

\begin{tabular}{@{\kern\tabcolsep}p{1.8cm}>{\baselineskip=10pt}p{1.4cm}>{\RaggedRight\arraybackslash\baselineskip=15pt}p{3.2cm}>{\baselineskip=15pt}p{2.0cm}>{\baselineskip=15pt\RaggedRight\arraybackslash}p{11cm}>{\baselineskip=10pt}p{1.1cm}>{\baselineskip=10pt}p{1.5cm}c@{}}
\toprule
\cellcolor{gray!20}Distance &
\cellcolor{gray!20}Origin &
\cellcolor{gray!20}Mode of Transport &
\cellcolor{gray!20}Travel time (one way) &
\cellcolor{gray!20}Itinerary&
\cellcolor{gray!20}Price (EUR)&
\cellcolor{gray!20}Emissions (kg \CdOe)
\\ \cmidrule{1-7}
  
$<600$ km&
Paris  & 
\cellcolor{Pythongreen!30}Train  &
\cellcolor{Pythongreen!30}3h15     &
\cellcolor{Pythongreen!30}\textbf{Out:} Mon 24th 08:18--11:29 
\textbf{In:} Fri 28th 14:29--17:42     & 
\cellcolor{Pythongreen!30}178    & 
\cellcolor{Pythongreen!30}\noindent 25
\\ \cmidrule{3-7}
&  & 
Flight 
ORY--GVA& 
$\sim$1 hr& 
\textbf{Out:} Mon 24th 08:20--09:25 
\textbf{In:} Fri 28th 19:05--20:15& 
98& 
235
\\  \cmidrule{3-7}
&  & 
Car&
5h42&
& &
116 \\  \cmidrule{1-7}
$>600$ km& 
\cellcolor{gray!10}Hamburg&
\cellcolor{Pythongreen!30}Train (2 changes)&

\cellcolor{Pythongreen!30}$\sim$13.5 hrs&
\cellcolor{Pythongreen!30}\textbf{Out:} Sun 23rd 20:50--10:18 (+1 day)
\textbf{In:} Fri 28th 18:15--07:54 (+1 day) &
\cellcolor{Pythongreen!30}258&
\cellcolor{Pythongreen!30} \noindent 46 \\ \cmidrule{3-7}
&\cellcolor{gray!10} &
\cellcolor{gray!10}Flight
HAM--GVA
(1 change)&
\cellcolor{gray!10}$\sim$3 hrs&
\cellcolor{gray!10}\textbf{Out:} Mon 24th 07:00--10:10
\textbf{In:} Fri 28th 19:10--22:35&
\cellcolor{gray!10}261&
\cellcolor{gray!10}497\\ \cmidrule{3-7}
&\cellcolor{gray!10} &
\cellcolor{gray!10}Car&
\cellcolor{gray!10}9h50&
\cellcolor{gray!10}&\cellcolor{gray!10}
 &\cellcolor{gray!10}
225 \\ \cmidrule{2-7}
&
London&
\cellcolor{Pythongreen!30}Train
(2 changes out; 1 change in)&
\cellcolor{Pythongreen!30}$\sim$8  hrs&
\cellcolor{Pythongreen!30}\textbf{Out:} Sun 23rd 15:31--23:29
\textbf{In:} Fri 28th 15:30--22:30&
\cellcolor{Pythongreen!30}288&
\cellcolor{Pythongreen!30} 25\\ \cmidrule{3-7}
& &
Flight  
LTN--GVA&
1h40&
\textbf{Out:} Mon 24th 08:00--10:45
\textbf{In:} Fri 28th 21:40--22:20&
80&
402\\ \cmidrule{3-7}
& &
Car&
8h32&
& &
 196 \\ \cmidrule{2-7}
  &
\cellcolor{gray!10} Rome&
\cellcolor{Pythongreen!30}Train 
(1 change) &
\cellcolor{Pythongreen!30}$\sim$8.5 hrs &
\cellcolor{Pythongreen!30}\textbf{Out:} Sun 23rd 15:25--23:54 
\textbf{In:} Friday 28th  13:39--21:40&
\cellcolor{Pythongreen!30}238&
\cellcolor{Pythongreen!30} \noindent 70\\\cmidrule{3-7}
& \cellcolor{gray!10}&
\cellcolor{gray!10}Flight
 FCO--GVA &
\cellcolor{gray!10}1h30&
\cellcolor{gray!10}\textbf{Out:} Mon 24th 09:00--10:30
\textbf{In:}  Friday 28th 18:45--20:20&
\cellcolor{gray!10}77&
\cellcolor{gray!10}392\\ \cmidrule{3-7} 
&\cellcolor{gray!10} &
\cellcolor{gray!10}Car&
\cellcolor{gray!10}$\sim$8 hrs&
\cellcolor{gray!10}&\cellcolor{gray!10} & 
\cellcolor{gray!10}183 \\ \cmidrule{2-7}
&
Barcelona&
\cellcolor{Pythongreen!30}Train&
\cellcolor{Pythongreen!30}7-8 hrs&
\cellcolor{Pythongreen!30}\textbf{Out:} Sun 23rd 08:15--16:35
\textbf{In:} Fri 28th 12:35--19:32&
\cellcolor{Pythongreen!30}147&
\cellcolor{Pythongreen!30}\noindent 18\\  \cmidrule{3-7}
& &
Flight
BCN--GVA&
$\sim$1.5 hrs&
\textbf{Out:} Mon 24th 08:40--10:20
\textbf{In:} Fri 28th 17:00--18:25&
83&
370\\  \cmidrule{3-7}
& &
Car&
7 hrs&
& &
164 \\  \cmidrule{1-7}
$>1,200$ km &
\cellcolor{gray!10}Warsaw&
\cellcolor{Pythongreen!30}Train
(2 changes)&
\cellcolor{Pythongreen!30}22.5--24.5 hrs&
\cellcolor{Pythongreen!30}\textbf{Out:} Sat 22nd  19:49--18:18 (+1 day) 
\textbf{In:} Fri 28th  18:42--19:15 (+1 day)&
\cellcolor{Pythongreen!30}319& 
\cellcolor{Pythongreen!30} \noindent 176\\  \cmidrule{3-7}
&\cellcolor{gray!10} &
\cellcolor{gray!10}Flight 
WAW--GVA&
\cellcolor{gray!10}2h20&
\cellcolor{gray!10}\textbf{Out:} Mon 24th  07:20--09:40 
\textbf{In:} Fri 28th  19:45--21:55&
\cellcolor{gray!10}185&
\cellcolor{gray!10}531\\  \cmidrule{3-7}
&\cellcolor{gray!10} &
\cellcolor{gray!10}Car&
\cellcolor{gray!10}12.5 hrs&
\cellcolor{gray!10}&\cellcolor{gray!10} &
\cellcolor{gray!10}398\\\bottomrule
\end{tabular}
\caption[Comparison of modes of travel to CERN from different origins]{Comparison of modes of travel to CERN from different origins.  The mode giving rise to the lowest emissions for each origin is highlighted in green.}
\label{tab:CERNtravel}
}
\end{landscape}

\clearpage


\begin{casestudy}[Comparative study of travel emissions for ICHEP conferences (2012--2020)\label{case:ICHEP}]{Comparative travel emissions for ICHEP conferences}%
Based on the study of the annual meetings of the American Geophysical Union (AGU) in Ref.~\cite{RefAGU}, and the methodology and software tools employed therein, we undertake a survey of the past five editions of the ICHEP with the aim of assessing the \acrshort{ghg} emissions of conference travel to ICHEP, as well as the (geographical) diversity of participants.

The International Conference for High Energy Physics (ICHEP) is a biannual conference with a large and steadily growing participation, of order 1,000 researchers, and a location that alternates mainly between Europe, America and Asia.  We study the 5 most recent instances, with locations in Melbourne, Australia (2012); Valencia, Spain (2014); Chicago, United States (2016); Seoul, Korea (2018); and Prague, Czech Republic (2020, fully virtual).\footnote{At the time of writing, the 2022 conference, held in Bologna, had not yet begun.}

\paragraph{Methodology}

Participant details were taken from the Indico conference system registration pages~\cite{indico}. The departure location for each participant was assumed to be the city of their affiliation, save for cases where it was clear that the participant was based in Geneva, as is often the case for members of LHC collaborations. Direct travel to and from the conference was assumed.  Distances were calculated as the great-circle distance using coordinates obtained with Nominatim~\cite{Nominatim} from the OpenStreetMap data base~\cite{OSM}. Rail, car or bus travel was assumed for all journeys with distances of less than 400 km, with air travel assumed for longer distances.  `Short-haul' was defined as travel distances of less than 1,500 km; distances up to 8,000 km are `long-haul'; and longer distance still were classified as `super long-haul'.\\~
\\
\bigskip
\begin{center}
{\scriptsize
\ra{1.1}
\captionsetup{type=table}
\begin{tabular}{@{}p{2.3cm}>{\baselineskip=10pt}p{1.8cm}>{\baselineskip=10pt}p{1.5cm}>{\baselineskip=10pt}p{1.5cm}>{\baselineskip=10pt}p{1.5cm}>{\baselineskip=10pt}p{1.4cm}>{\baselineskip=10pt}p{1.7cm}c@{}}\toprule
&AGU Fall Meeting 2019 & 
ICHEP Melbourne 2012 & 
ICHEP Valencia 2014 &
ICHEP Chicago 2016 & 
ICHEP Seoul 2018 &         
ICHEP Prague 2020 (virtual) \\ \cmidrule{2-7}
    
Number of participants&
24,009       & 
764     &
966     &
1,120    & 
1,178    & 
2,877    \\ \midrule

GHG emissions per participant [kg \CdOe]     & 
2,883     & 
8,432     & 
1,902     & 
2,699     & 
2,648     & 
0 \\
\bottomrule
\end{tabular}
\caption[Travel GHG emissions for participants of recent ICHEP conferences.]{Total number of participants of recent ICHEP conferences and the GHG emissions per participant.
The corresponding numbers for the American Geophysical Union (AGU) Fall Meeting~\cite{RefAGU} are shown for reference.}
\label{tab:icheppart}}
\end{center}

Table~\ref{tab:icheppart} shows the average GHG emissions per participant for the ICHEP editions alongside those for the 2019 AGU Fall Meeting for reference. With the exception of the 2012 Melbourne edition of ICHEP, the per-capita emissions were significantly lower for ICHEP, which is a ``travelling'' conference, as compared with the stationary AGU Meeting, which always takes place in San Francisco. This indicates that moving a conference series between continents naturally reduces the travel-related emissions as participants tend to wait for the conference to be held near them to make the trip.  Comparing the geographical distribution of home institutes for each conference reinforces this conclusion.  Note that ICHEP Melbourne (2012) was the first and only ICHEP conference taking place in Oceania.

The emissions for two typical ICHEP conferences, one in Europe (Valencia) and one in Asia (Seoul) are displayed as a function of travel distance in \fref{fig:EmmPerDistance}.  A large fraction of attendees at the Seoul conference had to fly super long-haul, giving rise to the majority of the emissions.  Emissions for the remaining half of the attendees was nearly negligible.  This was not the case for Valencia, where as many attendees travelled short haul or less.  It is also clear that the bulk of the emissions is due to long-haul or super long-haul air travel.

\begin{center}
    \vspace{1.5em}
    \captionsetup{type=figure}
    {\includegraphics[width=.49\textwidth]{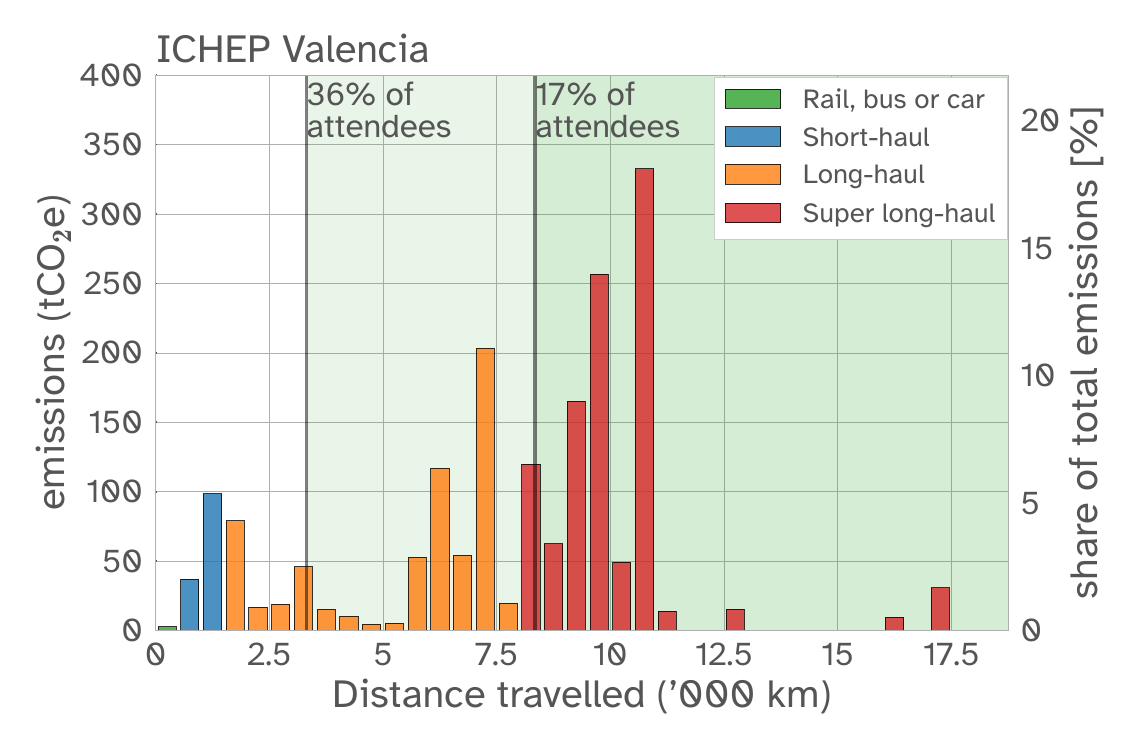}}
    {\includegraphics[width=.49\textwidth]{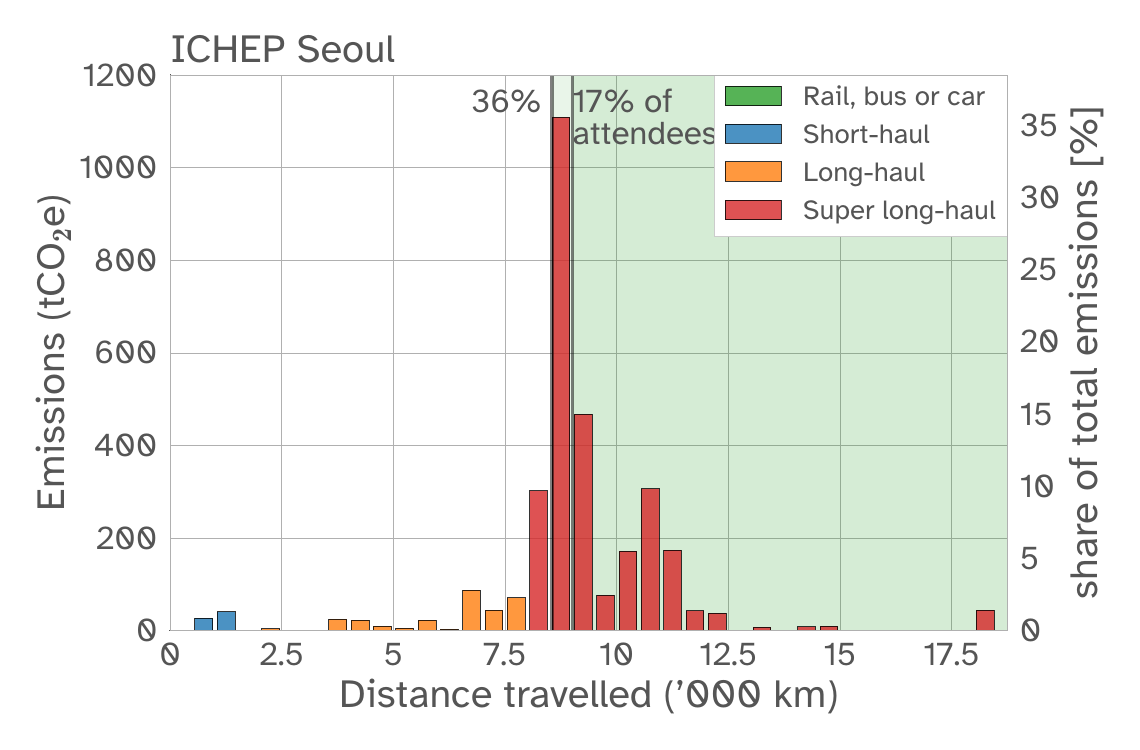}}\\
    \caption[Emissions per distance for two different ICHEP editions]{Emissions per distance for ICHEP Valencia and ICHEP Seoul shown in \tCdOe\ (left axis) and as share of the total emissions. Additionally, the emissions caused by the 17\% and 36\% of participants travelling farthest are shaded in green.}\label{fig:EmmPerDistance}
\end{center}

Reference~\cite{RefAGU} investigated possible optimisations of the conference location for the given participant distribution in order to reduce emissions.\footnote{Optimisations were carried out with a grid spacing, and hence resolution, of 1 degree longitude and latitude}  Note that this is a slightly artificial construction because of the basin of attraction phenomenon discussed above, where participant distribution is self-selecting, based on the conference location. Unlike the AGU example, where moving the conference location to the middle of the country, rather than on a coast, significantly decreased the travel-related emissions, we found that the ICHEP locations were already pre-optimised, and further optimisation yielded at most a 10.2\% reduction of GHG emissions.  (The real outlier again was Melbourne, where the majority of participants had to fly super long-haul, and for which a 70.7\% reduction would be achievable given the same participants by changing the location).  If the location was optimised using participants from all 5 ICHEPs, the optimal location would be close to Amsterdam.

Further emissions reductions are only possible with a hub-based conference, and mandatory virtual participation above a certain distance from the hubs. Reference~\cite{RefAGU} trialled hubs in Chicago, Seoul and Paris, with virtual attendance for all participants with origins greater than 2,000 km from the hubs.  Having found that Chicago, Seoul and Paris were not far from the optimal locations for the respective ICHEP conferences, we did the same, for the total ICHEP participation over the 5 conferences.  Simply using a 3-hub model can reduce the carbon footprint of the conference to around 15--35\% of a traditional one.  Adding compulsory virtual participation for more distant participants reduces the carbon footprint further by 5--15\% of a traditional conference with 10--25\% of the participants attending virtually.  As a test case, and without any prior optimisation, we chose Rio de Janeiro, Johannesburg, and Kolkata as alternative hubs. This, however, increased virtual participation to 95\%¸ mainly due to the strong European participation in HEP and the remoteness of Johannesburg from Western Europe. Switching Paris for Johannesburg reduced the footprint to about 10\% of the nominal one, with 40\% of participants attending virtually. While the virtual fraction is still relatively high, it might be acceptable in a bid to include more remote High Energy Physics (HEP) communities (like Melbourne), while keeping the emissions low.

Finally, one might expect a fully virtual conference to be more inclusive than in-person ones, especially for underserved participants, such as those with care-giving responsibilities, limited travel funding, or visa problems.  We studied this by classifying participants by the human development index (HDI)~\cite{hdiref} of their country of affiliation, and dividing them into four categories (low, medium, high and very high HDI).\footnote{Examples of countries with very high HDI are Norway,  Malaysia, Kuwait and Serbia, high HDI are, \eg Trinidad and Tobago, Albania, Egypt and Vietnam. Medium HDI countries include Morocco and Pakistan, while low HDI countries are \eg Nigeria, Chad and Niger. A brief overview of the categories can be found in Ref.~\cite{hdiref}.}  The share of participants in these categories for each of the ICHEP conferences is shown in \fref{fig:DevelopmentIndex}. Indeed, in addition to enjoying the largest number of participants (by a factor of 2), the virtual ICHEP in Prague had the largest proportion of participants from countries with high or medium human development index, although it was not clear how much of this increase was due to its virtual nature, as opposed to a steady increase in physics participation from high and medium-HDI countries.  There was virtually no participation from low HDI countries in any of the ICHEP conferences studied.


\begin{center}
    \centering
    \captionsetup{type=figure}
    \includegraphics[width=1.\textwidth]{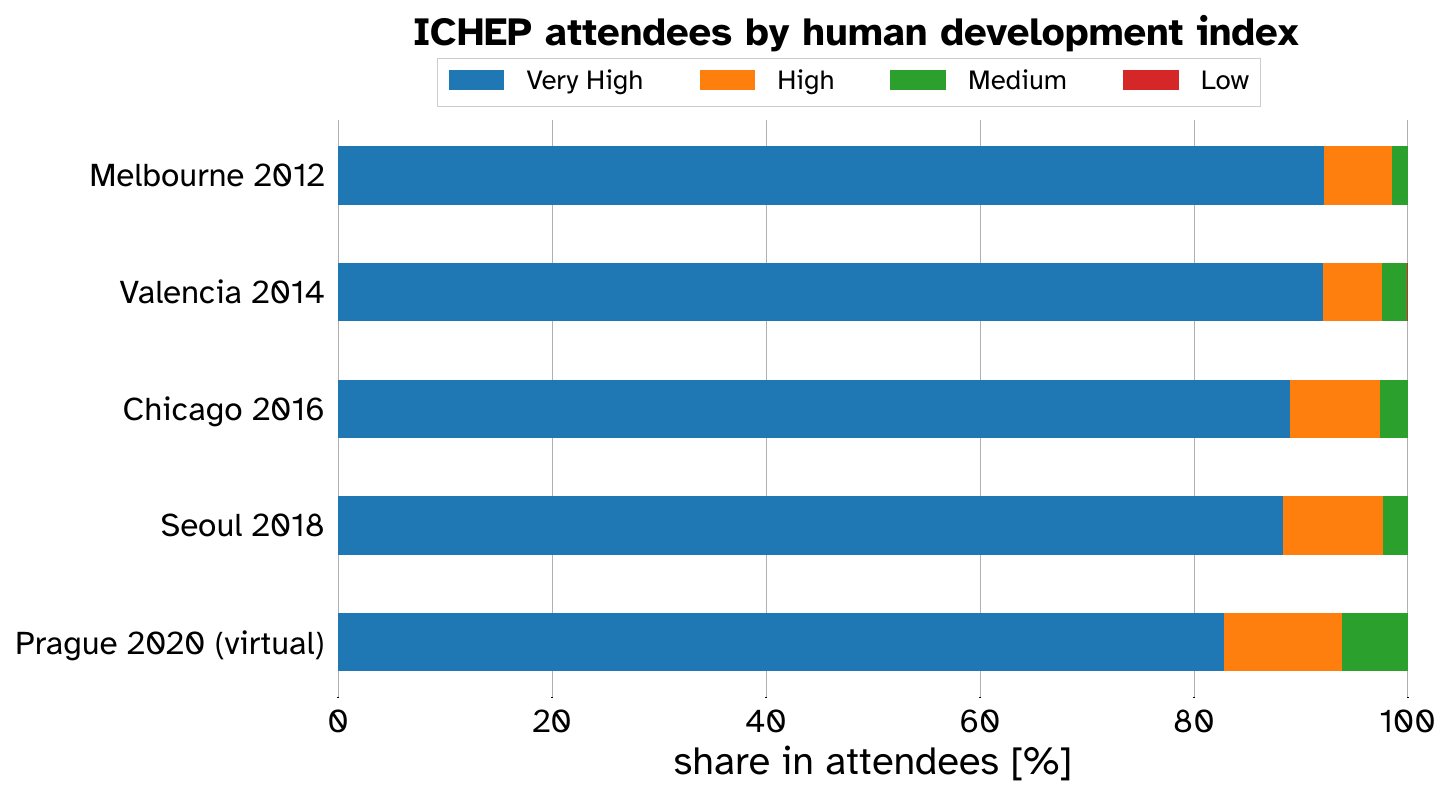}
    \caption[ICHEP participants by development index]%
        {The fraction of participants, categorised by human development index (HDI)~\cite{hdiref} attending the last 5 instances of ICHEP.\label{fig:DevelopmentIndex}}
\end{center}


\end{casestudy}


\begin{bestpractice}[Green travel top-ups on Erasmus+\label{bp:Erasmus}]{Green travel top-ups on Erasmus+}%
The EU mobility and training programme Erasmus+ has implemented funding top-ups for environmentally sustainable travel, which is more costly than point-to-point air travel in many instances, in particular between hubs for low-cost airlines.  See \tref{tab:ErasmusGreenSupplement} for exact supplements for participants who receive travel funding, as excerpted from the 2022 programme guide~\cite{Erasmus+}. Green travel over large distances can be more time-consuming. The programme allows for this by providing travel support for an additional 4 days of travel.\\
\bigskip
\ra{1.05}
\centering
\captionsetup{type=table}

\begin{tabular}{@{}>{\baselineskip=10pt\centering\arraybackslash}p{3cm}>{\baselineskip=10pt\centering\arraybackslash}p{3cm}>{\baselineskip=10pt\centering\arraybackslash}p{3cm}@{}}
\toprule
    Travel distance\newline (km) & Standard travel \newline (\EUR{}/participant) & Green travel \newline (\EUR{}/participant)\\
    \midrule
    10--99 & 23  & \\
    100--499  & 180  & 210 \\
    500--1,999  & 275  & 320 \\
    2,000--2,999  & 360  & 410 \\
    3,000--3,999  & 530  & 610 \\
    4,000--7,999  & 820  &  \\
    >~8,000  & 1,500  &  \\
\bottomrule
\end{tabular}
\caption[Green travel supplements for Erasmus+ participants]{Green travel supplements for Erasmus+ participants. Reproduced with permission from Ref.~\cite{Erasmus+}.}
\label{tab:ErasmusGreenSupplement}
\end{bestpractice}


\begin{bestpractice}[Internal regulations to reduce the impact of business travel at DESY~\cite{DESYsustainableReport2022, privateKSchulz}\label{bp:DESY_travel}]{DESY sustainable travel regulations}%
    The regulations at DESY have been based on the goal of preserving the excellence of science and career opportunities while reacting to the necessity to save \CdO\ and other emissions. With its directive for business trips adopted in 2021, DESY relies on the climate policy principle:\ Avoid --- Reduce --- Compensate.

    \textbf{Avoid:} The number of business trips will be reduced by 30\% compared to the situation before the start of the COVID-19 pandemic. This means that all travel planning is reviewed to identify whether the trip is needed to achieve the intended purpose, whether a virtual meeting could be just as beneficial, whether rotational changes between presence and digital are possible, and whether and how appointments can be bundled. In addition to \CdO\ savings and travel time, it should also be considered how much of the time spent traveling and for travel planning can actually be used as working time, and what costs are incurred or saved. Ultimately, digital meetings also contribute to a better flexibility between personal and work life. They also reduce travel-related risks.

    \textbf{Reduce:} For some time now, it has already been possible to use the train instead of the plane, even if the costs were higher.  With the new directive, the use of the train is now mandatory if the destination can be reached within six hours total travel time. It should also be noted that the usable working time during the trip for rail travel is given as at least 50\% of the travel time (depending on the transfer frequency) and for flying it is assumed to be about 25\% of the travel time.

    \textbf{Compensate:} Until recently, compensation was not possible under the Federal Travel Expenses Act. However, since September 2020, there is a new regulation on the reimbursement policy for carbon offsets from the German ministry of science, whereby also grant recipients like DESY are allowed to offset their \CdO\ emissions for business trips. Starting in 2021, DESY will compensate the consequences of unavoidable air travel (see \sref{sec:impelling}). 
\end{bestpractice}


\begin{casestudy}[Cosmology from Home\label{cs:CfH}\\{\footnotesize \noindent This case study is, in part, adapted from the Cosmology from Home website~\cite{CfHwebsite}.}]{Cosmology from Home}%
Cosmology from Home (CFH)~\cite{CfHwebsite} is an ``online by design'' conference series that has been run on a yearly basis since 2020. It exploits the advantages of digital communication to accomplish things that have no analogue in traditional conference formats, while staying true to the dynamic and social nature of traditional conferences.

CfH is spread over two weeks, with two days per week dedicated to plenary and parallel talks. Talks are pre-recorded and shared with the conference participants ahead of the start of the conference. Other than a strict time limit, CfH places few constraints on the format of the talks. Participants are expected to watch the talks before the scheduled online discussion events. The scheduled discussions last a maximum of three hours per day to mitigate online fatigue, but multiple sessions can be scheduled to accommodate different time zones.

Other days are reserved for themed discussions, which are proposed by the participants.  Topics have included content related directly to cosmology, to scientific research more generally (i.e., technical or computational aspects), and social aspects, such as inclusivity and outreach in science.

The live discussions are hosted in virtual conference venue spaces to create a social atmosphere in which participants can wander around and join conversations. Example conference spaces are Sococo~\cite{Sococo}, Welo~\cite{Welo} and Gather~\cite{Gather}. In case the number of participants is such that the conference spaces are too small to host everyone at once, break-out rooms of a suitable video-conferencing platform can be used. In the specific case of CfH so far, this has been \cite{Zoom}.

CfH is coordinated asynchronously through an online message board, and a large part of the conference is dedicated to asynchronous text-based discussions. These discussions can start before the official opening of the conference and continue long after the conference has finished. In previous years, CfH used Slack~\cite{Slack}. (Alternatives include MatterMost~\cite{MatterMost}, Zulip~\cite{Zulip}, Microsoft Teams~\cite{Teams} or Discord~\cite{Discord}.) Discussion channels are grouped according to the scheduled live-talk discussion sessions (see below). Participants can discuss points in dedicated threads in these channels. Once a given discussion has grown sufficiently and has branched out into different sub-discussions, it can be given its own dedicated channel.

In addition to scheduled talk discussions, CfH implements scheduled themed discussions. These are moderated, workshop-style discussions on cosmological themes that are of interest to the conference participants. The topics are suggested, voted on and chosen via the Slack workspace. People can connect to the session via the conference space, but the main session is hosted in Zoom break-out rooms.

The final live discussion format featured by CfH is composed of spontaneous and unscheduled\footnote{“Unscheduled” means that the discussions are not part of the conference program. The participants can and do schedule these discussions on the asynchronous platform.} “informal” live discussions. The time allocated to the scheduled live discussions is limited. The informal, breakaway sessions allow the participants to engage in more detailed explorations of the topics brought up by the live and asynchronous discussions.

The final live component of CfH consists of social events and interaction formats, such as social games and casual get-togethers, which complement the scientific discussion sessions on all days of the conference. These activities aim to reproduce (at least partially) the evening interactions of in-person conference social events.

One of the main advantages of the CfH format is that, compared to an in-person conference, a much longer time can be spent debating the talks of the participants. This allows for a much greater dissemination and understanding of the research that is presented. Additionally, it removes the need for travel. This keeps the participation costs low and offers the potential for the conference to be carbon neutral or carbon negative (achieved, at present, through carbon offsetting of residual emissions).  The online format also makes the conference accessible to researchers from all over the world, subject to the availability of a suitable device and a stable internet connection by which to connect. The geographical distribution of participants in the previous three CfH conferences are shown in \fref{fig:CfH1}.


\begin{center}
    \vspace{1.5em}
    \centering
    \captionsetup{type=figure}
    \includegraphics[width=1.\textwidth]{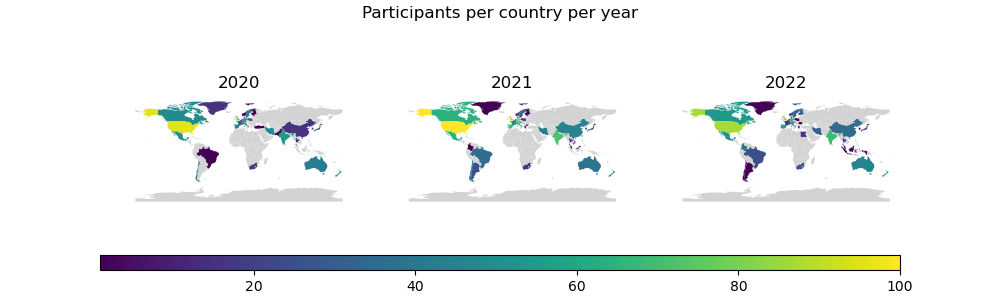}
    \caption[Geographical distribution of CfH participants over time.]%
        {Geographical distribution of Cosmology from Home participants for each of the installments by year.\label{fig:CfH1}}
\end{center}

Additionally, participants can watch the talks according to their own schedules and personal obligations. Participants are able to pause and restart the talks, take time to digest them and to look up background material. They are also able to prepare and raise points for discussion in any of the conference environments. The asynchronous discussions can be tailored exactly to the schedule of the conference participants.

In the first CfH (2020), participants needed time to adjust to the format. This was to be expected, and the organisers actively encouraged participants to partake of the various aspects of the conference and actively modelled the expected social norms. The activity and enthusiasm of participation increased year on year. Speakers created innovative and accessible talk records, and participants regularly referred to these in the live and text-based discussions. Participants organised watch parties and impromptu discussions, made use of the various breakout rooms, and gathered in the virtual environments. The themed discussions proved to be particularly popular, and parallel sessions were often necessary to accommodate the high number of topic suggestions.

All participant feedback has been constructive and positive, and CfH has been well attended. The number of CfH participants were 255, 427 and 275 in 2020, 2021 and 2022, respectively~\cite{CfHwebsite}. The format is easily tailored to various topics: a conference featuring this format in HEP ran in 2024~\cite{neutrinosfromhome}.

\end{casestudy}

\RaggedRight
\sloppy
\newpage


\section{Research Infrastructure and Technology}
\label{sec:Technology}


\begin{center}
    \includegraphics[width=\SDGsize]{Sections/Figs/Common/SDG_6_CleanWater.png}~%
    \includegraphics[width=\SDGsize]{Sections/Figs/Common/SDG_7_CleanEnergy.png}~%
    \includegraphics[width=\SDGsize]{Sections/Figs/Common/SDG_9_IndustryInnovation.png}~%
    \includegraphics[width=\SDGsize]{Sections/Figs/Common/SDG_11_SustainableCities.png}~%
    \includegraphics[width=\SDGsize]{Sections/Figs/Common/SDG_12_ResponsibleConsumption.png}~%
    \includegraphics[width=\SDGsize]{Sections/Figs/Common/SDG_13_ClimateAction.png}~%
    \includegraphics[width=\SDGsize]{Sections/Figs/Common/SDG_15_LifeOnLand.png}
\end{center}


\exSum

\noindent \ACR\ research areas rely on big science infrastructure. These particle accelerators, large-scale collider experiments, observatories and associated buildings infrastructure have a lifetime environmental impact from cradle to grave. This is recognised in Section 8, ``Sustainability considerations'', in the Accelerator R\&D Roadmap of the European Strategy for Particle Physics~\cite{EuropStrategyPP}. It divides the topic into three aspects:
\begin{itemize}
    \item Energy efficient technologies,
    \item Energy efficient accelerator concepts, and
    \item General sustainability aspects.
\end{itemize}
The first two focus on the biggest impact of accelerators:\ the energy consumption during their operation. One aspect focuses on the current technology and its energy efficiency, the other on the development of new accelerator concepts with smaller energy requirements. These topics are discussed in depth in other sections of the Strategy. The third aspect is more broadly defined and considers sustainability beyond energy~\cite{EuropStrategyPP}:
\begin{quotation}
    ``A carbon footprint analysis in the design phase of a new facility can help to optimise energy consumption for construction and operation. For cooling purposes accelerator facilities typically have significant water consumption. Cooling systems can be optimised to minimise the impact on the environment. For the construction of a facility environment-friendly materials should be identified and used preferably. The mining of certain materials, in particular rare earths, takes place in some countries under precarious conditions. It is desirable to introduce and comply with certification of the sources of such materials for industrial applications, including the construction of accelerators. A thoughtful life cycle management of components will minimise waste.''
\end{quotation}

In the case of astronomy research, Ref.~\cite{Kn_dlseder_2022} argues that emissions due to research infrastructure dominate the carbon footprint of an astronomer.

A number of initiatives have already been formed to consider the manifold technical challenges of improving the environmental sustainability of research infrastructure and associated technologies. Three examples are listed below, (for others see \sref{sec:other_initiatives} and references therein):
\begin{itemize}

    \item The International Committee for Future Accelerators (ICFA) has the specific panel ``Sustainable Accelerators and Colliders''~\cite{SustainableAcceleratorsICFA}.
    
    \item Every 2 years since 2011, the Energy for Sustainable Science at Research Infrastructures (ESSRI) workshop~\cite{ESSRI5} takes place.
    
    \item Innovation Fostering in Accelerator Science and Technology (I\.FAST)~\cite{IFAST} is a HORIZON 2020 Research Innovation Action in which the “WP 11 – Sustainable concepts and technologies” is aimed to increase sustainability. The current participating institutes are CERN, DESY in Germany, the European Spallation Source (ESS) in Sweden, the GSI in Germany, the Paul Scherrer Institut (PSI) in Switzerland, and the Science and Technology Facilities Council (STFC) in the United Kingdom.

\end{itemize}
Environmental sustainability is also being considered by individual experiments, and \csref{case:LHCb} provides a summary of efforts by the LHCb experimental collaboration to assess and mitigate the environmental impact of both the experiment and work practices, more generally.

In this section, we consider the following aspects of environmental sustainability in research infrastructure: life cycle assessment (LCA), (carbon) accounting, and technological developments, particularly in the context of accelerator technologies and detector gases. While the discussions of technological developments provide concrete examples, the primary focus of this section is the need for critical life cycle analysis for all research infrastructure projects to assess and limit their environmental impacts. The impacts of mining and processing of materials is also considered in~\sref{subsec:Resources}, wherein complementary aspects of the LCA are also discussed.

\clearpage
\begin{reco2}{\currentname}
{
\begin{itemize}[leftmargin=6 mm]
\setlength{\itemsep}{\recskip}
\item Seek out new innovations and best practice.

\item Rethink how the impact of frequently used equipment can be reduced, and reduce "over-design" by reassessing safety factors and other margins to reduce resource consumption.

\item Read section on resources and waste (Section~\ref{sec:Waste}).

\end{itemize}
}
{
\begin{itemize}[leftmargin=6 mm]
\setlength{\itemsep}{\recskip}
\item  Ensure that environmental sustainability is an essential consideration at all stages of projects, from initial proposal, design, review and approval, to assembly, commissioning, operation, maintenance, decommissioning and removal, using life cycle assessment and related tools.

\item Engage with industrial partners who exemplify best practice and sustainable approaches.

\item Appoint a dedicated sustainability officer to oversee project development, and institute regular meetings with a focus on environmental sustainability.

\end{itemize}
}
{
\begin{itemize}[leftmargin=6 mm]
\setlength{\itemsep}{\recskip}
\item Critically assess the environmental impact of materials, construction and the operational life cycle as an integral part of the design phase for all new infrastructure.

\item Provide training opportunities, required tools and technical support to assess and improve the environmental sustainability of project life cycles.

\item Recognise and reward innovations that minimise negative environmental impacts.

\item Promote knowledge exchange on sustainability initiatives between groups and institutions, including decision-makers, designers and operators of projects, setups and infrastructure.
\end{itemize}
}

\end{reco2}


\subsection{Accounting and Reporting}

The methodology of a life cycle assessment can be used to analyse the environmental impact of resources used to build, run and decommission an accelerator, observatory or experiment, see \sref{subsec:lca} for further details. Such assessments have already been undertaken by a number of facilities, including:
\begin{itemize}
    \item The European Southern Observatory (ESO)~\cite{ESO}.
    \item The Giant Radio Array for Neutrino Detection (GRAND) Project, a multi-decade astrophysics experiment~\cite{Aujoux_2021} --- This led to a full issue of the Nature Astronomy Journal on climate change~\cite{NatureClimateIssue}.
    \item The Relativistic Ultrafast Electron Diffraction and Imaging (RUEDI) facility at STFC Daresbury Laboratory ~\cite{Shepard}.
    \item The Compact Linear Collider (CLIC), the proposed linear accelerator complex at CERN, is planning to conduct an assessment ~\cite{privateBList}.
    \item The ISIS-II project, the next generation of the ISIS neutron and muon source, is planning to conduct life cycle analyses for the project and various design options ~\cite{ISISII}.
\end{itemize}

There is currently limited availability of data on estimated emissions and resources consumption for basic research infrastructure, and, where it is available, its presentation is not standardised. This makes overall assessments of sustainability and comparisons of individual technologies challenging. Implementation of effective life cycle assessment across the \ACR\ community could provide the impetus for standardised reporting that will provide the data needed for ongoing assessment of current and future technologies and research infrastructure projects, such as any future collider concept (see \csref{case:FutureColliders} and \bpref{bp:CCC}). The Labos1point5 working group has proposed a standardised carbon accounting procedure and associated assessment tool for research laboratories. This programme is described in detail in \bpref{BP:l1p5}.

For illustration, \bpref{bp:SiWafer} provides an example of a partial LCA for the production phase of a monocrystalline silicon wafer, as summarised from the ProBas library for life cycle assessment~\cite{ProBas}.  Wafers of this purity are used as sensors for particle tracking and calorimetry in detectors, and as a starting point for the custom ASIC and FPGA chips used in detector readout. Purification of silicone is a low-yield process, which is energy- and water-intensive~\cite{boyd2011life}. Particle detectors use a large wafer surface area and are therefore more sensitive to small local defects, resulting in more wastage~\cite{Hartmann:2017gzy}.  The total surface area covered by these detectors is large e.g., the CMS endcap calorimetry upgrade for the HL-LHC will require more than 600 m$^2$ of silicon sensors~\cite{PAULITSCH2020164428}, while the inner detector tracking upgrades of ATLAS~\cite{ATLASITKTDR} and CMS~\cite{CMSITKTDR} require just under 400 m$^2$ combined.  Production of detector readout chips is more similar to the commercial manufacture of microchips for the computing industry, requiring additional processing steps (oxidation, photolithography, doping, annealing, metallization and dicing) that increase GHG emissions up to an order of magnitude~\cite{boyd2011life}.  Suppliers for the sensors and chips used within the HECAP+ community are limited, and their processes proprietary, making up-to-date evaluation of their environmental costs difficult.

\begin{bestpractice}[Ecological impacts due to production of a silicon wafer\label{bp:SiWafer}]{Ecological impacts due to production of a silicon wafer}%

\noindent Excerpts from the life cycle assessment for production of a 1 ${\rm m}^2$ monocrystalline silicon wafer (thickness 300 \textmu{}m, weight 0.7 kg) in Germany in 2005, are summarised here from Refs.~\cite{ProBasSi1,ProBasSi2,ProBasSi3,ProBasSi4}.  Table~\ref{tab:InputLCA} contains selected elementary flows for the silicon process chain, from production of metallurgical-grade (MG) silicon from sand, through manufacture of high purity electronic-grade (EG) silicon, to the final wafer.  A complete LCA for wafer production traces all such nesting process chains for fabrication of the complete set of input components in these elementary flows, collating their outputs and impacts, shown in \tref{tab:OutputLCA}. Third-party tools and software exist that can automate this complex process.  This data is for illustrative purposes only; LCA data undergoes rapid obsolescence, particularly in the context of swiftly developing technologies, where up-to-date information on processing methods are more likely to be proprietary. \\
{\ra{1.01}
\centering
\footnotesize
\captionsetup{type=table}
\begin{tabular}{m{0.2\textwidth}m{0.18\textwidth}}
\multicolumn{2}{l}{Metallurgical-grade silicon}\\
\toprule
Input & Quantity\\
\midrule
sand & 5.2 kg\\
petroleum coke & 0.7 kg\\
liquid O${}_2$ & 0.04 kg \\
electricity & 89 MJ\\
carbon anode & 0.16 kg\\
\midrule
Product output & 1.79 kg MG Si\\
\bottomrule
\\
\\
~\vspace{1.5em}~
\\
\multicolumn{2}{l}{Silicon wafer}\\
\toprule
Input & Quantity\\
\midrule
EG Si & 1.58 kg\\
distilled water & 65.0 kg\\
electricity & 522 MJ \\
argon & 6.20 kg\\
potable water & 0.0060 kg\\
detergent & 0.24 kg\\
steel & 1.49 kg\\
SiC (high purity) & 2.16 kg \\
quartz crucible & 0.36 kg\\
heat & 77 MJ\\
propylene glycol & 2.93 kg\\
NaOH (50\%) & 0.029 kg\\
HCl & 0.0027 kg\\
glass & 0.0098 kg\\
acetic acid (high purity) & 0.039 kg\\
\midrule
Product output & 0.70 kg Si wafer\\
\bottomrule
\end{tabular}\raisebox{6.9em}{\begin{tabular}{c}\Huge{$\rightarrow$}\\ \\ \\ \\ \\ \vspace{0.7em} \\ \Huge{$\leftarrow$
}\end{tabular}}
\raisebox{9em}{\begin{tabular}{m{0.2\textwidth}m{0.18\textwidth}}
\multicolumn{2}{l}{Electronic-grade silicon}\\
\toprule
Input & Quantity\\
\midrule
MG Si & 1.79 kg\\
heat & 292.7 MJ\\
NaOH (50\%) & 1.04 kg \\
HCl & 1.06 kg\\
H${}_2$ & 0.16 kg\\
electricity & 626.5 MJ\\
\midrule
Product output & 1.58 kg EG Si\\
\bottomrule
\end{tabular}}
\caption[Selected elementary flows of silicon wafer production in the manufacturing stages]{Selected elementary flows for the production of a 1 ${\rm m}^2$ monocrystalline silicon wafer in Germany in 2005, illustrating one process chain used in the LCA through the manufacturing of metallurgical-grade (MG) and electronic-grade (EG) silicon, to the final silicon wafer. Data taken from Refs.~\cite{ProBasSi1,ProBasSi2,ProBasSi3}.}
\label{tab:InputLCA}
}

{\ra{1.01}
\centering
\captionsetup{type=table}
\footnotesize
\begin{tabular}{m{0.29\textwidth}m{0.15\textwidth}|m{0.29\textwidth}m{0.15\textwidth}}
\toprule
Output & Quantity & Output & Quantity \\
\midrule

monocrystalline Si wafer	&	0.70	kg	&	{\bfseries Emissions to air}	&			\\
\cellcolor{Pythongreen!30}{\bfseries Impact Assessment}	&	\cellcolor{Pythongreen!30}	&	particulates (PM10)	&	$1.71\times 10^{-2}$	kg	\\
\cellcolor{Pythongreen!30}energy renewable	&	\cellcolor{Pythongreen!30}213	MJ	&	SO${}_2$	&	0.137	kg	\\
\cellcolor{Pythongreen!30}energy total	&	\cellcolor{Pythongreen!30}4850	MJ	&	CFC-11	&	$1.3\times 10^{-5}$	kg	\\
\cellcolor{Pythongreen!30}GHG emissions	&	\cellcolor{Pythongreen!30}319 \kgCdOe&	PFC-14	&	$5.17\times 10^{-6}$	kg	\\
\cellcolor{Pythongreen!30}acidification	&	\cellcolor{Pythongreen!30}0.369	\kgSdOe&	PFC-116	&	$6.50\times 10^{-7}$	kg	\\
{\bfseries Waste}	&			&	dioxins	&	$1.18\times 10^{-11}$	kg	\\
FGD residues (landfill)	&	2.79	kg	&	lead	&	$1.51\times 10^{-5}$	kg	\\
production waste	&	2.32	kg	&	acid (as H+)	&	$4.60\times 10^{-4}$	kg	\\
sewage sludge	&	$4.89\times 10^{-3}$	kg	&	nitrogen oxides	&	0.309	kg	\\
ash	&	11.0	kg	&	cadmium	&	$9.8\times 10^{-7}$	kg	\\
waste (landfill)	&	1263	kg	&	nickel	&	$2.72\times 10^{-5}$	kg	\\
nuclear waste	&	$3.97\times 10^{-4}$	kg	&	ammoia	&	$9.66\times 10^{-4}$	kg	\\
{\bfseries Emissions to water}	&			&	nitrous oxides	&	$8.89\times 10^{-3}$	kg	\\
lead 	&	$3.28\times 10^{-10}$	kg	&	mercury	&	$3.24\times 10^{-6}$	kg	\\
phosphorus	&	$9.24\times 10^{-7}$	kg	&	fluorides	&	$8.65\times 10^{-4}$	kg	\\
nitrogen total (excluding N${}_2$)	&	$5.45\times 10^{-5}$	kg	&	polycyclic aromatic hydrocarbons	&	$9.38\times 10^{-10}$	kg	\\
mercury	&	$2.51\times 10^{-11}$	kg	&	chlorides	&	$1.47\times 10^{-2}$	kg	\\
chromium	&	$3.28\times 10^{-6}$	kg	&	H$_2$S	&	$6.08\times 10^{-6}$	kg	\\
chemical oxygen demand	&	0.352	kg	&	CO${}_2$ (fossil)	&	299	kg	\\
biological oxygen demand	&	$9.87\times 10^{-3}$	kg	&	CO (fossil)	&	6.27	kg	\\
arsenic	&	$2.06\times 10^{-11}$	kg	&	CH${}_4$ (fossil)	&	0.571	kg	\\
adsorbable organic halides	&	$8.68\times 10^{-8}$	kg	&	other volatile organic compounds	&	$1.79\times 10^{-2}$	kg	\\
inorganic salts and acids	&	0.014	kg	&	arsenic	&	$3.49\times 10^{-6}$	kg	\\
\bottomrule
\end{tabular}
\caption[Outputs of silicon wafer production]{Outputs for the production of a 1 ${\rm m}^2$ monocrystalline silicon wafer in Germany in 2005, obtained from following the process chain in the manufacturing stages of the LCA, cf.~the process(es) indicated above in Table~\ref{tab:InputLCA}. The Impact Assessment (shaded) summarises key impacts. Data taken from Ref.~\cite{ProBasSi4}. A pollutant with 1 \kgSdOe\ has the same acidification potential as a kilogram of sulphur dioxide.}
\label{tab:OutputLCA}
}

\end{bestpractice}


\begin{casestudy}[\label{case:FutureColliders}Sustainability of future colliders]{Sustainability of Future Colliders}%
\noindent The future of HEP includes decisions on future collider facilities to be built. \fref{fig:powerCollider} compares the energy needs of future electron-positron (e$^+$e$^-$) colliders. The projected grid power during operation is given, including for the laboratory, computer center and detector.

\begin{center}
    \captionsetup{type=figure}
    {\includegraphics[width=.8\textwidth]{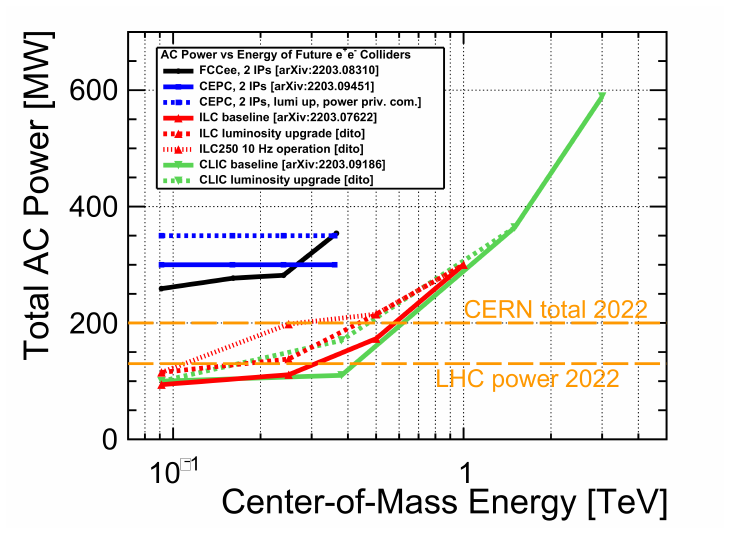}}
    \caption[Site power required for proposed $e^+e^-$ colliders]{Site power required for proposed electron-positron ($e^+e^-$) collider projects at different center-of-mass energies, compared to the average power consumption of CERN and the LHC during accelerator operation in 2022~\cite{privateJList,poweringCERN}.  Figure used with permission from the author~\cite{privateJList}. }\label{fig:powerCollider}
\end{center}

The environmental sustainability of future facilities will become an increasingly important and heavily scrutinised factor in the decision making process about which, if any, facility should be built. The life cycle assessment for such facilities is extremely complex, and must cover not only the accelerator and the detectors, but also the civil engineering, such as tunnels and caverns, buildings infrastructure, and computing needs. The impact of construction, then deconstruction (except for the tunnel) and disposal (including activated materials, which may constitute a radiation hazard) after a few years is not negligible.

The evaluation of the \CdO\ footprint due to the electricity consumption of a future collider is particularly challenging, since any estimate relies heavily on assumptions about future electricity mix. A conservative estimate might be given based on the current \CdO\ footprint for electrical energy, under the assumption that the grid will be decarbonised --- indeed it has to be to meet the world's climate goals --- by the time any future collider is commissioned. However, there is the opportunity to adapt to potential sources of renewable energy (see also \sref{sec:Energy}) in the design stages, and to go beyond arguments of ``electrical effectiveness'' based on scientific benchmarks, such as kWh/Higgs or kWh/fb$^{-1}$ compared to the required connected power.
\end{casestudy}

\begin{bestpractice}
[\label{bp:CCC} Sustainability strategy for the Cool Copper Collider\\{\noindent\footnotesize Edited contribution from Martin Breidenbach, Brendan Bullard, Alessandro Emilio, Dimitris Ntounis and Caterina Vernieri, SLAC and Stanford University.}]{Sustainability strategy for the Cool Copper Collider}%
\noindent Reference~\cite{Breidenbach:2023nxd} evaluates the carbon impact of constructing and operating the Cool Copper Collider (C3),
 a linear electron-positron collider designed to study the Higgs boson and search for new physics. The carbon footprint of C3
 is compared  to other proposed Higgs factory colliders: the International Linear Collider (ILC), the Compact Linear Collider (CLIC),  the Future Circular Collider (FCC), and  the Circular Electron-Positron Collider (CEPC). The framework for the comparison
 serves as an example of best practices for establishing holistic sustainability benchmarks for future HEP facilities.

The
 comparison is achieved by assuming a common carbon intensity of electricity production that will be available during the time of operation, which is supported by a detailed investigation of grid electrification projections. This is an improvement over previous
 analyses that reference current figures for carbon intensity, which are subject to significant reduction driven by high levels of national investment and industry R\&D. The comparison also includes the contribution of embodied carbon from construction for the
 first time. Common emission rates for cement and steel are taken for all facilities. These common inputs enable a direct comparison of the dominant emissions factors for all HEP facilities to be made.

In
 addition to the analysis of absolute carbon emissions,  a new metric is defined that weighs a facility’s total carbon footprint by the expected physics impact that each machine is expected to have. The latter is evaluated in terms of the ultimate sensitivity
 reached in the determination of Higgs couplings. This metric is more appropriate than other proxy metrics for physics output, such as the number of Higgs bosons produced, which essentially considers only the total luminosity of the facility. The metric targets
 directly the physics potential of each facility, folding in  the important contributions of beam polarisation and center of mass energy.

A
 highlight of this publication is an analysis of energy efficiency of the C3
 accelerator. Through a combination of careful tuning of the bunch structure, utilisation of high efficiency radiofrequency (RF) sources and pulse compression, an energy reduction of 63\% is achieved for the cryogenics and RF systems relative to the nominal
 operating parameters. This leads to an overall reduction in expected site power by 42\% and 45\% for the 250 and 550 GeV running modes.

The
 sustainability of C3
 is driven by beam energy optimization and the compact 8km footprint of the C3
 linac, which significantly reduces embodied carbon in building materials and enables a cut-and-cover construction approach that reduces construction complexity. As a result, C3
 emerges as an attractive choice for advancing our understanding of the Higgs boson within an environmentally conscious approach to facility design and evaluation.

\end{bestpractice}

\begin{bestpractice}
[Standardised accounting
of the carbon footprint of French research 
institutions: labos1point5 \label{BP:l1p5}]{Carbon footprint accounting with the Labos1point5 tool}%
\noindent \emph{Laboratoires de recherche}, loosely translated as research labs,
are the entities around which most of French research is organised.
They enjoy a relevant degree of autonomy, including 
aspects such as scientific goals and experimental designs. 
Access to research facilities, as well as 
a fraction of the annual budget, is also managed at the lab scale.
Hence, this makes it a relevant scale to tackle the question 
of the carbon footprint of academic research. 

This has motivated the creation of the \emph{labos1point5} 
working group (\emph{Groupement De Recherche}, GDR),\footnote{``1point5''
refers to the warming limit of the Paris accord}
gathering an interdisciplinary team of engineers and researchers from various research fields in France. 
One of the main outputs of this collaboration is the \emph{GES 1point5},\footnote{``GES''
is the French acronym for greenhouse gases.} a
standardised online tool for the accounting of the carbon 
footprint of French research labs. What follows is a brief
summary of the latter. We refer to Ref.~\cite{labos1p5}
for a publication describing it. Further information 
can be found on the website of the labos1point5 collaboration~\cite{labos1p5web}. The tool itself is available at Ref.~\cite{ges1p5}, 
while the open source code is hosted at Ref.~\cite{ges1p5git}.
The GDR labos1point5 is also active in helping labs in their
transition to a lower footprint and in developing new ways of 
teaching climate and ecological aspects to students, as well
as communicating to the general public. Finally,
there are  teams dedicated to the reflection on the role
of science in the climate crisis, and on fostering collaboration 
between arts and science in this context.

As explained in Ref.~\cite{labos1p5}, one of the main motivations for the 
creation of the GES 1point5 tool was the difficulty in aggregating or comparing the results of 
the many existing studies in the literature on the carbon footprint of academic research.
This difficulty was caused by the sensitivity of the footprint to the applied 
methodology which made comparisons extremely challenging, since discrepancies in results
could not be disentangled from methodological differences. 
The creation of the GES 1point5 was then intended to provide 
a tool specifically designed to estimate the carbon footprint of research with a transparent
and accessible methodology and a database of carbon footprints assessed with the same
methodology to enable a robust comparison of research carbon footprints across institutions,
contexts or disciplines.

More specifically, GES 1point5 allows research labs to
estimate their yearly emissions --- as of February 15, 2023,
628 laboratories have compiled 1,140 yearly carbon footprint 
determinations \cite{labos1p5web}.
Currently, GES 1point5 can estimate \acrshort{ghg} emissions 
due to the energy consumption and refrigerant gases of the laboratories' buildings, 
those attributed to the purchase of their digital devices, and to computing, 
commuting and professional travel, as well as the associated uncertainties.
A module estimating emissions from consumables purchases has recently been added.
In all cases, the estimation is based upon its established
standardised methodology and the database of emission factors, and turns out 
to be relatively straightforward for the end user.\footnote{In the case
of professional travel, a feature has been added to the internal management software
of the French National Centre for Scientific Research (CNRS) that can output the travel data --- origin, destination, means of transportation, etc.~---
for a given year in a ready-for-GES 1point5 format,
sparing the end user tedious data entry and conversion.} To give an example, 
commuting emissions are estimated by gathering data through an anonymised survey
sent to staff members, which can be answered in less than five minutes.
For each specified commute (up to two different ones per week can be entered), GES 1point5 multiplies the distance traveled via each means of transportation by
the specific emission factor from its database. These are collected in \fref{fig:emiMobility}.
The underlying routines are available at Ref.~\cite{1p5commute} for anyone to test their 
commutes; similarly, those used in the determination of professional travel emissions are 
available at Ref.~\cite{1p5travel} and those for purchases at 
Ref.~\cite{1p5purchases}.

Once the emissions have been estimated, GES 1point5 presents the results 
through its graphical interface, highlighting the main drivers of the carbon footprint. 
Finally, emissions reduction actions that may be undertaken after the evaluation of the 
footprint can be evaluated in the subsequent years, thanks to the reliance
on a standardised protocol. The latter also allows the aforementioned 
aggregation and comparison of GES 1point5-based carbon footprints.

As stated in Ref.~\cite{labos1p5}, while some aspects of GES 1point5 are specific 
to the context of French research, the tool  may be reused in research centers elsewhere, provided
the necessary adjustments --- e.g., carbon intensity of the grid --- are made. 
French- and English-language versions of GES 1point5 are built into the current version to ease deployment in any country. Contacts with several institutions outside France 
have been established.
\end{bestpractice}


\subsection{Technological Improvements}

There are ongoing efforts across the \ACR\ community to reduce the environmental impacts of the technologies implied in research facilities. Two examples in the case of accelerator technologies are provided in Best Practice~\ref{BP:EnergyRecoveryAccelerator} and Case Study~\ref{case:wakefield}, namely energy-recovery accelerators and plasma wakefield acceleration technologies. Further details of efforts by the LHCb collaboration are included in \csref{case:LHCb}. A detailed discussion of the impact of detector gases is provided in the following subsection.

\begin{bestpractice}[\label{BP:EnergyRecoveryAccelerator}Realisation of a multi-turn energy-recovery accelerator\\{\noindent\footnotesize Edited contribution from Tetyana Galatyuk on behalf of KHuK (Komitee f\"{u}r Hadronen- und Kernphysik)}.]{Multi-turn energy-recovery accelerator}%
\noindent The operation of particle accelerator facilities is inherently resource-intensive, and thus poses a challenge to sustainability. In line with acknowledging our responsibility for sustainable usage of energy resources, the development, establishment, and demonstration of a scalable multi-turn Energy Recovery Linac (ERL) with efficient energy recycling was implemented at the S-DALINAC accelerator at TU Darmstadt, Germany~\cite{Arnold:2020snn}. An efficient energy-recycling in multi-turn operation with a saving of up to 87\% of the beam power-consumption in the main LINAC has been recently demonstrated. This result, together with further developments on multi-turn ERLs is a promising basis for future high-power beams that truly support sustainability aspects. These examples include ER@CEBAF in the USA~\cite{Meot:2018yoo}; MESA ERL in Germany~\cite{MESA}; International PERLE Collaboration~\cite{PERLE, PERLECDR}; CBETA, the Cornell-BNL ERL Test Accelerator~\cite{CBETACDR}; for an overview, see Refs.~\cite{Klein:2022lgx, Hutton:2022kac}.
\end{bestpractice}

\begin{casestudy}[Sustainability of plasma wakefield acceleration technology for future accelerators\label{case:wakefield}\\{\noindent\footnotesize{Edited contribution from Nikola Crnkovi\'{c}.}}]{Plasma wakefield for future accelerators}%
\noindent  A promising technology, which would reduce both the material and energy cost of accelerating particles, and hence improve its environmental sustainability, is wakefield acceleration~\cite{wakefield1}.  This uses laser pulses (in the case of laser wakefield accelerators) or particle beam bunches (for plasma wakefield accelerators (PWFAs)) as the driver to accelerate plasma electrons, creating ion cavities.  This creates an electric field that pulls electrons back to their original positions, which they overshoot, creating waves in the plasma, known as the wakefield.  These plasma waves can accelerate electrons by transferring energy from the drive beam to electrons by putting electrons just behind the drive beam.  Wakefield technology routinely gives acceleration gains of $10$--$100$~GeV/m \cite{wakefield1}, thus resulting in a significant reduction of the resources required (materials, energy) to build particle accelerators.  For example, accelerating electrons to 1 TeV of energy using PWFA would require, \eg only a 21~km-long particle accelerator, while CLIC technology needs 52~km~\cite{wakefield2}.  Furthermore, there are indications that PWFA is more power efficient at high energies than conventional accelerator technology~\cite{wakefield2}.
\end{casestudy}

\subsubsection{Gases}

Significant quantities of GHGs are used for particle detection technologies and cooling systems across \ACR. As such, they are used as a resource, but can escape the detector volume into the atmosphere and turn into potentially dangerous waste gases. Table~\ref{tab:GasImpact} lists a number of GHGs, their chemical formulae, atmospheric lifetimes and global warming potentials (\acrshort{gwp}). All of these gases are used either for cooling purposes or as active ingredients in gaseous detector systems, where they are often added to noble gases to improve detector properties, such as drift charge velocities or diffusion coefficients~\cite{Sauli:2014cyf}.

While gases are used in a variety of detectors, such as time projection chambers, ring-imaging Cherenkov detectors or multi-wire proportional chambers, the main offenders in ecological terms are the detectors used in the muon systems of the LHC experiments, specifically resistive plate chambers (RPCs). This is due to their large areas of around 7,000 m$^2$ in total for both the CMS and ATLAS muon systems, and the gas mixtures used to cope with the large event rates at the LHC, which often feature HFC-134a as a main component. 

As shown in Figure~\ref{fig:Intro-ComparativeEmissions}, Scope 1 direct emissions made up about 25\% of CERN's carbon footprint in 2019. During the LHC Run 2 in 2018, it was about a factor of two larger.  92\% of these Scope 1 emissions are related to the activities of the large LHC experiments~\cite{Environment:2737239,envrep2020}. CERN and the LHC experiments are actively working on reducing this impact by continuously repairing gas leaks that are one of the main reasons for the large amount of waste gas. In addition, CERN has tested an HFC-134a recuperation plant showing an efficiency of close to 85\%, which is to be installed in the detectors to reduce the environmental impact~\cite{envrep2020}, and is actively researching alternative gas mixtures~\cite{Mandelli:2022qjc}. Future detector projects still plan to use RPCs (DUNE covering an area of about 860 m$^2$~\cite{DUNE:2016rla} and SHIP with about 100 m$^2$~\cite{SHiP:2021nfo}) but are testing their prototypes also with alternative gas mixtures~\cite{Albanese_2023}.

\begin{table}
\begin{tabular}{@{}lllc@{}}\toprule
Name& Chemical\,\,\; & 
Lifetime\,\,

& Global warming potential (GWP) \\ 
&  Formula& 

[years]
& [100-yr time horizon]\\ 
\midrule
Carbon dioxide &	CO$_2$ & -- & 1 \\
Dimethylether& CH$_3$OCH$_3$ & 0.015 & 1 \\ 
Methane & CH$_4$& 12& 25\\
Sulphur hexafluoride& SF$_6$ & 3,200 & 22,800\\
\midrule
\multicolumn{4}{c}{Hydrofluorocarbons (HFCs)}\\
\midrule
HFC-23 & CHF$_3$ &  270 & 14,800\\
HFC-134a & C$_2$H$_2$F$_4$ & 14 & 1,430\\
\midrule
\multicolumn{4}{c}{Perfluorocarbons (PFCs)}\\
\midrule
PFC-14 & CF$_4$ & 50,000 & 7,390 \\
PFC-116& C$_2$F$_6$ & 10,000  & 12,200\\
PFC-218&C$_3$F$_8$ & 2,600 & 8,830 \\
PFC-3-1-10 &C$_4$F$_{10}$ & 2,600 &8,860\\
PFC-5-1-14 &C$_6$F$_{14}$ & 3,200 & 9,300\\
\bottomrule
\end{tabular}
\caption[Warming potential of selected GHGs]{Environmental impact associated with GHGs, adapted with permission from Ref.~\cite{IPCC17technical}, and the EU regulations described in Ref.~\cite{EUFgasregulation}, and source data for calculations in the CERN environmental report~\cite{CERN:2723123}, .}\label{tab:GasImpact}
\end{table}

The gases responsible for about 80\% of CERN's Scope 1 direct annual GHG emissions are perfluorocarbons (PFC), hydrofluorocarbons (HFC) and sulphur hexafluoride (SF$_6$) in particle detection, and HFCs and PFCs for detector cooling. To put the emissions into context, CERN's PFC emissions are roughly of the same size as the Swiss emissions~\cite{PFCdata} and only reduce by about 30\% when there is no LHC run. For 2017 and 2018 (during LHC data taking), CERN's SF$_6$ emissions are about 5\% of Switzerland's, and of the same size as those of Luxembourg or Latvia~\cite{SF6data}. The HFC emissions are 6\% of the Swiss emissions, about twice the size of Luxembourg's and nearly half of Latvia's emissions, again looking at 2017--2018 data~\cite{HFCdata}. During 2020--2021, when the LHC, and more importantly its experiments, shutdown for upgrades and maintenance, SF$_6$ emissions were down to about a third, while HFC emissions were down to 25\%.  
All of these so-called F-gases have EU supply restrictions imposed on them since 2015~\cite{EUFgasregulation}, effectively phasing out their usage. A 2022 regulation proposal aims to extend EU regulations, \eg to reduce HFC usage down to 2.4\% by 2048 compared to 2015 levels~\cite{EUFgasregulation_new}. An additional significant factor in this proposal is the removal of certain sector exemptions for HFC usage, such as research. All of this could significantly impact the availability and cost of F-gases in the future and therefore affect the \ACR\ community, which should be reflected in the plans for ongoing and future experiments. Even independent of these EU regulations, the \ACR\ community should work with highest priority on abolishing problematic GHGs in existing and future detectors. Ideally, this would consist of replacing them with non-GHGs, or gases with low GWP, or gasless detector technology. 

For cooling of the LHC-experiments, concrete plans are in place to upgrade the future detectors (Phase-II Upgrades) to CO$_2$ cooling, reducing the total Scope-1 carbon footprint of the experiments during HL-LHC exploitation significantly~\cite{Barroca:2021hal,Zwalinski:2023ira}.

Solutions to the problem of gas use for particle detection are less straightforward. In some scenarios, low-GWP gases used as replacements for industrial applications are not suitable for detector applications and, as such, studies for alternative gases are currently ongoing. The difficulty in finding replacement gases originates from having to satisfy several factors: safety (non-flammable and low toxicity) and environmental impact (minimising GWP), while maintaining their detector performance (including preventing the ageing of the detectors, ensuring good quenching\footnote{Quenching is the prevention of secondary electron avalanche (and thus signals) caused by, \eg photon emissions of the positive ions in the gas when recombining with electrons. As the positive ions travel more slowly through the gas detectors, these secondary signals happen after the primary signal from the electrons are registered and cause significant dead-times, when the detector is not responsive to new signals.} and being radiation-hard)~\cite{Bloom:2022gux}.

Current gas mixture alternatives for particle detection are centred around tetrafluoropropene (chemical formula C$_3$H$_2$F$_4$ and industrially referred to as HFO-1234ze/R-1234ze). The mixture has zero ozone-depletion potential and a global warming potential below 1, over a span of 100 years. Simulations of RPCs that operate with various gas mixtures, which include R-1234ze, have shown encouraging results, although further studies are still required~\cite{Fan:2022azj_ref3,Proto:2021taf_ref4}.

Long term, it is crucial to design future detector systems with gas GWP in mind. Consequently, it is essential that current state-of-the-art and future detectors are compatible with this and, if not, R\&D is aimed at reducing the GHG emissions of such systems.

\begin{casestudy}[LHCb and sustainability\label{case:LHCb}\\\noindent {\footnotesize Edited from Framework TDR for the LHCb Upgrade II~\cite{LHCbU2FTDR}: Opportunities in flavour physics, and beyond, in the HL-LHC era, edited extracts from the U2 FDTR chapter on environmental impacts of the project, as contributed by Chris Parkes.}]{LHCb and sustainability}%
\noindent In a world with increasing demand on limited resources and undergoing climate change, the LHCb collaboration feels a responsibility to consider energy consumption, sustainability and efficiency when discussing our scientific proposals. To this end the Framework Technical Design Report of the next-generation LHCb Upgrade II experiment~\cite{LHCbU2FTDR} has included a dedicated chapter on these considerations analysing the current Upgrade I system and indicating directions for future investigation. This section reports some of the main elements.  

The 2020 update of the European Strategy for Particle Physics~\cite{EuropeanStrategy2020} reports: ``The environmental impact of particle physics activities should continue to be carefully studied and minimised. A detailed plan for the minimisation of environmental impact and for the saving and re-use of energy should be part of the approval process for any major project. Alternatives to travel should be explored and encouraged.'' 
As one of the major experimental infrastructures operating at the LHC,  the LHCb environmental protection strategy should be made in coordination with CERN guidelines, as described in the first CERN environment report~\cite{envrep2020}.

CERN has a formal objective to reduce direct emissions (``Scope 1'') by $28\%$ by the end of 2024. These are dominated by the activities of the LHC experiments, and in particular by the use of fluorinated gases for particle detection and detector cooling purposes, as shown in \fref{fig:cern_co2}. These emissions have to be carefully considered in the operation of the Upgrade I detector and in the design of its future upgrade. Other relevant aspects of the environmental impact of our project are the power consumption of the experimental infrastructure (indirect emissions, ``Scope 2''), the impact of digital technologies, and travel of the members of the collaboration. \fref{fig:lhcb_co2} shows the relative contribution of each of these sources to the CO$_2$ equivalent footprint of the experiment operations expected during Run 3. 

\begin{center}
    \captionsetup{type=figure}
    {\includegraphics[width=.8\textwidth]{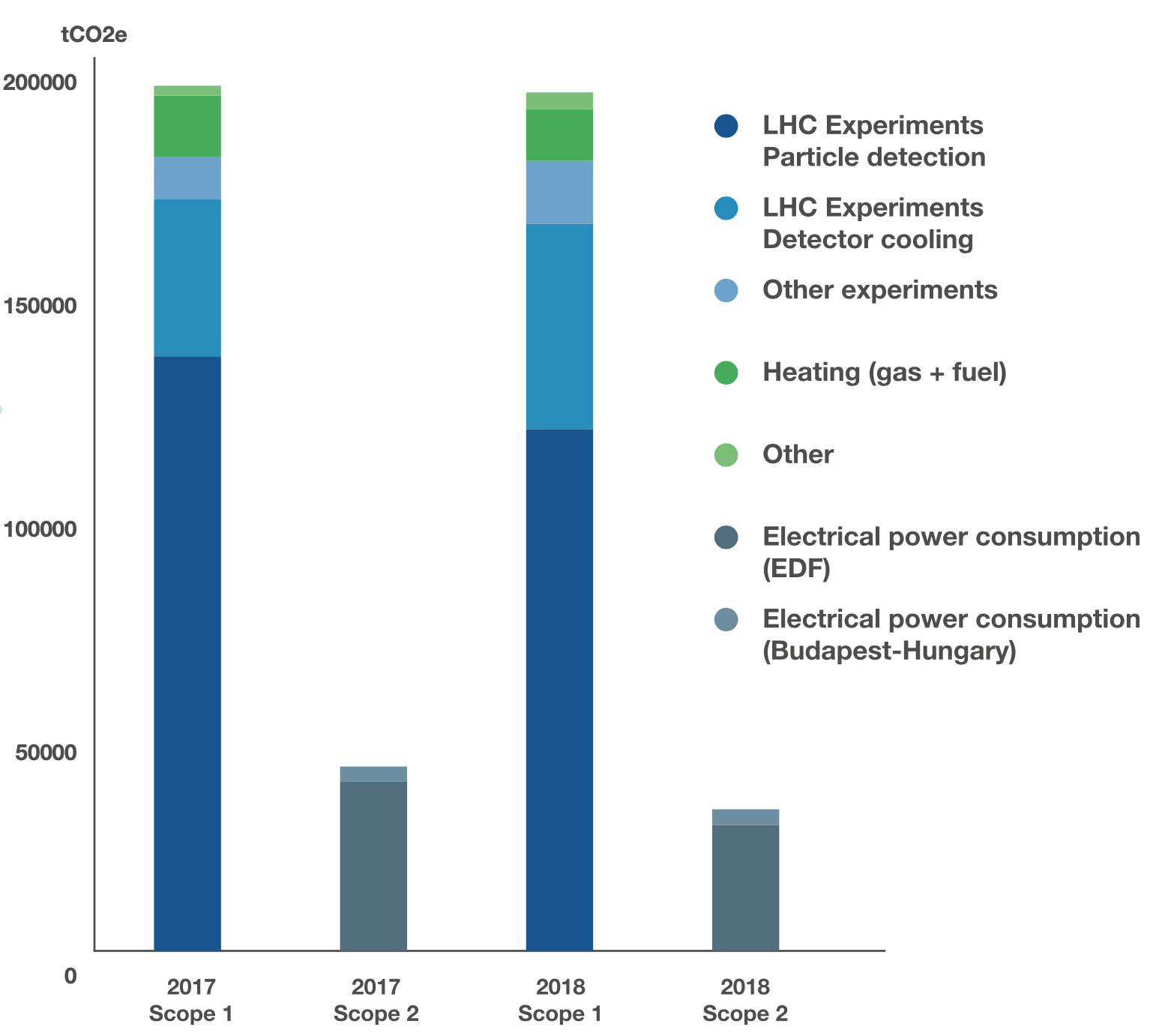}}
    \caption[CERN Scope 1 and Scope 2 emissions for 2017 and 2018]{CERN Scope 1 (direct) and Scope 2 (indirect, by electricity consumption) emissions for 2017 and 2018, in CO$_2$ equivalent tonnes, by category; `Other' includes air conditioning, emergency generators and CERN vehicle fleet fuel consumption.  `Budapest-Hungary' refers to electricity use at the (now inactive) Wigner data centre in Hungary. Figure reused from Ref.~\cite{envrep2020} under the terms of the \href{https://creativecommons.org/licenses/by/4.0/}{Creative Commons Attribution 4.0 International (CC BY 4.0) license}.\label{fig:cern_co2}}
\end{center}

\begin{center}
    \captionsetup{type=figure}
    {\includegraphics[width=.8\textwidth,trim=0.05cm 0.2cm 0.05cm 0.2cm , clip=true]{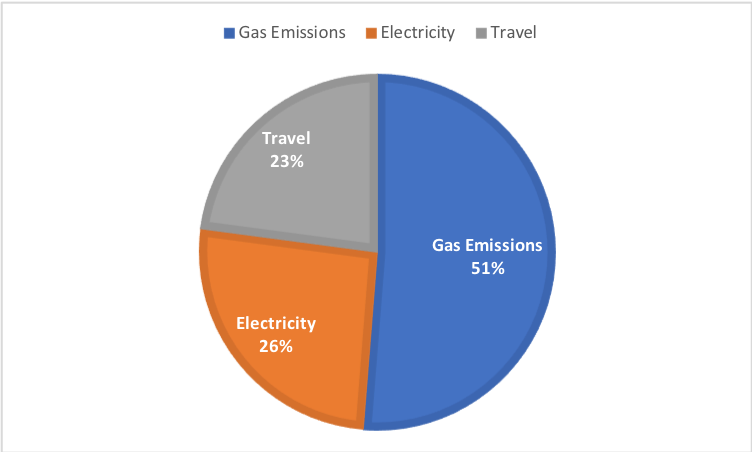}}
    \caption[Expected relative emissions from LHCb operations in Run 3]{Expected relative contribution to CO$_2$ equivalent emissions from LHCb operations in Run 3. The total emissions are estimated to be 4,400 tonnes CO$_2$ equivalent per annum. Figures and captions reproduced with permission from Ref.~\cite{LHCbU2FTDR}.}\label{fig:lhcb_co2}
\end{center}

\paragraph{Direct emissions}

Direct emissions from LHCb are dominated by losses of gases with sizeable global warming potential (GWP). The GWP is the heat absorbed by a GHG in the atmosphere as a multiple of the same mass of CO$_2$.  It thus allows conversion into CO$_2$ equivalent emissions. As some gases break down, they have time-dependent values, and we use the 100 year value~\cite{AR5}. The gases are utilised in LHCb in detector cooling systems and in the detection systems. Improvements made in the cooling systems of LHCb mean that emissions are now dominated by the detection system in Upgrade I. All systems are closed, with emissions being the result of losses.

In the original LHCb detector of Run 1 and 2, the gas C$_6$F$_{14}$ (GWP 7,910) was used in cooling plants. For upgrade I, ``Novec 649" (GWP 1) is planned to be used, along with 
increased use of low-impact CO$_2$ based cooling.
For Upgrade II, lower operating temperatures are foreseen, and the GWP of the cooling systems will be considered.

In the detector systems, the Ring Imaging Cherenkov Systems (RICH1 and 2) and Muon systems of Upgrade I use GHGs. The RICH2 system currently uses CF$_4$ (GWP 6,630) and RICH1 C$_4$F$_{10}$ (GWP 9,200) radiators. R\&D will be pursued for Upgrade II on alternative gases, RICH2 is looking at CO$_2$ use, where a test has already been performed, and leakless systems. Significant effort has been made to minimise leaks.  
In the original LHCb detector, GEM detectors (gas electron multiplier detectors) were utilised in a part of the muon system. The removal of these for Upgrade I reduces the detector system emissions by 40\%. 
Recirculating systems are used throughout. The study of alternative gas mixture will be conducted to reduce the CF$_4$ consumption in the proposed future muon systems.

The CO$_2$ equivalent emissions expected in Run~3 are shown in \fref{fig:LHCbCO2e}. These are taken from the average values of annual usage during Run 2 for the detector systems that are still present, or that have been replaced with similar systems for Run~3. 

\begin{center}
    \captionsetup{type=figure}
    {\includegraphics[width=.8\textwidth,trim=0.05cm 0.2cm 0.05cm 0.2cm , clip=true]{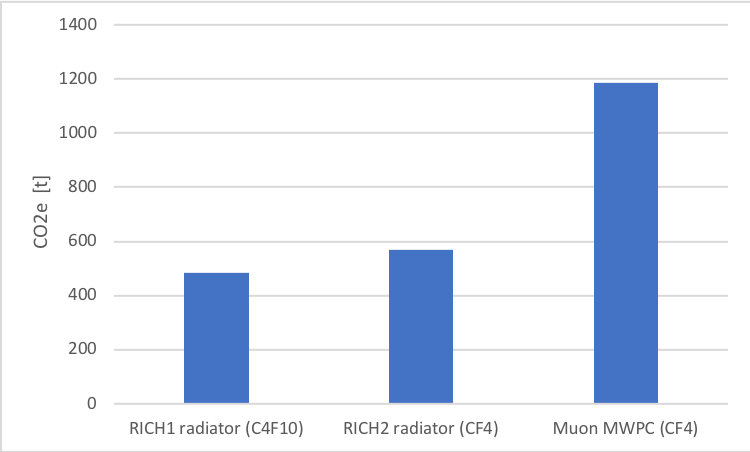}}
    \caption[Expected Scope 1 emissions from LHCb detector gas systems in Run 3]{Expected Scope 1 (direct)  emissions in CO$_2$ equivalent tonnes from the LHCb detector gas systems in Run 3. The data is taken from the annual emissions of the systems, or predecessors, during Run 2. Figures and captions reproduced with permission from Ref.~\cite{LHCbU2FTDR}.}\label{fig:LHCbCO2e}
\end{center}

\paragraph{Power consumption}
\label{sec:powerconsumption}
CERN peak power demand, with the full accelerator chain running, is about $180$~MW, which brings the total annual energy consumption to $1.2$~TWh. This very large energy demand is partially mitigated by the fact that the electricity procurement is mainly from France, whose production capacity is 87.9$\%$ carbon-free (2017-figures). This keeps the contribution from the electrical power to the total CERN emission budget below 20$\%$, as shown in \fref{fig:cern_co2}.\footnote{It was argued in \sref{sec:Energy} that French energy production is part of the common EU~market and that it would therefore be more appropriate to use a conversion factor for an EU mix, which is about a factor of five higher than that for France~\cite{EUmix}.} Nevertheless, guided by the Energy Management Panel, EMP, CERN is spending a large effort to improve energy efficiency, with special focus on the accelerator sector. As an example, in the transition to the \acrshort{hl-lhc}, with a tenfold increase in luminosity, the organisation’s immediate priority is to limit the increase in energy consumption to 5$\%$ up to the end of 2024.
 
LHCb during a normal data taking period of Run 2 had a peak power demand stably around 5.5~MW, of which 4.6~MW was from the experiment dipole magnet, and the rest was from the detector electronics and the online computing farm.  The Run 3 expectation is for an increase of $\sim$1.5~MW due to the increased demand of data processing power, which is to be compared to the five-fold increase in luminosity.
For Run 5 and beyond, the contribution of online computing is expected to increase substantially, as a consequence of a further order of magnitude increase in the data throughput.

For the power dissipated by the LHCb magnet, an important mitigation has been implemented very recently by CERN with the installation of a heat-recovery plant at the experimental site. This is intended to use the
hot water produced by the magnet and the machine cooling systems to heat a new residential area in the
town of Ferney-Voltaire next to the LHCb site. Thanks to this project, up to 8,000 people’s homes will be heated at a lower cost and with reduced CO$_2$ emissions, corresponding to $\sim$2.5\% of the total CERN emission budget per year. 

\paragraph{Digital technologies}
The power consumption of the online computing farm at LHCb has been about 530~kW on average during Run~2 of the LHC. To cope with the significantly increased computing needs after Upgrade~I, a new data centre has recently been installed at Point~8 and the power consumption for computing is going to increase to 2,000~kW for the upcoming data taking periods. 
The new computing data centre at Point~8 is located in a surface building and for practical reasons could not be included in the heat recovery project discussed above. However, great care has been put into the design to optimise its power efficiency, for example by implementing a state-of-the-art indirect free air cooling system with adiabatic assist~\cite{lbldcreport}.
A \acrshort{pue} of better than 1.08 has been achieved for the new data centre at Point~8, a value that compares favourably with other large computing centres~\cite{pue-data}. 

While it does not seem feasible to further improve the PUE of the data centre, energy savings could potentially be achieved by adjusting the operating mode to the actual computing needs at a given point in time.  Significant improvements in energy efficiency can be achieved by rewriting software so that it can efficiently exploit today's highly parallel computing architectures. LHCb has been doing this in preparation for Run~3 data taking, and the impact of these activities on the energy efficiency of our software has been documented in Ref.~\cite{hlt1-energy-eff}. In total, the energy efficiency of HLT1 (High Level Trigger 1) software has been improved by a factor $4.8$ on CPUs, with the improvements coming in roughly equal parts from physics optimisations and the rewrite of the underlying software framework. A further improvement in energy efficiency  can also be achieved by porting suitable algorithms from CPUs to more efficient technologies such as GPUs, field programmable gate arrays (FPGAs) or even custom-made application specific integration circuits (ASICs). LHCb has demonstrated this with the Allen project~\cite{Allen}, which implemented HLT1 on GPUs, leading to an overall improvement in energy efficiency of up to $19$ times compared to the Run~2 architecture. These improvements require significant effort and investment, above all in the training and retention of scientists able to effectively program across a range of modern computing architectures. 

The energy efficiency of the underlying computer hardware has also improved substantially over time. For example, the
AMD~7502~\cite{AMD7502} CPUs, which were evaluated as candidates for LHCb's Run~3 HLT, are $2.6$ times more energy efficient than 
the benchmark E5-2630~Xeon CPUs used by LHCb during Run~2. 
Within a given computing architecture, energy savings can also be achieved by purchasing more expensive, higher quality hardware. 
As an example from the world of CPU, the more expensive AMD~7742~\cite{AMD7742}
provides twice the number of CPU cores and threads as the cheaper AMD~7502~\cite{AMD7502},
while its specified power consumption is only 25\% higher. The energy consumption and carbon footprint from data transfer, data storage and offline computing are much harder to assess than those for online computing, due to the distributed nature of the computing model with data centres and users distributed over many different countries. The GRAND collaboration has performed pioneering work in this direction~\cite{Aujoux_2021}.

\paragraph{Mobility}
As an international collaboration operating in an international field of research, travel is an intrinsic part of how LHCb operates. We have estimated the environmental impact of travel in order to attend LHCb collaboration meetings and international conferences. We have not taken into account local commuter travel or travel related to on-site work at LHCb, such as shifts, although the latter is probably significant.

The impact of travel per participant for a typical LHCb collaboration week, pre-pandemic, corresponded to around 0.5 \acrshort{tco2e} with the average LHCb week in 2019 leading to travel-emissions of $\sim$ 180$\,\mathrm{tCO_2e}$. The Speakers' Bureau database provides a complete record of all LHCb conference talks, allowing us to estimate the environmental impact in terms of $\mathrm{tCO_2e}$ per year. LHCb weeks and conference travel contribute a total of approximately $1,000\,\mathrm{tCO_2e}$ per annum, a similar carbon footprint to the Run~3 experiment's projected electricity use due to online computing and the magnet (French energy mix). LHCb weeks contribute about three times as much to  LHCb's carbon footprint as conference travel.
 The carbon footprint of virtual conference attendance is calculated according to the life cycle and operating costs of endpoint devices estimates in Ref.~\cite{VideoConCO2}, and is small.

LHCb, in common with other HEP collaborations, had extensive experience with virtual meetings before COVID, and videoconferencing
technology has already helped to reduce travel-related emissions over the past decade.
However, the pandemic, as well as recent improvements to the videoconferencing software infrastructure, have shown us ways 
in which the organisation of virtual meetings can be improved and made more inclusive. 
At the same time, the pandemic has served as a reminder of the ongoing importance of in-person interaction,
not least to avoid fracturing the collaboration between those who can regularly travel to CERN in eco-friendly ways and those who cannot.
The collaboration has only just started to navigate this tension but is actively exploring ways to reduce its 
travel-related environmental impact.

\end{casestudy}

\RaggedRight
\sloppy
\newpage


\section{Resources and Waste}
\label{sec:Waste}


\begin{center}
\includegraphics[width=\SDGsize]{Sections/Figs/Common/SDG_3_GoodHealth.png}~
\includegraphics[width=\SDGsize]{Sections/Figs/Common/SDG_6_CleanWater.png}~
\includegraphics[width=\SDGsize]{Sections/Figs/Common/SDG_8_EconomicGrowth.png}~
\includegraphics[width=\SDGsize]{Sections/Figs/Common/SDG_11_SustainableCities.png}~
\includegraphics[width=\SDGsize]{Sections/Figs/Common/SDG_12_ResponsibleConsumption.png}~
\includegraphics[width=\SDGsize]{Sections/Figs/Common/SDG_13_ClimateAction.png}~
\includegraphics[width=\SDGsize]{Sections/Figs/Common/SDG_14_LifeBelowWater.png}~
\includegraphics[width=\SDGsize]{Sections/Figs/Common/SDG_15_LifeOnLand.png}
\end{center}


\exSum

\noindent Half of the world's \acrshort{ghg} emissions, and over 90\% of global water stress and biodiversity loss events, are due to the extraction and processing of raw materials~\cite{EURaw}. Although most extracted materials are slated for the energy or agriculture sectors, the small fraction associated with consumption of goods and services is responsible for 18\% of EU emissions~\cite{EURaw}. Mitigating the climate impacts of the extraction, processing and trade of raw materials is a priority for the resilience of the EU~\cite{EURaw}, and it should also be a priority for the world environmental agenda.

The generation of waste is a direct consequence of material consumption, and is aggravated by constraints in production, distribution, usage and repair, and disposal or recycling of consumables. Waste has severe impacts on life on land and at sea, often destabilizing local ecosystems.  It also damages the global ecosystem by contributing to climate change. Accumulations and inefficient disposal of waste products can result in pollution of ground water and air, thus directly affecting the health of individuals and communities at a large cost to society in terms of disease burden and lives lost. 

In an attempt to curb the footprint of waste generation, the concept of a circular economy has been  proposed~\cite{CircularEconomy}.  Any such proposal must be established in parallel with a will to reduce waste at source through sustainable procurement, repair and reuse, and used only as a transitional measure. Even a fully circular economy has some dissipation, and signatures of this energy waste need to be independently addressed and reduced~\cite{Forbes,SocialEurope,WRI}.

Procurement accounts for almost two-thirds of annual emissions at CERN \cite{CERNTownHall}, with a GHG footprint of the same order as its direct emissions in 2018, when the LHC was running \cite{Environment:2737239}. Although not yet fully included in reporting by other \ACR\ institutions, the environmental cost of procurement is likely proportionately large elsewhere.  Maximising the sustainability of the use cycle of resources should be a priority of the \ACR\ community. 

This section covers sustainable sourcing in \sref{subsec:Resources}, and reduction and treatment of waste, including E-waste, in \sref{subsec:Waste}. The use of materials in research infrastructure is also discussed in Section~\ref{sec:Technology}.


\clearpage
\begin{reco2}{\currentname}
{
\begin{itemize}[leftmargin=3.5 mm]
\setlength{\itemsep}{0.95\recskip}
\item Limit purchases and consider environmental credentials such as repairability and recyclability of products in purchasing decisions.

\item Service appliances regularly; share, repair, reuse and refurbish to minimise waste; sort and recycle.

\item Read the sections on computing (\sref{sec:Computing}), energy (\sref{sec:Energy}), food (\sref{sec:Food}), and research infrastructure and technology (\sref{sec:Technology}).
\end{itemize}
}
{\begin{itemize}[leftmargin=3.5 mm]
\setlength{\itemsep}{0.95\recskip}
\item Adopt life cycle assessments and associated tools to assess environmental impact of all projects, including infrastructure.

\item Institute sustainable purchasing, usage and end-of-life policies in the management of group consumables,  office supplies and single-use plastics, \eg in food service or conference events (see also Section~\ref{sec:CateringTableware} and Best Practice~\ref{bp:PlasticFreeConf}).
\end{itemize}
}
{
\begin{itemize}[leftmargin=3.5 mm]
\setlength{\itemsep}{0.95\recskip}
\item Prioritise suppliers instituting sustainable sourcing and operating policies, with a particular focus on the raw materials processing stage (see Best Practice~\ref{bp:SustainableRawMaterials}) and with the aim of creating demand for recycled (secondary) raw materials.

\item Provide an institutional pool of infrequently used equipment to avoid redundancy in purchasing.

\item Proceduralise and prioritise repair of equipment, and enable through provision of tools and know-how.

\item Assess resource use, waste generation and management for the design, operation and decommissioning of IT and infrastructure projects by right-sizing needs, establishing specific treatment channels for all waste categories, and setting recycling targets that include the recycling of all construction waste, see, \eg \bpref{bp:HERAShieldingBlocks}.

\end{itemize}
}
\end{reco2}

\subsection{Resources}
\label{subsec:Resources}

\ACR\ research can be resource-intensive, particularly in the building and maintenance of the often large experiments that drive progress in our fields.  These resources have an environmental impact over their entire life cycle, due to extraction of the raw materials used in their manufacture, their production and use, and their disposal once they become unusable or obsolete. Of these, the raw materials processing stage has been highlighted as having the greatest potential for emissions reduction (see, \eg Figure 2 of Ref.~\cite{EURaw}).  

The extraction of raw materials has important and extensive environmental costs~\cite{Dolega:2016}, mostly associated with the mining industry.  Acid mine drainage is the overriding problem and is a serious threat to water resources.  It results from water flow over ore creating sulphuric acid and leaching heavy metals from surrounding rock, thus contaminating groundwater and soil.  Mining operations can also deplete water resources, particularly in regions of limited water supply, severely restricting the availability of water to local consumers.  Fine particles and dust produced during mining operations and dispersed by winds affect air quality, and mining and its infrastructure leads to loss of agricultural land and even entire ecosystems through contamination or destruction of soil cover. Mining is the world's largest producer of waste, with copper, zinc, bauxite and nickel mining generating the largest ratios of waste to mined metal.  Disposal or storage of tailings, the waste products remaining after the extraction of valuable material from ore, is a major problem. These can be radioactive, and are sometimes illegally disposed of directly into rivers or seas.  Even when stored `responsibly' in tailings dams, incorrect geological siting of these dams, in tectonically active zones or regions of high rainfall, can lead to catastrophic loss of life and usable land~\cite{SILVAROTTA2020102119}.

Environmental sustainability aside, mining has a poor safety and human rights record~\cite{ResponsibleMiningIndex}, and is sometimes subject to dubious financing~\cite{MiningandMoneyLaundering}.  Mining of `conflict minerals', such as tin, tungsten, tantalum and gold, used in mobile phones and other everyday products, are sometimes used to finance armed conflict~\cite{EUConflictMinerals}.  An analysis of components used inside a smartphone and their impacts can be found in~Ref.~\cite{fairphone} by smartphone manufacturer Fairphone. Fairphone claimed to sustainably source 56\% of 8 of the materials used in its phones in 2020, and set a target of fair sourcing of 70\% of 14 materials by 2023~\cite{fairphone2}.  However, their phones are sold at a premium, making them unaffordable for many.

A ranked list of mined metals by overall environmental impact can be found in Table~\ref{tab:MetalImpact}.  This table was taken from the EU Raw Materials Information System~\cite{EURMIS}, with source data from Ref.~\cite{UNEP2010}.  Sourcing recycled metal from scrap produces significantly lower emissions.  Secondary aluminium, for example, was reported by the European Aluminium Association to reduce GHG emissions by 95\% compared to primary production~\cite{EURaw}. Other materials used in \ACR\ experiments (\eg cobalt for magnets, rare earths for permanent magnets, niobium) are produced under very difficult conditions, with a high environmental or societal cost~\cite{FARJANA2019150, EURare, ALVES2019275}.  Formal discussions of their use and impact have already begun in the \ACR\ community, most recently at a workshop on Rare Earth Elements organised by I.FAST at DESY~\cite{DESYRareEarth}.

Sustainability regulations, both externally imposed and voluntary, are slowly being incorporated into the raw materials supply chains (see, \eg the Voluntary Principles on Security and Human Rights~\cite{VoluntaryPrinciples}), albeit slowly, and in an inconsistent and sometimes superficial manner~\cite{ResponsibleMiningIndex,ResponsibleMiningFoundation}.  For examples of sustainability initiatives, in particular in relation to raw materials supply chains, see~\bpref{bp:SustainableRawMaterials}.

\bigskip
{\centering
\ra{1.1}
\captionsetup{type=table}
\begin{tabular}{@{}lll@{}}\toprule
Ranking& Impact per kg & 
Impact global production \\ 
\midrule
1 & Palladium & Iron \\
2 & Rhodium & Chromium \\
3 & Platinum & Aluminium \\
4 & Gold & Nickel \\
5 & Mercury & Copper \\
6 & Uranium & Palladium \\
7 & Silver & Gold \\
8& Indium & Zinc \\
9 & Gallium & Uranium \\
10 & Nickel & Silicon \\

\bottomrule
\end{tabular}
\caption[Environmental impact associated with primary metals]{Environmental impact associated with primary metals, ranked by impact per kg, and total impact due to global production.  Table excerpted from Ref.~\cite{UNEP2010}.}
\label{tab:MetalImpact}
}

\subsubsection{Life cycle assessment}
\label{subsec:lca}

Best practices in sustainable use and disposal of resources begins with a life cycle assessment (LCA): a cradle-to-grave accounting of all the environmental impacts of a resource.  As an example the ISO 14040 ~\cite{ISO14040} and ISO 14044~\cite{ISO14044} standards provide a systematic procedure for the analysis. Depending on the goal and scope of the analysis, the life cycle inventory comprises the quantification of all input and output flows. This includes raw materials, consumables, energy, products, waste, emissions, and groundwater and soil contamination. There are online tools and auditing agencies who provide help with the analysis, see, \eg Ref.~\cite{ProBas}.  For an LCA of a silicon wafer used in particle detectors, see \bpref{bp:SiWafer}.

\subsubsection{Sustainable sourcing \label{sec:sustainablesourcing}}

Purchasing policy can have a major impact on the environmental costs of procurement. 
The \ACR\ community should prioritise suppliers that implement sustainable thinking, sourcing, and operation. This could include voluntary provision of life cycle assessments for their products (see above), certification of, \eg proof of origin, and demonstration of a commitment to eliminating premature (or planned) obsolescence (now prohibited under French law~\cite{ObsolescenceBriefing}).
Sustainability requirements on suppliers could also be incorporated into tenders and purchasing regulations, allowing these considerations to be weighed in tandem with cost in the tendering process.  Since much of \ACR\ funding is public, purchasing regulations, which are influenced by funding agencies, an additional important stakeholder in this process, must be reassessed.  For examples of best practice in sustainable procurement, see~\bpref{bp:SustainableRawMaterials}.  A strategic approach to sustainable purchasing has been outlined in ISO 20400~\cite{ISO20400}.  

CERN is in the process of defining a new environmentally responsible procurement policy in implementation since 2023~\cite{CERNTownHall, CERNProcurementPolicy2023}.  Key measures being considered include requiring sustainability certification from suppliers, with a focus on those with highest potential to drive sustainability issues~\cite{CERNTownHall}.  For further sustainable procurement and waste policies being explored by CERN, see~\bpref{bp:CERNSustainableProcurement}.  

\begin{bestpractice}[\label{bp:SustainableRawMaterials}Sustainability in raw materials supply chains\\{\noindent\footnotesize Edited contribution from Enrico Cennini, IPT (Industry, Procurement and Knowledge Transfer), CERN, summarised from Ref.~\cite{EURaw}.}]{Sustainability in raw materials supply chains}%
\noindent Sustainable procurement requires sustainability in all phases of the supply chain, from producers to processors and traders.  Hallmarks of sustainability among suppliers include the following (with the relevant phase(s) shown in parentheses):
\begin{itemize}
    \item Compliance with sustainability standards set and certified by non-profit, multi-stakeholder organisations. (Producers, processors, traders) \\ 
    {\small \eg Aluminium Stewardship Initiative, Responsible Steel certification, Responsible Jewelry Council (precious metals, stones).}
    \item Voluntary implementation of certified energy- and environmental-management systems, such as ISO 50001 and ISO 14001. (Producers, processors, traders)  
    \item Setting of voluntary sustainability targets by sectoral industry associations. (Producers, processors, traders) \\
    {\small \eg steel-making industry and ammonia producers investigating low-carbon sources of hydrogen for imminent adoption.}
    \item Demonstrable investment in technological solutions for improving energy efficiency of operations, such as employing `Best Available Techniques' (BATs) and keeping to `Associated Environmental Performance Levels' (BAT-AEPLs). (Producers, processors, traders) \\ 
    {\small \eg transition to lower carbon sources for power supply, implementation of `ventilation-on-demand' (gold and potash mining); replacing diesel fleet with hybrid electric, innovative and energy efficient loading and haulage systems (mining); continuous monitoring and improvement of transportation methods.}
    \item Requiring third-party verified sustainability certificates from upstream suppliers. (Processors, traders) \\
    {\small \eg some aluminium manufacturers; selected wood and copper processors' business partner code of conduct encodes supplier sustainability requirements.}
\end{itemize}
\end{bestpractice}

\begin{bestpractice}[\label{bp:CERNSustainableProcurement}CERN sustainable procurement and waste policy\\{\noindent\footnotesize Edited contribution from Quentin Salvi, Waste Management Project Coordinator, CERN.}]{CERN sustainable procurement and waste policy}%
\noindent Avenues being actively explored by CERN to improve sustainability in procurement and waste include the following:
\begin{itemize}
    \item Implementing a circular economy, both internally and externally.      
\item Knowledge-sharing with partners and the local authority. 
\item Consolidation of CERN equipment, 
behavioural change in industrial practices and management. 
\item Optimisation of services based on CERN data. 
\item User-awareness.
\end{itemize}
\end{bestpractice}

\subsection{Waste}
\label{subsec:Waste}
Around 3\% of global GHG emissions is due to solid waste disposal, the organic component of which decomposes in wastewater and landfills, producing methane and nitrous oxide~\cite{owidco2andothergreenhousegasemissions}.  This is despite a 60\% decrease in the amount of waste landfilled in the EU in the past two decades, due partly to an increased legislative focus on alternative treatment methods, such as recycling and composting, as well as more widespread landfill gas recovery~\cite{Eurostat}.

\begin{figure}
    \centering
    \includegraphics[width=0.9\textwidth]{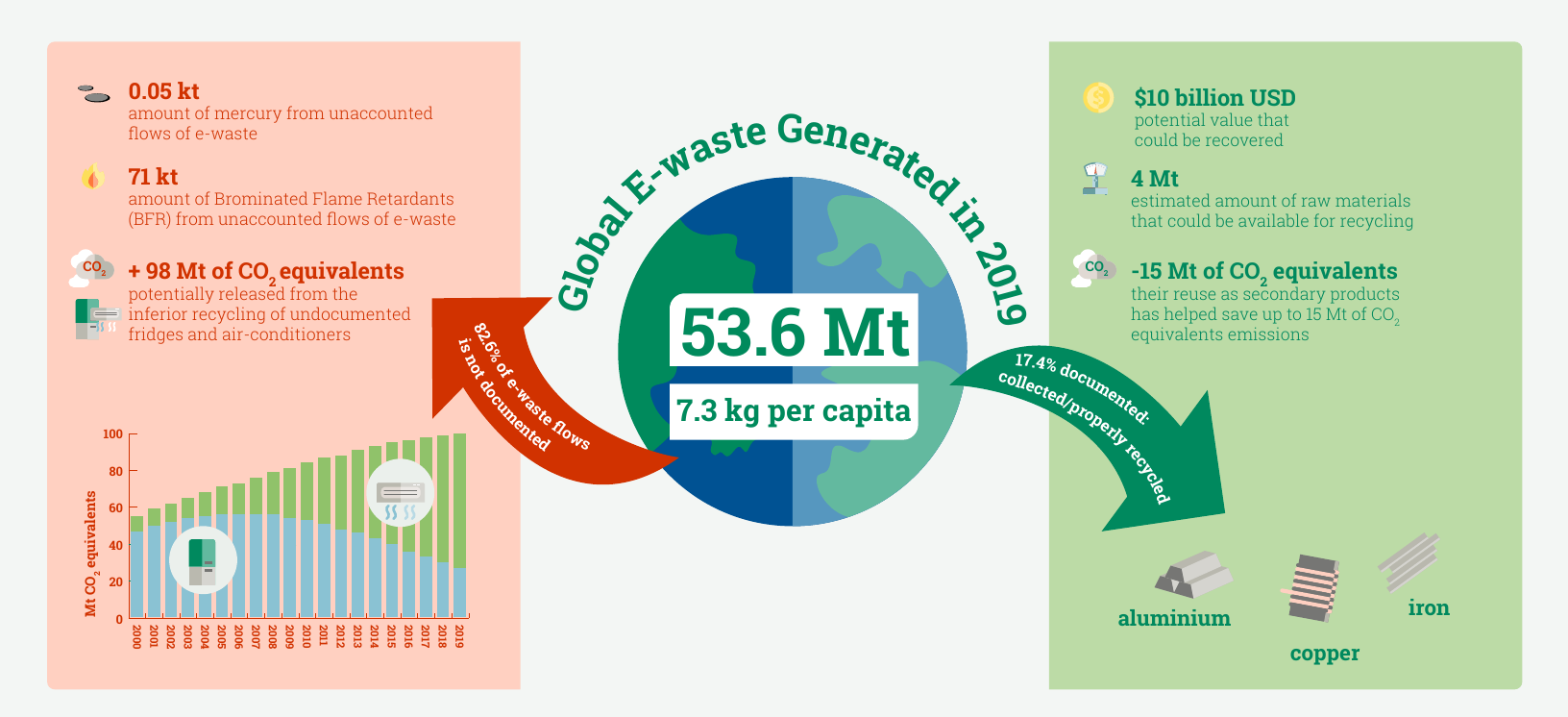}
    \caption[Global E-waste generated in 2019]{Global E-waste flows in 2019~\cite{Forti2020}. In 2022, the total mass of generated E-waste grew to 62 billion kg, of which approximately 80\% was managed outside of formal waste management channels. E-waste generation is growing 5 times faster than than formal collection and environmentally sound recycling schemes~\cite{Balde2022}. Reproduced, with permission, from Ref.~\cite{Forti2020}.\label{fig:ewaste}}
\end{figure}

The fastest-growing portion of EU waste output is E-waste~\cite{EUNews}: powered products with electrical components that are discarded into the waste stream. In addition to releasing hazardous
chemicals into the environment, improperly treated E-waste contributes to global warming through failure to recuperate valuable mined materials, and direct release of GHGs including refrigerants (see \fref{fig:ewaste} for statistics on global E-waste generation and disposal in 2019). Moreover, E-waste can often be shipped illegally to developing countries without infrastructure for safe recycling~\cite{Forti2020, OECDLib}. Exploding demand for digital devices, fuelled by their difficulty to repair and premature obsolescence, has also given rise to a boom in the mining industry.  This boosts the economy in resource-rich countries, where working conditions are usually unsafe and unpleasant, and causes deforestation and pollution~\cite{OECDLib}.

Increasing the life-span of consumer electronics through more comprehensive right-to-repair 
legislation is an integral part of the EU's strategy for a circular economy~\cite{EUNews,EUR2RSummary}.  According to a 2018 European Commission study, however, the cost of repair was the most frequent reason consumers chose to replace four common household electrical and electronic goods~\cite{ECConsumerStudy}. 
Right-to-repair advocates are pushing for more comprehensive and far-reaching legislation, including policies that will empower consumers to make simple repairs themselves, as well as government incentives to repair through, \eg financial aid~\cite{R2RPolicy}. A trial scheme for the latter has already been implemented in France~\cite{FrenchRepairAid}.  Unfortunately, today's legislation is formulated piecemeal and is slow to take effect.  In addition to instituting sustainable purchasing, use and de-inventorising/disposal policies for electrical and electronic goods (see Recommendations), \ACR\ institutions could encourage the sustainable use of personal electronic devices by making repair equipment freely available, with guidance from experts on a volunteer basis.

\begin{bestpractice}[Re-purposing shielding blocks\label{bp:HERAShieldingBlocks}]{Re-purposing shielding blocks}%
\noindent Initially, the heavy concrete blocks in the halls of the now-decommissioned HERA experiment had served
as protection against radiation. Five hundred of these discarded shielding blocks were stored, unused,  for years on the DESY campus in Hamburg, Germany. 6,000 tonnes of this heavy concrete were shredded in 2020. The concrete rubble is already being used as a new building material for campus renovations~\cite{DESYsustainableReport2022}.
\end{bestpractice}

\subsubsection{Plastic Waste}
Plastic is a versatile product; efficient, cheap, stable, and infinitely mouldable.  Its unique properties have resulted in our increasing reliance on it over decades, making it very hard to live without. 

GHG emissions from production, use and disposal of conventional (fossil-fuel-based) plastics is significant, and growing.  It is expected to increase to $\sim$2.1 \acrshort{tco2e} per capita by 2040, accounting for almost 20\% of the global carbon budget~\cite{UNPollutionSolution}.  Less than one tenth of plastic waste produced to date has been recycled, with the remainder being lost to the environment, landfilled or incinerated~\cite{UNPollutionSolution}.  This plastic waste contaminates the natural environment.  It is slow to degrade (biodegradable plastics included), releasing potentially harmful chemicals into the environment in the process, eventually breaking down to micro- and nano-plastic particles that infiltrate the food chain.  Microplastics are also found in waste water, due to washing of synthetic textiles, and in cosmetic and personal care products~\cite{Microplastics}.  Although the full risks of ingesting microplastics are still under investigation, ingestion of the chemicals used in their production are known to provoke serious health conditions, ranging from neurodevelopmental to hormonal and respiratory disorders~\cite{UNPollutionSolution}.

A ban on single-use plastics, and plastics that are not at least 99\% recyclable both within the food industry and outside it would greatly limit microplastic pollution. 

\subsubsection{Conference Waste\label{sec:ConferenceWaste}}

The amount of waste generated at conferences can be significantly reduced by replacing printed timetables and welcome packs with a well-designed conference app, and keeping conference gifts digital (\eg e-vouchers/discounts for local restaurants or activities).    Sustainable stationery should be distributed on a need-only basis, and banners, posters and name tags made plastic-free and reusable.  For waste-minimising initiatives implemented at the plastic-free 2019 conference of the Australian Marine Society, see~\bpref{bp:PlasticFreeConf}.  For sustainability concerns in conference catering, see Section~\ref{sec:CateringTableware}.

\begin{bestpractice}[Plastic-free 2019 conference of the Australian Marine Sciences Association\\
{\noindent\footnotesize Edited excerpt from Ref.~\cite{PlasticFreeConf}.}\label{bp:PlasticFreeConf}]{Plastic-free conference, Australian Marine Sciences Association}%
\noindent In response to the growing problem of plastic pollution, the Australian Marine Sciences Association undertook to make their 2019 conference 100\% plastic free. The specific measures that they implemented for their roughly 600 delegates included:
    \begin{itemize}
        \item plastic-free cardboard name badges with bamboo lanyards and metal clips
        \item complimentary fabric tote bags with conference logo
        \item no printed envelopes for registration packs, no printed conference abstracts
        \item necessary printing done on sustainably sourced paper, using a solar-powered printer
        \item sustainably sourced pencils instead of pens, with sharpening stations provided
        \item no packaged sweets
        \item delegates asked to bring reusable water bottles, or pre-register to buy them at the conference
        \item water jugs with glassware provided at back of each presentation room
        \item reusable, washable plates, cups silverware and glassware for all meal and coffee breaks
        \item vegetarian catering for tea breaks
    \end{itemize}
    These measures were implemented without affecting the budget, although some solutions reportedly took significant planning and forethought, and clear communication with the event organiser and providers.
\end{bestpractice}

\subsubsection{Catering Tableware}
\label{sec:CateringTableware}
A life cycle analysis (see Sections \ref{sec:Technology} and \ref{subsec:lca}) by the UN Environment Programme concludes that ``reusable tableware consistently outperforms single-use tableware in all the studies and across all environmental impact categories (with water use being the exception, because of washing). The case for reusable tableware is strengthened in countries where renewable energy makes up a high proportion of the grid mix and where end-of-life treatment options are not well developed''~\cite{UNEP2021}. 

In outdoor or remote environments or ‘pop-up’ events with no fixed catering facilities, where reusable tableware is impractical, single-use biodegradable tableware is preferable to other single-use tableware if it is industrially composted mixed in with food waste~\cite{UNEP2021}.\footnote{Industrial composting of household food waste is currently not the norm in most geographical locations within the USA~\cite{EPAWasteFoodMgmt}, and many existing industrial composters do not accept biodegradable plastic waste~\cite{BioplasticsAtIndustrialSites}.} 

Unlike emissions due to reusables, which are dominated by their use phase due to repeated washing, the  main impact of biodegradable tableware is due to its production. For conventional plastic, a significant role is also played by end-of-life management.  Quantitative analysis of their relative emissions is thus strongly dependent on assumptions about manufacture and disposal, including the material demand.  On the practical side, to minimise this impact when planning conference catering one should always choose the lightest-weight disposable tableware fit for purpose, preferably manufactured in a country with a significant proportion of renewables in its energy and electricity mix.  For a comparison of emissions due to different choices of disposable catering tableware for pop-up catering, see~\csref{cs:conferencetableware}.

\begin{casestudy}[Comparing tableware for pop-up catering\label{cs:conferencetableware}]{Comparing tableware for pop-up catering}%
\noindent The production and disposal of single-use tableware has a significant impact across all environmental factors, including acidification, eutrophication, human health, land use and water depletion.  However, we will focus here only on its GHG emissions.

    We will consider for benchmarking purposes a large-scale conference with 1,000 attendees and informal lunchtime catering (\ie with no dishwashing capability), and compare the life cycle emissions due to tableware made from conventional plastics, which are disposed of by a combination of incineration and landfill according to the European average (presumed food remnants making them unsuitable for recycling), and from biodegradable bioplastics, which are industrially composted along with the food remnants.

    We assume each set of tableware consists of a dinner plate and cup, a knife and fork, and a paper napkin and tray mat, all manufactured to the same size and thickness, but with possible differences in weight due to their respective material densities.  A full list of assumptions and details of the analysis can be found in the original article~\cite{Fieschi2018}.   

    The total emissions for 1,000 sets of conventional polystyrene tableware is 221 kg \acrshort{co2e}, as compared with 109 kg \CdOe\ for the biodegradable bioplastic tableware, a saving of 112~kg \CdOe, around the emissions of a flight from Paris to Geneva.  
    For the purposes of comparison, we include here the emissions cost of 1,000 dishes and cups from a 2015 study by Italian plastics company Pro.mo~\cite{PROMO2015}.  They put the total emissions due to reusables at 26 kg \CdOe, with the emissions due to conventional plastic dishes and cups (polypropylene in this case) at 79 kg \CdOe.
Note that these figures are specific to the electricity and energy mix of the European market, which has a large impact on the dominant emissions in all cases.

\end{casestudy}

\RaggedRight
\sloppy
\newpage


\section{Outlook}
\label{sec:Outlook}

Since this document was conceived at the first Sustainable HEP Workshop~\cite{SustHEP2021} in June 2021, its first release in June 2023, and its publication in March 2025, developments have been rapid. Assessments have continued of the severity and extent of changes to the global climate, of extremes of weather, of damage to the world's delicate ecosystems, and of the manifest climate injustices being wreaked across the globe, all with increasing alarm.

In the same period, however, positive shifts within the HECAP+ community have continued:\ The Sustainable HEP Workshops are heading for their sixth edition, with the 2nd~\cite{SustHEP2022}, 3rd~\cite{SustHEP2024}, 4th~\cite{SustHEP2025} and 5th~\cite{SustHEP2026} International Workshops on Sustainable High Energy Physics (Sustainable HEP) taking place online from 5 to 7 September 2022, 10 to 12 June 2024, 12 to 15 May 2025, and 8 to 10 July 2026. The most recent instance of the International Conference in High Energy Physics (ICHEP) series, hosted in Prague in July 2024, included a parallel track and plenary talk on sustainability for the first time~\cite{ICHEP2024}, and the track will continue at ICHEP 2026 in Sao Paulo, Brazil~\cite{ICHEP2026}. The labos1point5 programme featured in \csref{BP:l1p5} has significantly expanded its provision of seminar programmes, resources and tools (with modules to account for the per-institution share of large research infrastructure, including CERN), all aimed at bringing about the changes needed to secure a sustainable future for basic research. Additional programmes have emerged, including SPARKHub~\cite{SPARKHub} and NetDrive~\cite{NetDrive}. Life cycle assessments are being more widely implemented, e.g., the CLIC and ILC assessment conducted by CERN and KEK in collaboration with ARUP~\cite{CLIC_ILC_LCA}, and attention continues to be drawn to the sustainability of future collider programmes~\cite{Bloom:2024ixe}, with life-cycle studies for the proposed Future Circular Collider (FCC)~\cite{FCCLCAZenodo, FCCFaesibilityStudy} now published. Guidance has been collated on sustainable accelerator design~\cite{Wakeling:2025uqc}, efforts to develop detector cooling systems that do not rely on high-GWP gases continue~\cite{CMS:2025lht}, and focuses on sustainable research computing have increased~\cite{ATLAS:2025sgg}, including emphases on data sharing~\cite{LHCREIWGOpenEventGenerationTaskForce:2026dvu}. The most recent update to the European Strategy in Particle Physics has seen significant attention given to environmental sustainability~\cite{ECFAGuidance}, across grassroots~\cite{Boisvert:2025ayi}, national and consortium inputs (including from the astroparticle theory community~\cite{EuCAPTRoadmap}), and the mandated sustainability working group of the Laboratory Directors Group~\cite{LDGWorkingGroup:2025xhf, Bloise:2025buj}. This has seen an expanded and strengthened commitment to considering and limiting the negative impacts of future particle physics projects in the deliberation document~\cite{European:2957411} and recommendations~\cite{CERN-ESU-2025-002}. The Rubin Observatory has been awarded funding for an initial solar power generation system that may provide up to 40\% of energy needs at the summit site~\cite{RubinSolar}, and the newly proposed TAMBO neutrino observatory aims to achieve socially and environmentally responsible siting, collaborating directly with local social scientists~\cite{TAMBO2026}. ESA's Annual Space Environment Reports (see Ref.~\cite{ESASpaceReport} for the 2025 report) continue to extend the reach of environmental considerations into Earth orbit.  Efforts in astronomy predate those of this reflection document, driven in part by Astronomers for Planet Earth~\cite{A4E}, and recent focuses include research infrastructure~\cite{knodlseder2025environmentalimpactsastronomicalresearch, freour2025visiongroundbasedastronomy2030s} and the wider role that astronomers can play in addressing the climate crisis~\cite{10.1093/astrogeo/atae065} (with implications for scientists more generally).

Although beyond the scope of this document, it is important to emphasise that efforts have not been limited to HECAP+ research areas. Examples of similar reflections on research sustainability span analytical chemistry~\cite{ByrneKilkoyne}, biomedical science~\cite{Kelly2023}, computational science~\cite{Lannelongue2023}, earth and environmental science~\cite{10.1371/journal.pstr.0000135}, and medical imaging~\cite{10.1162/IMAG.a.36}, with proposals for reporting frameworks on research-related GHG emissions~\cite{Sweke2022}.

This progress is encouraging, but the central message of this document remains the same:
\begin{quotation}
{\bfseries Assessing, reporting on, defining targets for, and undertaking coordinated efforts to limit our negative impacts on the world's climate and ecosystems must become an integral part of how we plan and undertake all aspects of our research.}
\end{quotation}
Achieving this relies on individual-, group- and institution-level actions, and concrete suggestions are made in this document. This is a call to reflect with humility on these suggestions. This is a call to action, and a call to consider the opportunities that reassessing the environmental sustainability of our work practices also offers for addressing systemic barriers to inclusivity and accessibility.

This document is, however, incomplete, and a number of important topics were neglected. These include the economics of basic research and their supporting institutions, rebound effects as they apply to the HECAP+ fields, the sociology of behaviour and cultural change, the wider systems and drivers within which research is embedded, and the direct positive impacts that research undertaken by members of the HECAP+ community, or facilities managed by them, has on climate, environmental, biological and social sciences. Moreover, at the original time of writing, generative artificial intelligence was not yet widely available, and attention now turns to the critical assessment of and mitigation against the environmental impacts of these new tools as their deployment in scientific research grows~\cite{bakhtiarifard2026positionneglectingsustainabilityai}. There is much work still to be done to cover these topics, to address remaining geographic imbalances, and to better represent voices from every region of the global scientific community.

It must therefore be understood that this and similar documents are only the beginning:\ The \ACR\ community, and the research commmunity more broadly, must come together to secure a sustainable future for our fields.

\newpage


\section*{Acknowledgments}
\label{sec:Acknowledgments}
\addcontentsline{toc}{section}{Acknowledgments}
\RaggedRight
\sloppy

This initiative was conceived at the workshop “Sustainable HEP”, which was organised by Niklas Beisert, Valerie Domcke, Astrid Eichhorn and Kai Schmitz, and hosted by CERN and held online 28 to 30 June 2021.

The authors acknowledge input received during:~the Sustainable HEP Mini-Workshop, 18 January 2022; the Sustainability \& Inclusion Panel Discussion at the 21st String Phenomenology Conference, hosted by the University of Liverpool, 4 to 8 July 2022; the Education and Outreach track of the International Conference on High Energy Physics 2022 (ICHEP2022), Bologna, 6 to 13 July 2022; the second Sustainable HEP workshop, 5 to 7 September 2022.

The authors thank Shehu Shuaibu Abdulassalam, James Alvey, Nicolas Arnaud, Gabriela Barenboim, Till Bargheer, Niklas Beisert, Gianfranco Bertone, Soumyajit Bhar, V\'{e}ronique Boisvert, Martin Breidenbach, Brendan Bullard, Philip Burrows, Micah Buuck, Samuel Calvet, Enrico Cennini, Yann Coadou, Nikola Crnkovi\'{c}, Valerie Domcke, Alessandro Emilio, Astrid Eichhorn, John Ellis, Daniel Errandonea, Stefan Fredenhagen, Elina Fuchs, Tetyana Galatyuk, Spencer Gessner, Gian Francesco Giudice, Clare Gratrex, Roberto Guida, Beate Heineman, Aaron Held, Ren\'{e}e Hlo\v{z}ek, Andreas Hoecker, Tomas Kasemets, Stavros Katsanevas, Yves Kemp, Jinsu Kim, Ben Krikler, Valerie Lang, Paul Laycock, Sharachchandra Lele, Roberto Lineros, Benno List, Jenny List, Daniel Maitre, Sudhir Malik, Eric Mieland, Beatrice Mandelli, Zachary Marshall, Pablo Mart\'{i}nez-Mirav\'{e}, Dimitris Ntounis, Chris Parkes, Kalpita Paul, Laurie Pederson, Michael Peskin, Ruth Pöttgen, Melissa Quinnan, Salvatore Rappoccio, Melissa Ridel, Arjen van Rijn, Rachel Rosten, Jan Rybizki, Filippo Sala, Wayne Salter, Quentin Salvi, Sabrina Schadegg, Kai Schmitz, Heather Sidman, Kathrin Schulz, Jason St John, Julia Steinberger, Lindsay Stringer, Swiss National Computing Centre, Caterina Vernieri, Denise V\"{o}lker, Tien-Tien Yu, and Sebastian Zell for invaluable input.

The authors thank Jessie Muir for producing the individual, group and institution icons for the recommendations, and Shreyasi Acahrya for helping in the preparation of the third arXiv version.

The work of JG was supported by the Agence Nationale de la Recherche (France), under grant ANR-22- CE31-0018 (AUTOTHERM). CG acknowledges funding via the SWIFT-HEP project (grant numbers
ST/V002562/1 and ST/V002627/1). The work of MK is supported by the Deutsche Forschungsgemeinschaft (DFG, German Research Foundation) under Germany’s Excellence Strategy – EXC 2121 “Quantum Universe” – 390833306. The work of KL was supported by the Science and Technology Facilities Council (STFC) [Grant No.~ST/W000547/1]; and by the European Research Council [Grant Agreement~GA715871].  The work of PM was support by a United Kingdom Research and Innovation (UKRI) Future Leaders Fellowship [Grant Nos.~MR/V021974/1 and~MR/V021974/2]; the Science and Technology Facilities Council (STFC) [Grant No.~ST/X00077X/1]; and a Nottingham Research Fellowship from the University of Nottingham. The work of KP was supported by a United Kingdom Research and Innovation (UKRI) Science and Technology Facilities Council (STFC) Ernest Rutherford Fellowship [Grant No.~ST/T004568/2]. The work of RS was supported by Polish NAWA Bekker program No. BPN/BEK/2021/1/00342. The work of MvdW was supported by the Science and Technology Facilities Council via LOFAR-UK [ST/V002406/1]. The work of PZ is supported by the \textit{Atracci\'on de Talento Investigador} program of the Comunidad de Madrid (Spain) No.~2022-T1/TIC-24024 and the Spanish \textit{Ministerio de Ciencia e Innovaci\'on} grant No.~PID2022-136510NB-C31; and the European Union Horizon 2020 research and innovation program, grant agreement Num.~824093 (STRONG-2020).  This project has received funding from the European Union’s Horizon 2020 research and innovation program under the Marie Skłodowska-Curie grant agreement No 891169.

\appendix

\RaggedRight
\sloppy
\clearpage


\section{Supplementary Data for \fref{fig:Intro-ComparativeEmissions}}
\label{sec:DataforFig1.4}

Tables \ref{tab:ComparativeEmissionsData} and \ref{tab:ComparativeEmissionsDenominator} contain the raw data that was used to produce \fref{fig:Intro-ComparativeEmissions}.  Each set of data was taken from a publicly available environmental report issued by (members of) the institution in question; the original documents are referenced below.  Institution acronyms are defined in \sref{sec:Introduction}.

Our approach differs from existing estimates of the \acrshort{ghg} footprint per researcher in the divisor used to compute this quantity.  We shared the emissions per resource equally by the total number using that resource, whether it be the total number of employees, or research staff, or in the case of large laboratories like CERN, the number of Users, rather than using the same divisor throughout.  For instance, while we divide the commuting emissions for each institute by the total number of employees, we assign the business travel emissions solely to the research staff, assuming the support staff have negligible long-distance travel.  For concreteness, we have colour-coded the per-researcher estimates in Table \ref{tab:ComparativeEmissionsData} by the denominators used in their computation, with the colour key provided in Table \ref{tab:ComparativeEmissionsDenominator}. 

\begin{table*}[ht]
\centering
\ra{1.05}
{\setlength\tabcolsep{2pt}
\footnotesize
\begin{tabular}{>{\kern-\tabcolsep}ccccccccccccc<{\kern-\tabcolsep}}\toprule
\multirow{3}{*}{Sector} & \multicolumn{10}{c}{Emissions (\tCdOe)}\\
\cmidrule{2-13}
& \multicolumn{2}{c}{CERN } &  \multicolumn{2}{c}{MPIA } &
\multicolumn{2}{c}{ETHZ DPHYS}& \multicolumn{2}{c}{Nikhef}& \multicolumn{2}{c}{FNAL} & \multicolumn{2}{c}{LANL}\\
& Inst.  & Res.\ & Inst. & Res.\ & Inst. & Res.\ & Inst. & Res.\ & Inst. & Res.\ & Inst. & Res.\ \\
    \midrule
Scope 1 (direct) & 78,169 & \cellcolor{Pythonblue!30}4.9 & 446 & 1.4 & 0 & 0 & 150 & \cellcolor{Pythonblue!30}0.4 & 326 & 0.2 & 371,892& 31.7\\
Scope 2 (indirect) & 10,672& \cellcolor{Pythonblue!30}0.7 & 779 & \cellcolor{Pythongreen!20}5.2 & 570\footnote{This corresponds to the total ETHZ Scope 2 emissions rescaled by the fraction of employees working in the Department of Physics.} & 0.9 & 0 & 0 & 143,687\footnote{\label{REC}Includes purchased renewable energy credits.} & \cellcolor{Pythonblue!30} 29.4 & \multicolumn{2}{c}{(scopes 1+2)}\\
Travel (business) & 3,330 & \cellcolor{Pythongreen!20}0.9 & 1,280 & \cellcolor{Pythongreen!20}8.5 & 1,449 & \cellcolor{Pythongreen!20}3.2 & 785 & \cellcolor{Pythongreen!20}3.3 & 2,658 & \cellcolor{Pythongreen!20}2.3 &  & \\
Travel (commuting) & 5,836 & 1.1 & 139 & 0.4 & 1,700 & 0.2 & 146 & 0.7 & 5,393 & 2.9 & & \\
Food & 738 & 0.1 & 16 & 0.1 & & & & &   &  & 77,586 & 6.6\\
Procurement & 178,010 &\cellcolor{Pythonblue!30}11.2 & 64 & \cellcolor{Pythongreen!20}0.4 & 497 & \cellcolor{Pythongreen!20}0.3 &  & & & &\multicolumn{2}{c}{(total scope 3)}\\
Waste treatment & 2,194& 0.4 & & & & &  & & 259 & 0.1 & &\\
Other & 2,619 & 0.5 & & & & & & & 6,158\textsuperscript{\ref{REC}} & \cellcolor{Pythonblue!30}1.3 & & \\ 
\bottomrule
\end{tabular}}\\
\scriptsize{CERN data for 2019 is taken from Refs.~\cite{Environment:2737239,CERN-HR-STAFF-STAT-2019,CERN:2723123,CERNTownHall}, MPIA data for 2019 from Ref.~\cite{Jahnke2020}, ETH Zurich Department of Physics (DPHYS) data from 2018 taken from Ref.~\cite{Beisert2020}, NIKHEF data from 2019 from Ref.~\cite{Nikhef,NikhefStats}, Fermilab data from Refs.~\cite{FermilabEnvReport2019,FNALPrivate,FNALPrivate2}, and Los Alamos National Laboratory data from Refs.~\cite{LANLEmissions,LANLEmployeeDemo}. Scope 3 estimates incomplete for all but CERN. Almost all of Nikhef's emissions estimates include employees of other universities who work at Nikhef.}
\caption[Average annual GHG emission data for \ACR\ institutions]{Average annual GHG emissions (\acrshort{tco2e}) for researchers at various \ACR\ institutions, by sector.  Colour-coding corresponds to the key in Table~\ref{tab:ComparativeEmissionsDenominator} for the staff type that was used in the divisor to compute the emissions per researcher. The abbreviations `Inst.' and `Res.' are used to indicate institute and per-researcher emissions, respectively. 
 Blank cells indicate missing data.\label{tab:ComparativeEmissionsData}} 
\end{table*}

\begin{table*}[ht]
{\footnotesize
\ra{1.05}
\centering
\begin{tabular}{>{\kern-\tabcolsep}ccccccc<{\kern-\tabcolsep}}
\toprule
Employee type& CERN & MPIA & ETHZ DPHYS & Nikhef & FNAL & LANL\\
\midrule
Staff & 5,235 & 320 & 630 & 210 & 1829 & 11,743\footnote{Data as of 30 September 2018, not including contractors and guard force.}\\
\cellcolor{Pythongreen!20}Research staff &\cellcolor{Pythongreen!20}3,509 &\cellcolor{Pythongreen!20}150 &\cellcolor{Pythongreen!20}450 &\cellcolor{Pythongreen!20}235 &\cellcolor{Pythongreen!20} 1162\footnote{Includes technical staff.} & \cellcolor{Pythongreen!20}\\
\cellcolor{Pythonblue!30}`Users' + research staff & \cellcolor{Pythonblue!30}15,937& \cellcolor{Pythonblue!30}&\cellcolor{Pythonblue!30} & \cellcolor{Pythonblue!30}350& \cellcolor{Pythonblue!30}4,887 &\cellcolor{Pythonblue!30} \\
\bottomrule
\end{tabular}}\\
\scriptsize{Employment statistics: CERN~\cite{CERN-HR-STAFF-STAT-2019}, MPIA~\cite{Jahnke2020}, ETHZ DPHYS~\cite{Beisert2020}, Nikhef~\cite{NikhefStats}, FNAL~\cite{FNALPrivate2}, LANL~\cite{LANLEmployeeDemo}}
\caption[Employee statistics for \ACR\ institutions]{Institute employee statistics, colour-coded by type.  The same colour codes are used in the researcher numbers above to indicate which staff statistics were used as the divisor in each case.  Nikhef emissions statistics in some cases include staff employed by external institutions who work at Nikhef, c.f.~`research staff' and `User' numbers above.\label{tab:ComparativeEmissionsDenominator}}
\end{table*}

\RaggedRight
\sloppy
\newpage


\section{Forewords}
\label{sec:Forewords}
\addcontentsline{toc}{section}{Forewords}

\textcolor{Pythongreen}{\rule{2cm}{3pt}}

\begin{quotation}
What should we do to limit the negative environmental impacts arising from our scientific community?
 
Productive discussions on this crucial issue are often hindered by a number of roadblocks. Climate change scepticism is fortunately not very frequent in our community, but the same cannot be said, for instance, of \textit{climate change whataboutism}  ---  ``What about other communities who produce way more greenhouse gas emissions than us? Why should we make efforts, when others are not doing enough?''. 

But even for those of us who recognize the importance of doing something, a general lack of data and detailed information often prevents us from aligning our actions with our ethical values --- ``Does my Carbon footprint increase more by flying to conferences a couple of times a year or by eating red meat a couple of times a week? By computing or by commuting?''.
 
This timely and thought-provoking paper, arising from the grassroots initiative \textit{Striving
towards Environmental Sustainability in High Energy Physics, Cosmology and
Astroparticle Physics}, provides a much welcome reflection on how to remove these and other common roadblocks towards sustainability, and how to put in place concrete actions to limit our environmental impact.
 
The reader will find here a synthesis of current data, best practices, and research in climate science and sustainability, and a set of concrete recommendations that clearly show how we can empower individuals and the broader community to take direct action and responsibility for mitigating climate change.
 
I encourage every member of our community to reflect upon the contents of this document and to actively engage in the ongoing dialogue surrounding this crucial topic. I hope it will spark broader change and promote a culture of sustainability in our community and beyond.
\end{quotation}
\begin{flushleft}
Prof.\ Gianfranco Bertone\\
University of Amsterdam\\
Director, European Consortium for Astroparticle Theory
\end{flushleft}

\textcolor{Pythongreen}{\rule{2cm}{3pt}}

\begin{quotation}
The conclusions from the IPCC are clear: global heating and climate change are an existential threat to human civilization. All aspects of society need to follow the recommended guidelines and eliminate greenhouse gas emissions as quickly as possible. The research fields of particle physics, cosmology and astroparticle physics have a role to play in this great transition. Not only are the emissions associated with those fields relatively large compared to other areas of daily life, but as researchers and scientists we understand the science and we can use our creativity and ingenuity in finding solutions. By placing sustainability at the forefront of our scientific approaches, we will provide guidance to other research fields and convey to society the importance of this topic. It is our duty to act now such that future generations can also enjoy the wonders of exploring the secrets of the universe. We used to pride ourselves with being great innovators at the cutting edge of technology. Let’s be ambitious again and help tackle all of the challenges associated with a net zero economy. In the process we will improve our health and make our research fields more diverse and accessible to all regions of the Earth. As scientists ourselves we can not ignore the science, there is a climate emergency and everyone needs to play their part.
\end{quotation}
\begin{flushleft}
Prof V\'{e}ronique Boisvert\\
Centre for Particle Physics Group Leader at Royal Holloway, University of London\\
Founder and Co-coordinator of the ATLAS Sustainability Forum\\
Co-coordinator of the Snowmass 2021 Topical Group:\ Environmental and Societal Impacts
\end{flushleft}

\textcolor{Pythongreen}{\rule{2cm}{3pt}}

\begin{quotation}
The climate crisis is one of the most pressing problems facing humanity today:\ the long-term survival of our species, and countless others, will be impacted critically by when and how we choose to address it. This is not an issue we can ignore or put off; action needs to be taken in the very near-term --- the next few decades --- if we are to avoid a critical rise in mean global temperature that is predicted to result in major changes to climate and weather patterns, the first signs of which we are already experiencing. Of course tackling such a monumental issue requires global collaboration and coordinated action by world governments.  But as individuals, as well as via our memberships of teams, groups, institutes, laboratories and research organisations, each of us can, and must, take our individual responsibility for helping to ensure the vitality of our planetary environment and sustainability of its resources.

For those of us working in the scientific fields represented, this report provides a clarion call for action on sustainability. The report both draws attention to the environmental impacts of pursuing our scientific endeavours in a ‘business-as-usual’ fashion and, moreover, highlights actions that we can take to reduce significantly these impacts. In many cases these involve straightforward changes to behaviour and practice that will yield benefits almost immediately. In other cases more effort will be required and the benefits may be realised later. The point is: we owe it to ourselves to take action, and we can start now. This report is timely and it contains many constructive recommendations; the findings are widely applicable within the broader scientific community, and well beyond.
\end{quotation}
\begin{flushleft}
Philip Burrows \\
Professor of Physics, University of Oxford\\
Director, the John Adams Institute for Accelerator Science
\end{flushleft}

\textcolor{Pythongreen}{\rule{2cm}{3pt}}

\begin{quotation}
In the context of the climate crisis, the scientific community is part of the solution --- through research, teaching and evidence-based policy advice. However --- and this aspect has traditionally been much less in focus --- the scientific community is also part of the problem, through the emissions that it produces through its own operations.
This report illustrates starkly, how large that problem is:\ for some researchers, emissions caused by their work are not just somewhat higher than the current average global per-capita emissions or the per-capita budget to 2050, they are roughly an order of magnitude higher (see Fig.~\ref{fig:Intro-ComparativeEmissions}, p.~\pageref{fig:Intro-ComparativeEmissions}). In the face of such data, inaction can cost the HECAP+ community i) the credibility and trust it enjoys both from policy-makers and society at large, ii) the enthusiasm for HECAP+ topics that turns young students (many of whom care passionately about solving the climate crises) into next generation’s researchers and iii) even its research freedom, because it is likely that societal pressure will increase as the climate crisis progresses and policy makers will ultimately step in to regulate carbon-intensive sectors much more strictly.

Yet, there is ground for hope. As the report illustrates, there are already best practise examples that others in the HECAP+ community can learn from. Beyond the existing examples, the report provides actionable recommendations at three critical levels:\ the level of the individual, the level of research groups and the level of institutions. These three levels are necessary to achieve a change in culture towards a climate sustainable research community. Such a change in culture will happen, if individual behaviours and framework conditions (often set at the institutional level) that determine norms and incentives, change. The recommendations in this report provide first steps that the HECAP+ community can take to achieve this cultural change. Everyone in the community should read about, reflect on and, wherever possible, implement such steps.
\end{quotation}
\begin{flushleft}
Prof. Dr. Astrid Eichhorn\\
Professor in Theoretical Physics at the University of Southern Denmark\\
Chair of the ALLEA (European Federation of Academies) Working Group on Climate Sustainability in the Academic System
\end{flushleft}

\textcolor{Pythongreen}{\rule{2cm}{3pt}}

\begin{quotation}
All of us have roles to play in combating the climate crisis that is bearing down upon us and the generations to come. Some of these roles are individual, such as our personal lifestyle choices, whereas others are collective, linked to our activities in society. As scientists studying fundamental physical laws, astroparticle physics and cosmology, our activities burden us with particular responsibilities. For example, the scales of our research facilities imply that we consume orders of magnitude more resources than is sustainable for the bulk of humanity, so we must strive to research as efficiently as possible. This implies minimising the wall-plug power and other resources consumed by our accelerators, experiments and data analyses. Moreover, the international nature of our research teams implies that we travel more than most, so we should strive to travel as sustainably as possible, e.g., by train, and as little as possible, e.g., by using teleconferencing tools such as Zoom or Teams. On the other hand, our research can provide humanity with valuable tools for living more sustainably. For example, the World-Wide Web, which CERN placed in the public domain 30 years ago, has enabled information to be shared and the planet’s business to be conducted more efficiently and therefore sustainably. Moreover, instruments that we develop can provide tools for monitoring and possibly mitigating the effects of climate change. Finally, through our international contacts we can both learn and disseminate best practice. Individually and collectively we have unique responsibilities and opportunities to follow the right paths: let us take them.
\end{quotation}
\begin{flushleft}
John Ellis\\
James Clerk Maxwell Professor of Theoretical Physics,\\
King’s College London, formerly at CERN
\end{flushleft}

\textcolor{Pythongreen}{\rule{2cm}{3pt}}

\begin{quotation}
Climate change is the biggest challenge that humanity is facing today, and science is called on to help society in identifying measures to avert the environmental catastrophe that looms over our future. While science can provide solutions, we must recognise that scientific research is also part of the problem: our projects, infrastructures, computing facilities and work habits are energy-intensive and waste producers. More than ever, it is mandatory for scientists today to assess carefully the environmental impact of their activities and prove to society that even the most advanced scientific projects can be carried out in environmentally sustainable ways. This document lucidly presents the challenges, substantiating them with facts and data, and paves the way towards realistic and effective solutions. It testifies to the unrelenting commitment to societal and environmental problems deeply felt within the scientific community.
\end{quotation}
\begin{flushleft}
Gian Francesco Giudice\\
Head of the CERN Department for Theoretical Physics\\
\end{flushleft}

\textcolor{Pythongreen}{\rule{2cm}{3pt}}

\begin{quotation}
In the past century the ever increasing resource demands of humans have a devastating impact on the climate of our planet. The resulting heat waves, droughts, strong rain falls, violent storms, melting ice and rising sea levels are posing an existential threat to many people world-wide already today, and even more in the future. This situation demands action from all of us individually, and as groups and institutions, to do whatever we can to reduce (or ideally eliminate) the emission of \CdO  (equivalent) gases, and more generally to preserve the resources of the planet. This report illustrates well that scientists working in the fields of high energy physics, cosmology, astroparticle physics, hadron and nuclear physics are using on average more resources than what will be acceptable, and action is needed immediately. Many recommendations are proposed for individuals, groups and/or institutions. Some are rather easy while others are challenging and require core habits to change. Given the very international nature of the research performed in these fields, there is a large potential to propagate the actions to 100s of institutions in countries of all continents and thereby increasing the impact further. I very much hope this document will help us embark on the right path.
\end{quotation}
\begin{flushleft}
Prof.~Dr.~Beate Heinemann\\
Director for Particle Physics at DESY, Professor of Physics at Albert-Ludwigs-Universitaet Freiburg\\
former Deputy Spokesperson for the ATLAS Collaboration at CERN
\end{flushleft}

\textcolor{Pythongreen}{\rule{2cm}{3pt}}

\begin{quotation}
Many of us who study astrophysics and cosmology do so out of a sense of awe and wonder at the universe. That same sense of wonder should compel us to consider the impact of our work on our planet. We work in areas that are driven by data, and the data on climate change shows clearly how humans continue to impact the climate.  This extensive report presents a summary of the ways in which our work contributes to increases in emissions of greenhouse gases, collectively through our international projects and individually through personal choices. It is a sobering look at our impact, and provides recommendations for how we can effect change collectively within our research communities, collaborations and institutions. 

International collaboration drives progress in the large, complex projects that we undertake to unravel the secrets of the cosmos. These projects often involve significant infrastructure investment and require a large computing budget. On the largest scale, the report challenges us to examine how we can reduce the environmental impact of the projects as a whole. This will require from us a renewed prioritisation of energy-efficiency, and strategic thinking to balance our research needs with these time-critical actions.  At an individual level, it suggests ways to reduce impact on the climate by considering how we can reduce our international travel, while considering solutions that are inclusive of all members in our collaborations, regardless of geographic location. This will again require creativity — but I am confident that as researchers who have been trained in solving difficult problems we can rise to the challenge.
\end{quotation}
\begin{flushleft}
Ren\'{e}e Hlo\v{z}ek\\
Associate Professor, Dunlap Institute and the Department of Astronomy and Astrophysics at the University of Toronto\\
Spokesperson-elect of the LSST Dark Energy Science Collaboration (DESC)
\end{flushleft}

\textcolor{Pythongreen}{\rule{2cm}{3pt}}

\begin{quotation}
The challenges posed by the human-made climate crisis, the largest, most rapidly developing and pervasive threat to the natural environment and societies the world has yet experienced, are enormous. They are created by ever-growing human consumption, bolstered by rising energy production and its inefficient use and waste. This energy is predominantly generated by the massive combustion of fossil fuels, predictably leading to the greenhouse effect we are experiencing. While fossil fuels are the dominant source of greenhouse gases, direct release of these gases also contributes to global climate change. As scientists we have the responsibility to transparently expose the extent to which we contribute to the problem, as discussed in this document, and help develop solutions. Among the biggest challenges for current experiments is the emission of greenhouse gases by detectors developed and built decades ago, in some cases using gases that were at that time thought to be environmentally friendly alternatives to even more harmful gases. Energy efficiency throughout all our instruments, including considerations of their embodied carbon, is another important challenge. All these challenges require and receive our immediate attention. Future high-energy colliders and experiments will need to be 'green': they will have to rely on decarbonated energy sources and employ detector technologies that avoid the use of greenhouse gases. I am confident that our field will meet these challenges and contribute to building a sustainable future.
\end{quotation}
\begin{flushleft}
Dr.~Andreas Hoecker\\
CERN Research Physicist \\
Spokesperson, ATLAS Collaboration
\end{flushleft}

\textcolor{Pythongreen}{\rule{2cm}{3pt}}

\begin{quotation}
Climate change does not necessarily threaten the survival of our planet in the bigger picture of the solar system and the universe; rather, it threatens our own survival and the survival of the bio- and eco-spheres that we rely on for our subsistence in this cosmos. We have no alternative planet which will provide for us, so we better not make our current one uninhabitable.
Climate research has been clear on the effects of climate change, and slowly the rising sea levels, the disappearance of glaciers, the frequency of hundred year floods, the storms, droughts, and heat waves beat the message home: climate change is happening, human action is the cause, and we better counter-act it immediately.
As fellow scientists from the areas of high energy physics, cosmology, and astroparticle physics, we are trained to understand and consider scientific results in our daily work. We know how to interpret statistics and draw conclusions from data. 
This document is a start to take the conclusions from climate research seriously and put them into action in our own fields of research. Collecting and reflecting on available results, together with recommendations for the implementation --- from easy to hard --- is an important first step, but only a first step. Let's use this document and get started in transforming our field of research into a sustainable field of research --- for the benefit of our planet and our own futures. The data has been clear for a while, now is the time to act on it.
\end{quotation}
\begin{flushleft}
Dr.~Valerie Lang \\
Chair of the management board of the young High Energy Physicists (yHEP) association, Germany\\
Researcher at the Albert-Ludwigs-Universitaet Freiburg, Germany\\Member of the ATLAS Collaboration at CERN
\end{flushleft}

\textcolor{Pythongreen}{\rule{2cm}{3pt}}

\begin{quotation}
The Intergovernmental Panel on Climate Change (IPCC) Working Group II in their Sixth Assessment Report (2022) underscored that “the science is clear. Any further delay in concerted global action [on climate change] will miss a brief and rapidly closing window to secure a liveable future”. We are also rapidly advancing towards the 2030 deadline to achieve the 17 Sustainable Development Goals: an international call to end poverty, protect the Earth and deliver peace and prosperity for all.

This report, which calls for climate and sustainability actions within the international scientific community, is both timely and welcome. It provides clear, actionable recommendations that can contribute to the collective effort to deliver positive changes across six key areas: i) Computing, ii) Energy, iii) Food, iv) Mobility, v) Research Infrastructure and Technology and vi) Resources and Waste. Ensuring both our science and the way we live are as sustainable as possible is an incredibly important undertaking, and scientists need to lead the way, and “walk the [sustainability] talk”. The global reach of this report offers a substantial opportunity to reorient the scientific community along a more sustainable trajectory. I encourage all who read it to commit to delivering its aspirations and best practices. 
\end{quotation}
\begin{flushleft}
Prof. Dr. Lindsay C. Stringer\\
Professor in Environment and Development at the University of York, UK\\ 
Director of the York Environmental Sustainability Institute, University of York, UK\\
IPCC Scientist (Working Group II)
\end{flushleft}

\textcolor{Pythongreen}{\rule{2cm}{3pt}}

\clearpage
\RaggedRight
\sloppy
\phantomsection
\addcontentsline{toc}{section}{Bibliography}
\bibliographystyle{IEEEtran-SHEP} 
\bibliography{IEEEabrv,HECAPdoc}

\newpage
\RaggedRight
\sloppy
\printglossary[title=Acronyms and Abbreviations, toctitle=Acronyms and Abbreviations, type=\acronymtype]

\newpage
\phantomsection 
\addcontentsline{toc}{section}{Best Practices}
\label{list:bestpractice}
\listofBestpractice
\makeatletter
\let\BP@pre\relax
\let\BP@post\relax
\makeatother
\newpage

\newpage
\phantomsection
\addcontentsline{toc}{section}{Case Studies}
\label{list:casestudy}
\listofCasestudy
\makeatletter
\let\CST@pre\relax
\let\CST@post\relax
\makeatother

\newpage
\renewcommand\listfigurename{Figures}
\phantomsection
\addcontentsline{toc}{section}{\listfigurename}  
\listoffigures 

\newpage
\renewcommand\listtablename{Tables}
\phantomsection
\addcontentsline{toc}{section}{\listtablename} 
\listoftables
\newpage

\newpage

\newpage


\section*{Authors}
\phantomsection
\label{AuthorsList}
\addcontentsline{toc}{section}{Authors}
\renewcommand{\arraystretch}{1.5}
\noindent\begin{longtable*}{@{}m{0.3\textwidth} p{0.7\textwidth}}
Shankha Banerjee & CERN, Switzerland; Institute of Mathematical Sciences, Chennai, India\\ 

Thomas Y.\ Chen & Department of Computer Science, Columbia University, USA\\

Claire David & York University, ON, Canada; Fermi National Accelerator Laboratory, IL, USA; AIMS South Africa\\

Michael Düren & Center for International Development and Environmental Research, Justus Liebig University Giessen, Gießen, Germany\\ 

Harold Erbin & 
Université Paris-Saclay, CEA, CNRS, Institut de physique théorique, 91191 Gif-sur-Yvette, France;
Center for Theoretical Physics, Massachusetts Institute of Technology, Cambridge, MA 02139, USA;
NSF AI Institute for Artificial Intelligence and Fundamental Interactions\\ 

Jacopo Ghiglieri & SUBATECH, Nantes Université, IMT Atlantique, IN2P3/CNRS, 4 rue Alfred Kastler, La Chantrerie BP 20722, 44307 Nantes, France\\

Mandeep S. S. Gill & Minnesota Institute for Astrophysics,  University of Minnesota; Fermi National Accelerator Laboratory, IL, USA\\

L Glaser & University of Vienna, Boltzmanngasse 5, 1090 Vienna, Austria \\ 

Christian G\"utschow & Department of Physics \& Astronomy,
University College London,
Gower Street, WC1E 6BT, London, United Kingdom\\

Jack Joseph Hall & University of Leeds, Woodhouse Lane,
Leeds, West Yorkshire, LS2 9JT,
United Kingdom\\

Johannes Hampp & Center for International Development and Environmental Research, Justus Liebig University Gießen, Hesse, Germany\\ 

Patrick Koppenburg & Nikhef, National Institute for Subatomic Physics, Amsterdam, Netherlands\\

Matthias Koschnitzke & Deutsches Elektronen-Synchrotron DESY, Notkestr. 85, 22607 Hamburg, Germany; II. Institut für Theoretische Physik, Universität Hamburg, Luruper Chaussee 149, 22761 Hamburg, Germany \\

Kristin Lohwasser & School of Mathematical and Physical Science, Physics and Astronomy, University of Sheffield, Hounsfield Road, Sheffield S3 7RH, United Kingdom\\

Rakhi Mahbubani & Division of Theoretical Physics, Rudjer Boskovic Institute, Bijenicka cesta 54, 10000 Zagreb, Croatia\\

Viraf Mehta & Institut f\"{u}r Astrophysik, Georg-August Universit\"{a}t, Friedrich-Hund-Platz 1, G\"{o}ttingen, Germany\\

Peter Millington & Department of Physics and Astronomy, University of Manchester, Oxford Road, Manchester M13 9PL, United Kingdom\\

Ayan Paul & The Institute of Experiential AI, Northeastern University, 360 Huntington Ave, Boston MA 02115 USA; Channing Division of Network Medicine, Brigham and Women’s Hospital, Harvard Medical School, 188 Longwood Ave. Boston MA 02115, USA\\

Frauke Poblotzki & Deutsches Elektronen-Synchrotron DESY, Notkestr. 85, 22607 Hamburg, Germany \\

Karolos Potamianos & Department of Physics, University of Warwick, Coventry CV4 7AL, United Kingdom \\

Nikolina ~\v{S}ar\v{c}evi\'c &
School of Mathematics, Statistics and Physics, Newcastle University, United Kingdom \\ 

Prajval Shastri & Raman Research Institute, C. V. Raman Avenue, Bengaluru 560080, India; International Centre for Radio Astronomy Research, University of Western Australia, Crawley, WA 6009, Australia\\

Rajeev Singh & Center for Nuclear Theory, Department of Physics and Astronomy, Stony Brook University, Stony Brook, New York, 11794-3800, USA\\

Hannah Wakeling & John Adams Institute for Accelerator Science, Department of Physics, University of Oxford, Keble Road, Oxford OX1 3RH, United Kingdom; ISIS Neutron and Muon Source, Rutherford Appleton Laboratory, Fermi Avenue, Didcot OX11 0QX, United Kingdom\\

Rodney Walker & Fakultät für Physik, Ludwig-Maximilians-Universität München, München, Germany\\

Matthijs van der Wild & Centre for Extragalactic Astronomy, Department of Physics, Durham University, Durham DH1 3LE, United Kingdom\\ 

Pia Zurita & Departamento de F\'isica Te\'orica \& IPARCOS, Universidad Complutense de Madrid, E-28040 Madrid, Spain

\end{longtable*}
\renewcommand{\arraystretch}{1}

\newpage


\section*{Endorsers}
\phantomsection
\addcontentsline{toc}{section}{Endorsers}

To register to and view the list of endorsers, please visit  \href{https://indico.cern.ch/event/1293733/}{https://indico.cern.ch/event/1293733/}.  

To make an institutional endorsement, please make direct contact.  


\end{document}